\documentclass[fleqn,usenatbib]{mnras}

\usepackage[T1]{fontenc}
\usepackage{ae,aecompl}
\usepackage[percent]{overpic}
\usepackage{pdflscape}
\usepackage{rotating}
\usepackage{float}
\usepackage[final]{changes}

\usepackage{graphicx}	
\usepackage{amsmath}	
\usepackage{amssymb}	
\usepackage[switch]{lineno}
\usepackage{longtable}

\usepackage{newtxtext,newtxmath}

  {\def\@captype{table}}
  {}

\title[Search for Galactic PeVatrons with Clouds]{Using Interstellar Clouds to Search for Galactic PeVatrons: \\ Gamma-ray Signatures from Supernova Remnants}

\author[A. M. W. Mitchell et al.]{
A. M. W. Mitchell,$^{1,2,3}$\thanks{E-mail: alison.mw.mitchell@fau.de}
G. P. Rowell,$^{4}$ 
S. Celli, $^{5,6}$
S. Einecke $^{4}$ 
\\
$^{1}$Department of Physics, University of Zurich, CH-8057 Zurich, Switzerland \\
$^{2}$Department of Physics, ETH Zurich, CH-8093 Zurich, Switzerland \\
$^{3}$Now at Erlangen Centre for Astroparticle Physics, Friedrich-Alexander-Universit\"at Erlangen-N\"urnberg, D-91058 Erlangen, Germany \\
$^{4}$School of Physical Sciences, University of Adelaide, Adelaide, SA 5005, Australia\\
$^{5}$Dipartimento di Fisica dell'Universit\`a La Sapienza, P.~le Aldo Moro 2, 00185 Rome, Italy \\
$^{6}$Istituto Nazionale di Fisica Nucleare, Sezione di Roma, P.~le Aldo Moro 2, 00185, Rome, Italy \\
}

\date{Accepted XXX. Received YYY; in original form ZZZ}

\pubyear{2020}

\begin{document}
\label{firstpage}
\pagerange{\pageref{firstpage}--\pageref{lastpage}}
\maketitle

\begin{abstract}
Interstellar clouds can act as target material for hadronic cosmic rays; gamma rays subsequently produced through inelastic proton-proton collisions and spatially associated with such clouds can provide a key indicator of efficient particle acceleration. 
However, even in the case that particle acceleration proceeds up to PeV energies, the system of accelerator and nearby target material must fulfil a specific set of conditions in order to produce a detectable gamma-ray flux. In this study, we rigorously characterise the necessary properties of both cloud and accelerator. 
By using available Supernova Remnant (SNR) and interstellar cloud catalogues, we produce a ranked shortlist of the most promising target systems, those for which a detectable gamma-ray flux is predicted, in the case that particles are accelerated to PeV energies in a nearby SNR. We discuss detection prospects for future facilities including  CTA, LHAASO and SWGO; and compare our predictions with known gamma-ray sources. The four interstellar clouds with the brightest predicted fluxes $>$100 TeV identified by this model are located at (l,b) = (333.46,-0.31), (16.97,0.53), (110.43,1.89) and (336.73,-0.98).
These clouds are consistently bright under a range of model scenarios, including variation in the diffusion coefficient and particle spectrum. On average, a detectable gamma-ray flux is more likely for more massive clouds; systems with lower separation distance between the SNR and cloud; and for slightly older SNRs.  

\end{abstract}




\section{Introduction}
\label{sec:intro}

Understanding the origin of Galactic cosmic rays (CRs) continues to be a very active research area, with many open questions \citep{BlasiCRreview,GrenierCRreview}. The search for evidence of Galactic PeVatrons - astrophysical accelerators capable of achieving the CR knee at $\sim1$\,PeV ($10^{15}$eV) - is a major focus of several present day astrophysical experiments. SNRs have long been thought to provide the bulk of the CR flux at these energies, however the gamma-ray spectra observed from SNR to date typically cut-off at energies of a few tens of TeV. Cosmic rays at PeV energies, however, would produce $\sim$100\,TeV gamma rays. These CRs are sufficiently energetic to escape the SNR shock and travel through the interstellar medium (ISM). 

Interstellar clouds provide suitable targets for CR interactions generating pions (among other particles), that may subsequently produce a detectable gamma-ray flux through the decay of neutral pions. Similarly, a detectable neutrino flux may be produced through the decay of charged pions. Both gamma rays and neutrinos are neutral messengers, providing a direct indication of their origin. The arrival direction of charged cosmic rays typically does not indicate the direction of origin, due to deflection by interstellar magnetic fields. In this paper, we focus on the gamma-ray signatures left by SNRs on nearby clouds, a scenario that was first investigated in the search for PeVatrons by \cite{GabiciAharonianPeVatron07}. 

The amount of gamma-ray flux produced grows both with the mass of the cloud, corresponding to the amount of target material available for interactions, and with the CR over-density. A local over-density of CRs may be present due to CR production from a source in the vicinity of the cloud, such as an SNR; or due to a strong suppression of the diffusion coefficient within a specific region. 
Gamma-ray emission coincident with interstellar clouds is typically invoked as evidence of hadronic particle acceleration. Lack of detectable emission may either disfavour hadronic scenarios, or may be due to gamma-ray absorption in surrounding radiation fields, leading to opaque sources (see e.g. \citet{celli17}). However, it is not necessarily the case that a gamma-ray flux detectable by current instruments is produced, even if PeV CRs are present and the source is transparent to high energy radiation, as the flux is highly dependent on the specific properties of the interacting systems. 

Previous modelling of SNRs and clouds of target material have enabled hadronic emission to be confirmed through the characteristic pion-bump signature in several SNRs, including IC\,443, W\,44, W\,28 and W\,51C \citep{FermiPionSNR13,W51C16,W2818}. For this study, we adopt a baseline model adapted from \cite{AA96} and \cite{Kelner06}, to describe the evolution of the particle flux during transport through the ISM and the production of gamma rays upon reaching the cloud respectively. The model is described in detail in section \ref{sec:model}.

Using this model, we quantify which combinations of accelerator and cloud properties achieve detectable levels of gamma-ray flux under standard assumptions. 
We present this information through plots demonstrating the favourable phase space for interstellar clouds, in an analogous manner to the well-known phase space for astrophysical sites of CR acceleration \citep{Hillas84}. In particular, this can be indicated with curves of constant gamma-ray flux, such as the 50 hour sensitivity of the future Cherenkov Telescope Array (CTA) at a gamma-ray energy of 100\,TeV \citep{2019scta.book.....C}. 

CTA will outperform the current generation of Imaging Atmospheric Cherenkov Telescope (IACT) arrays, such as H.E.S.S., MAGIC and VERITAS, towards 100\,TeV; yet data from these experiments can already be used to place constraints on the flux above 10\,TeV \citep{HESSCrab2006,MagicCrab2014,VeritasCrab_2015ICRC}. Water Cherenkov Detector (WCD) based facilities, such as HAWC, provide a complementary view of the very high energy gamma-ray sky - with improved sensitivity around 100\,TeV at the cost of a degraded angular resolution \citep{HAWCcrab17}. Operating as wide field-of-view survey instruments, WCD based experiments cover large areas of the sky reducing the bias in source detection. The future facilities LHAASO (already producing first results in a partial configuration) and SWGO (planned for the Southern hemisphere) may detect multiple such interstellar clouds illuminated by nearby SNRs \citep{LHAASOwhitepaper,sgsowhitepaper}.

Using current catalogues of known SNRs and interstellar clouds, we explore which combinations of accelerators and clouds are the most promising targets to look for evidence of PeVatron activity with current instruments, such as H.E.S.S. or HAWC, and for future detectability with CTA, LHAASO or SWGO. By applying our model to different plausible combinations of SNRs and nearby interstellar clouds, we produce a ranked shortlist of the most promising candidate targets.

\section{Model}
\label{sec:model}

\begin{figure}
\includegraphics[trim=2cm 2cm 2cm 2cm,clip,width=\columnwidth]{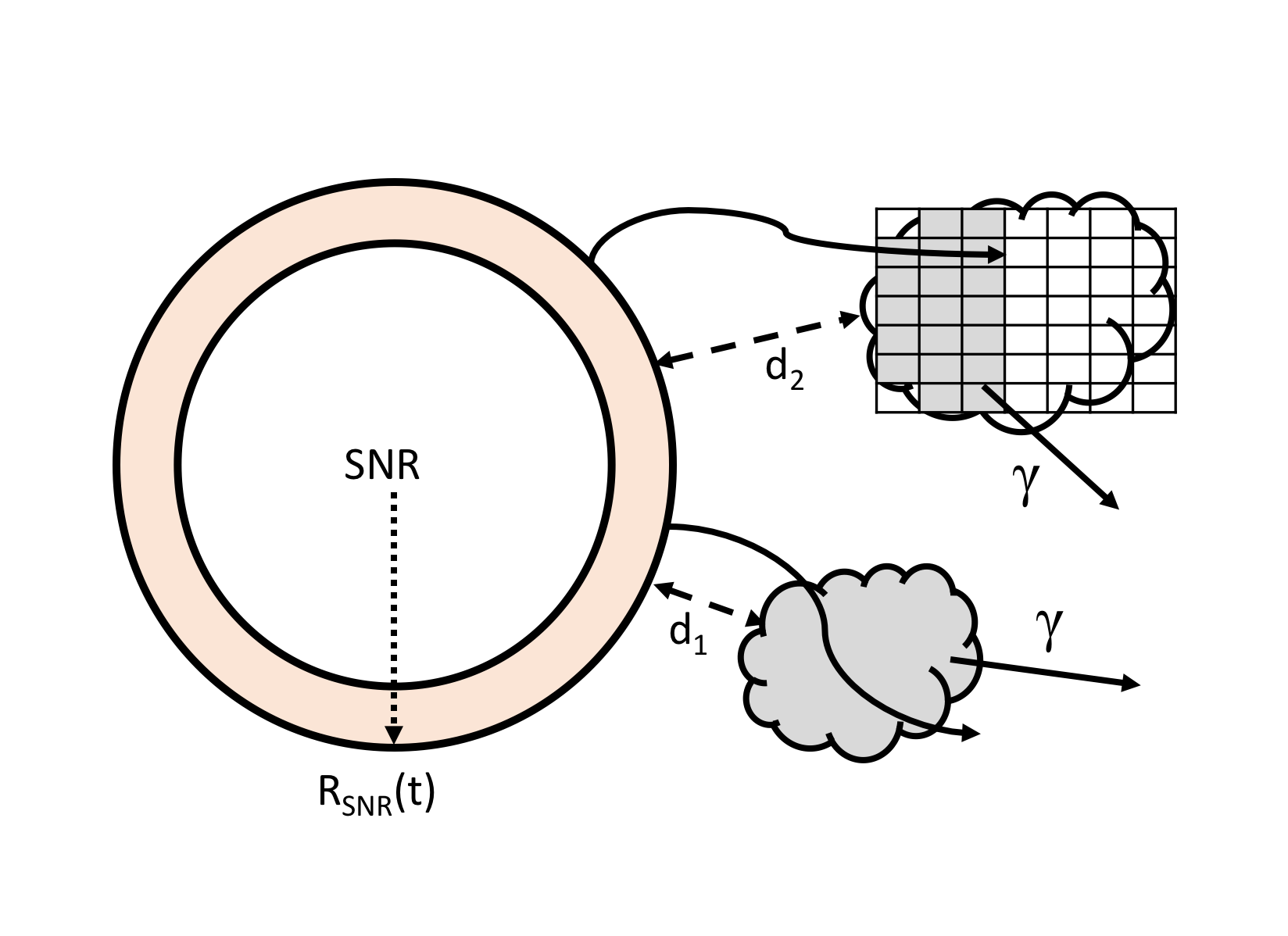}
\caption{Schematic showing the geometry assumed in the model. Particles are released from the time-dependent SNR radius $R_{\mathrm{SNR}}(t)$, propagate through the ISM and penetrate nearby clouds to produce gamma rays. For a nearby cloud located at a distance $d_1$, the cosmic rays fully traverse the cloud; whereas for a larger cloud located further away at a distance $d_2$, the cosmic rays are only able to partially traverse the cloud within the available time. A cell based approach is used to treat this case, as described in section \ref{sec:model}.} 
\label{fig:schematic}
\end{figure}

In this study, we model the injection of particles (treated as a pure proton flux) from SNRs. Figure \ref{fig:schematic} illustrates the scenario under consideration schematically. Three key aspects comprise the model: SNR evolution and particle escape; particle transport through the ISM; and particle interactions with interstellar clouds to produce gamma rays.
We adopt a dynamical description for the injection of particles into the ISM, which allows us to treat SNRs as impulsive accelerators. In fact, particles of different energies are released at different times, with high energy particles being able to leave the shock before the low energy ones \citep{Celli19}.
While details of the escape process are still poorly understood, expectations are such that the turbulence which confines particles inside the SNR decays as the shock expands, hence the most energetic particles with larger mean free path are able to leave the system earlier than lower energy ones. 
As the SNR expands, this also modifies the distance travelled by the particles to reach the target interstellar cloud. Less energetic particles, released later in the SNR evolution, will have a reduced distance to travel to the target cloud.

The propagation of particles from accelerator to target and subsequent modification to the particle spectrum arriving at the interstellar cloud are treated following \cite{AA96}. 
The gamma-ray flux produced due to particle interactions in the interstellar clouds are then calculated from \cite{Kelner06}. 

This enables us to explore the phase space of different accelerator and interstellar cloud properties and the resulting dependence of the gamma-ray flux at a given energy on these properties, assuming the same initial particle spectrum. 

\subsection{Particle spectrum reaching the target}
\label{sec:spec} 

For an impulsive source with a power law acceleration spectrum of slope $\alpha$, we use the following expression to determine the number of protons of energy $E$ at a distance $R$ from the centre of the SNR at the age $t_{\mathrm{SNR}}$ of the SNR:

\begin{equation}
    J_\mathrm{p}(E,R,t) = N_0 \cdot E^{-\alpha} \cdot f(E,R,t)\ ,
\end{equation}

\noindent where the function $f(E,R,t)$ describes the probability density function of the particles. The normalisation $N_0$ over the particle spectrum is obtained from: 

\begin{equation}
    N_0 = 
    \begin{cases}
        E_\mathrm{CR}/\left(\ln{E_\mathrm{max}} - \ln{E_\mathrm{min}}\right) & \alpha = 2\\
        E_\mathrm{CR} \cdot (2-\alpha)/\left(E_\mathrm{max}^{2-\alpha} - E_\mathrm{min}^{2-\alpha}\right) & \mathrm{otherwise}
    \end{cases}
\end{equation}

\noindent with $E_\mathrm{CR}$ the total energy budget available to cosmic rays from the supernova explosion, which we take to be $E_\mathrm{CR}=10^{50}$~erg for a canonical supernova explosion with kinetic energy $E_\mathrm{SN}=10^{51}$~erg. Considering those particles which have escaped the SNR and are diffusing into the ISM, the probability density function is \citep{AA96}:

\begin{equation}
f(E,R',t') \approx \frac{f_0}{\pi^{3/2}R_d^3}\exp\left[ -\frac{(\alpha-1)t'}{\tau_{pp}} 
- \frac{R'^2}{R_d^2}\right]~(t > t_{\mathrm{esc}}(E))~,
\label{eq:pflux}
\end{equation}

\noindent where $f(E,r,t')$ has units of $\mathrm{cm^{-3} GeV^{-1}}$, and the diffusion radius $R_d$ is given by:

\begin{equation}
R_d(E,t') = 2\sqrt{D(E)t'\frac{\exp(t'\delta/\tau_{pp})-1}{t'\delta/\tau_{pp}}}~,
\label{eq:rdiff}
\end{equation} 
\noindent $\delta$ being the slope with energy of the diffusion coefficient. 
The variables $t'$ and $R'$ correspond to the time spent and distance travelled in the ISM respectively, after CRs escape from the SNR. We note here that equation \eqref{eq:pflux} from \cite{AA96} is a Green's function solution which already takes the 3D isotropic spherical expansion into account. However, it does not include the energy losses that the particles undergo while the remnant expands. Such adiabatic losses are not relevant for the highest energy particles (at about 1~PeV) that we are interested in, as they leave the system when the remnant still has a well-contained size, as shown in equation \eqref{eq:snrrad}.

\noindent This expression \eqref{eq:rdiff} can be used for diffusion both through the ISM and within the denser target material. However, within the ISM p-p collisions are irrelevant due to the low ambient density ($\tau_{pp} \gg t'$) and the simplification $R_d \approx 2\sqrt{D(E)t'}$ can be used. 

The proton lifetime $\tau_{pp}$ against particle interactions is given by the expression:

\begin{equation}
\tau_{pp} = \left[ nc\kappa\sigma_{pp}(E)\right]^{-1}\,,
\end{equation}

\noindent where $n$ is the number density of the ambient gas and $\kappa \approx 0.45$ is the interaction inelasticity in the energy range of interest, which is approximately energy independent at GeV to TeV energies. The cross-section $\sigma_{pp}(E)$ for particle interactions follows an energy dependence described in \cite{Kafexhiu14}. 

\begin{eqnarray}
\sigma_{pp}(E) &=& \left[30.7 - 0.96\log\left(\frac{E}{E^{th}}\right) + 0.18\log^2\left(\frac{E}{E^{th}}\right) \right] \nonumber \\
& \times & \left[1 - \left(\frac{E^{th}}{E}\right)^{1.9}\right]^3\,\mathrm{mbarn}~,
\label{eq:ppcross}
\end{eqnarray}

\noindent where $E$ is the proton kinetic energy and $E^{th} =0.2797$\,GeV the threshold kinetic energy for pion production. 
For the diffusion coefficient we consider a power-law dependence on energy as: 

\begin{equation}
D(E) = \chi D_0 \left(\frac{E /\mathrm{GeV} }{B(n)/3\mathrm{\mu G}}\right)^{\delta}~,
\label{eq:diffcoeff}
\end{equation}

\noindent with index $\delta$, suppression factor $\chi$ (which relates to the level of turbulence within the propagation region) and reference diffusion coefficient $D_0$ at 1\,GeV, given in table \ref{tab:constants}. 
Note that two values for the normalization of the diffusion coefficient are considered, with the higher value corresponding to fast diffusion, i.e. a small diffusion time $\tau_d$ over a distance $d$; $\tau_d(E) = d^2/D(E)$. Such a value is consistent with Boron over Carbon measurements, namely with the average time spent by CRs in our Galaxy. This should be considered the reference situation, with $\chi = 1$ within the ISM. A reduced diffusion coefficient will be eventually considered (when explicitly mentioned), in order to investigate the case for enhanced magnetic turbulence around CR sources, as induced e.g. by the presence of CRs themselves, which can modify the diffusion \citep{dangelo,Inoue2019ApJ...872...46I}.
The magnetic field $B(n)$ has a dependence on the density of the cloud $n$ where \citep{Crutcher10}:

\begin{equation}
B(n) = \begin{cases}
10\,\mathrm{\mu G} & (n < 300 \mathrm{cm}^{-3})~,\\
10\left(\frac{n}{300}\right)^{0.65}\,\mathrm{\mu G} & (n \geq 300 \mathrm{cm}^{-3})~.
\end{cases}
\label{eq:bfield}
\end{equation}

We note that the estimation of the dependence of the magnetic field on cloud density relies on the Zeeman splitting effect which is significant only for dense clouds. Although a constant value of $B$ is assumed for $n < 300\,\mathrm{cm}^{-3}$, in practice the magnetic field is likely to vary with $n$, albeit a parameterisation of this relation is highly uncertain at the present time. Measurements of $B$ in lower density clouds must use other approaches, such as Faraday tomography \citep{VanEckFaradayTomo17,CrutcherReview}.
For the ISM, a constant value of $B=3\mu G$ is assumed \citep{Jansson_ISMB}. 

As the original formulation in \cite{AA96} refers to particle released from a point (such as the centre of the SNR), the normalisation factor $f_0$ (see Eq.~\eqref{eq:pflux}) is used to account for release of particles at a time dependent radius $R_{\mathrm{SNR}}(t)$. This was derived from the requirement that $\int f(E,R,t)~ \mathrm{d}V = \int f(E,R,t)~ 4\, \pi R^2 \,\mathrm{d}R = 1$, leading to the solution: 

\begin{equation}
    f_0 = \frac{\sqrt{\pi}R_d^3}{\left(\sqrt{\pi}R_d^2+2\sqrt{\pi} R_{\mathrm{SNR}}^2\right)R_d+4R_{\mathrm{SNR}}R_d^2}\,.
\label{eq:f0}
\end{equation}

\noindent 
A list of the benchmark values used for constants in the model (unless otherwise specified) is given in table \ref{tab:constants}. Results will be presented for the case of slow diffusion as default, with fast diffusion occasionally shown as indicated for comparison. 

\begin{table}
\caption{Parameters of the model and their values, unless otherwise specified. The `slow case' value of D$_0$ is taken from \citep{Gabici07}.}
\centering
\begin{tabular}{lcl}
Note & Parameter & Value \\
\hline
Slope of particle spectrum $f(E)$ & $\alpha$ & 2 \\
Slope in energy of $D(E)$ & $\delta$ & 0.5 \\
(Inverse) Slope in momentum of $t_{\mathrm{esc}}(p)$  & $\beta$ & 2.5 \\
Suppression factor of D(E) in clouds & $\chi$ & 0.05 \\
Normalization of $D(E)$ (fast case) & $D_0$ (1 GeV) & $3\times 10^{27}$ cm$^{2}$s$^{-1}$\\
Normalization of $D(E)$ (slow case) & $D_0$ (1 GeV) & $3\times 10^{26}$ cm$^{2}$s$^{-1}$\\
ISM density & $n$ & $1\mathrm{cm}^{-3}$ \\ 
Sedov time & $t_{\mathrm{sed}}$ & 1.6\,kyr
\end{tabular}
\label{tab:constants}
\end{table}

To find the particle spectrum reaching the target interstellar cloud, we first need to know the time at which particles escaped the SNR, $t_{\mathrm{esc}}(p)$, such that the time spent in the ISM is $t'=t-t_\mathrm{esc}$ with $t$ the time elapsed since the supernova event occurred (i.e. the remnant age). We consider the following equation to describe the momentum-dependence of escape time $t_{\mathrm{esc}}(p)$:

\begin{equation}
    t_{\mathrm{esc}}(p)=t_{\mathrm{sed}}\left(\frac{p}{p_{M}}\right)^{-1/\beta}~,
	\label{eq:tesc} 
\end{equation}

\noindent where $t_{\mathrm{sed}}\sim 1.6$\,kyr is the Sedov time for core collapse supernovae of 10 solar mass ejecta expanding into a uniform medium of density 1 proton/cm$^3$; and $p_M$ is the maximum momentum of the particles at the Sedov time, set to 1\,PeV/c in order to simulate the PeVatron phase of the SNR \citep{Celli19}. Note that for type Ia supernovae a shorter Sedov time is expected $t_{\mathrm{sed}}\sim250$\,yr. We adopt the normalisation for core collapse supernovae as default. The parameter $\beta$ shown in Eq.~\eqref{eq:tesc} is one of the main factors of uncertainty in the model: in fact, its value depends on the temporal evolution of the magnetic turbulence, which in turn is affected by the physical mechanisms from which it originates. Throughout this work we fixed $\beta=2.5$, consistently with gamma-ray observations of middle-aged SNRs \citep{Celli19}.
The particle momentum $p$ is related to the kinetic energy $E$ through the standard relativistic energy-momentum relation. 


\noindent From equation \eqref{eq:tesc} we can find the time $t'$ available for the particles to travel, namely the difference between the system age $t$ and particle escape time $t_{\mathrm{esc}}(p)$.
The distance to the interstellar cloud is calculated between the centres of both the SNR and the cloud. Meanwhile, the SNR expands to a non-negligible size; so we find the radius of the SNR, $R_{\mathrm{SNR}}$ at the time $t$:

\begin{equation} 
  R_{\mathrm{SNR}}(t) = 0.31\left(\frac{(E_{SN}/10^{51} \mathrm{erg})}{(n/1\mathrm{cm}^{-3})(\mu_1/1.4)}\right)^{1/5} \left(t/\mathrm{yr}\right)^{2/5}~\mathrm{pc}~,
\label{eq:snrrad}
\end{equation} 

\noindent assuming that the SNR is undergoing adiabatic expansion within the Sedov-Taylor phase, which typically lasts for $\sim40$\,kyr \citep{TrueloveMcKee,2008ARA&A..46...89Reynolds}. 
The radius at which the particles of energy $E$ escape from the SNR is therefore given by:

\begin{equation}
    R_{\mathrm{esc}} = R_{\mathrm{SNR}}(t=t_{\mathrm{esc}}(E))~,
    \label{eq:snresc}
\end{equation}

\noindent such that the distance travelled within the ISM is $R'=R-R_\mathrm{esc}$ with $R$ measured to the centre of the SNR. 
As this study focuses on the PeV particles, which are released at the start of the Sedov-Taylor phase, we do not take further evolutionary stages into consideration. 
Consequently, the distance travelled by the particles to reach the target also varies with energy and is found by the total distance less the SNR radius at $t_{\mathrm{esc}}$ as given by equation \eqref{eq:snresc}. 

In exploring the properties of cloud and accelerator which are ideal for tracing PeVatrons, we will focus on PeV particles and the corresponding 100\,TeV gamma-ray flux. 
The escape time for PeV particles (corresponding to $\sim100\,$TeV gamma rays) is 1.6\,kyr for core collapse supernovae, corresponding to the Sedov time as in equation \eqref{eq:tesc} \citep{Celli19}. \footnote{The alternative case of 250\,yr for type Ia supernovae will be explored in the discussion.}

\subsection{Gamma-ray flux produced: Two-step model}
\label{sec:twostep}

The gamma-ray flux produced is now evaluated in a two-step model. The expression for the particle flux, equation \eqref{eq:pflux}, is applied to propagation of the particles through the ISM to reach the target material. Treatment of particle propagation through and interaction with the interstellar cloud takes place in a second step. 

\subsubsection{Step 1: Interstellar Medium }

The time available for particle propagation through the ISM $t'$ is given by the accelerator age $t_{\mathrm{SNR}}$ minus the escape time $t_{\mathrm{esc}}$ for particles of a given energy, as determined by equation \eqref{eq:tesc}. Similarly, the distance that particles of a given energy must travel in order to reach the cloud is given by the total distance $d$ from the stated central position of the SNR and cloud, less the cloud radius $R_c$ and less the radius of the SNR at the escape time of those particles, $R_{\mathrm{esc}}(E)$, equation \eqref{eq:snresc}. 

For this propagation step, an average density $n=1$\,cm$^{-3}$ is assumed for the ISM. 
This first step is evaluated using equation \eqref{eq:pflux} with $R' = d - R_c - R_{\mathrm{esc}}$ through the ISM and $t'$ is either the available time $t_{\mathrm{SNR}}-t_{\mathrm{esc}}$ or the time needed to travel through the ISM $\tau_{\mathrm{ISM}} = d^2 / D(E)$, whichever is the smaller. 

\begin{figure}
\includegraphics[width=\columnwidth]{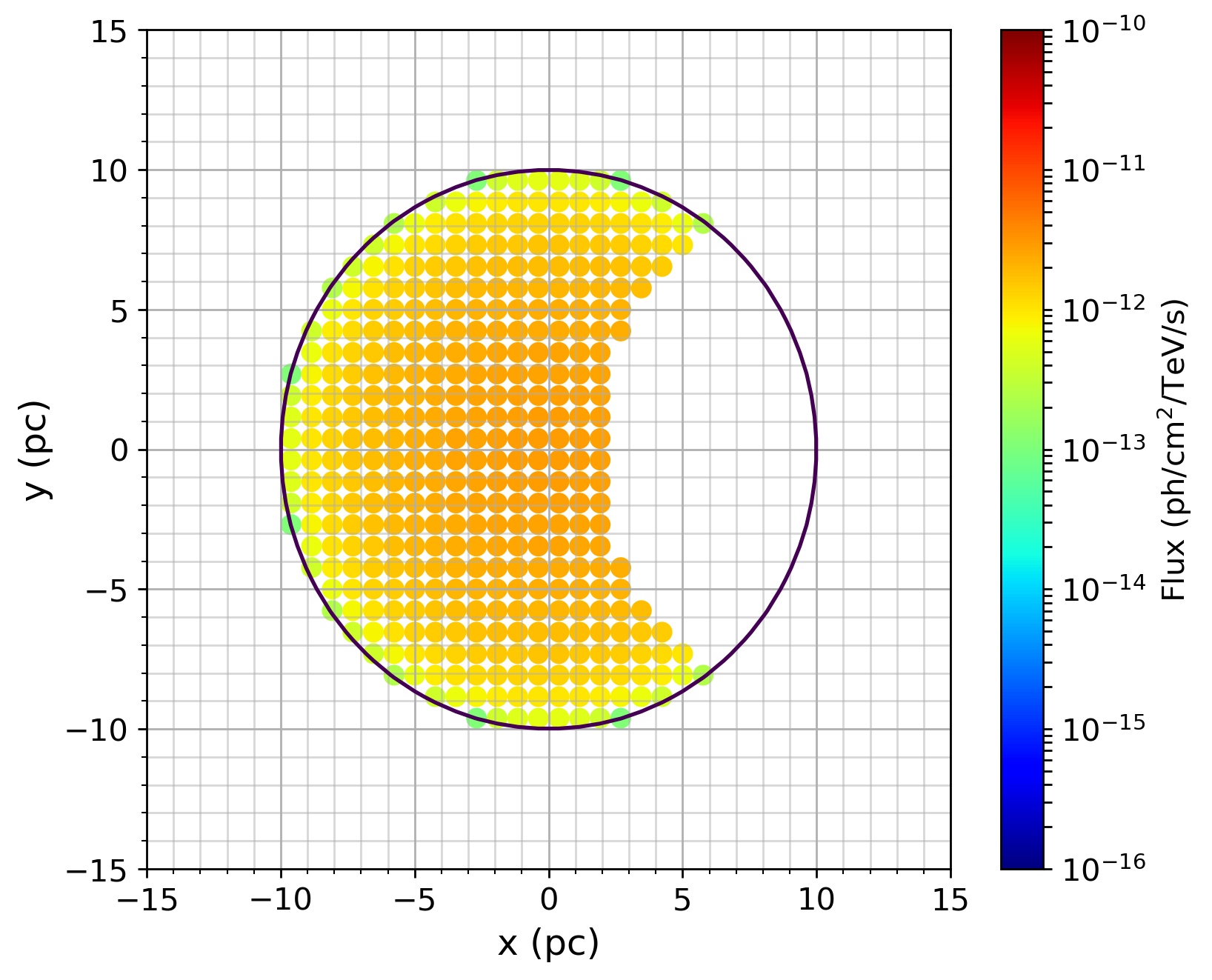}
\caption{Illustration of the cell based approach used to calculate the total flux from a cloud in cases when particles do not fully traverse the cloud. The particles arrive from the left hand side and penetrate the cloud to a constant depth (with constant density assumed throughout the cloud). The flux shown is integrated along the line of sight depth, with spherical geometry assumed. }
\label{fig:cloudcells}
\end{figure}

\begin{figure*}
\centering
\includegraphics[width=\columnwidth]{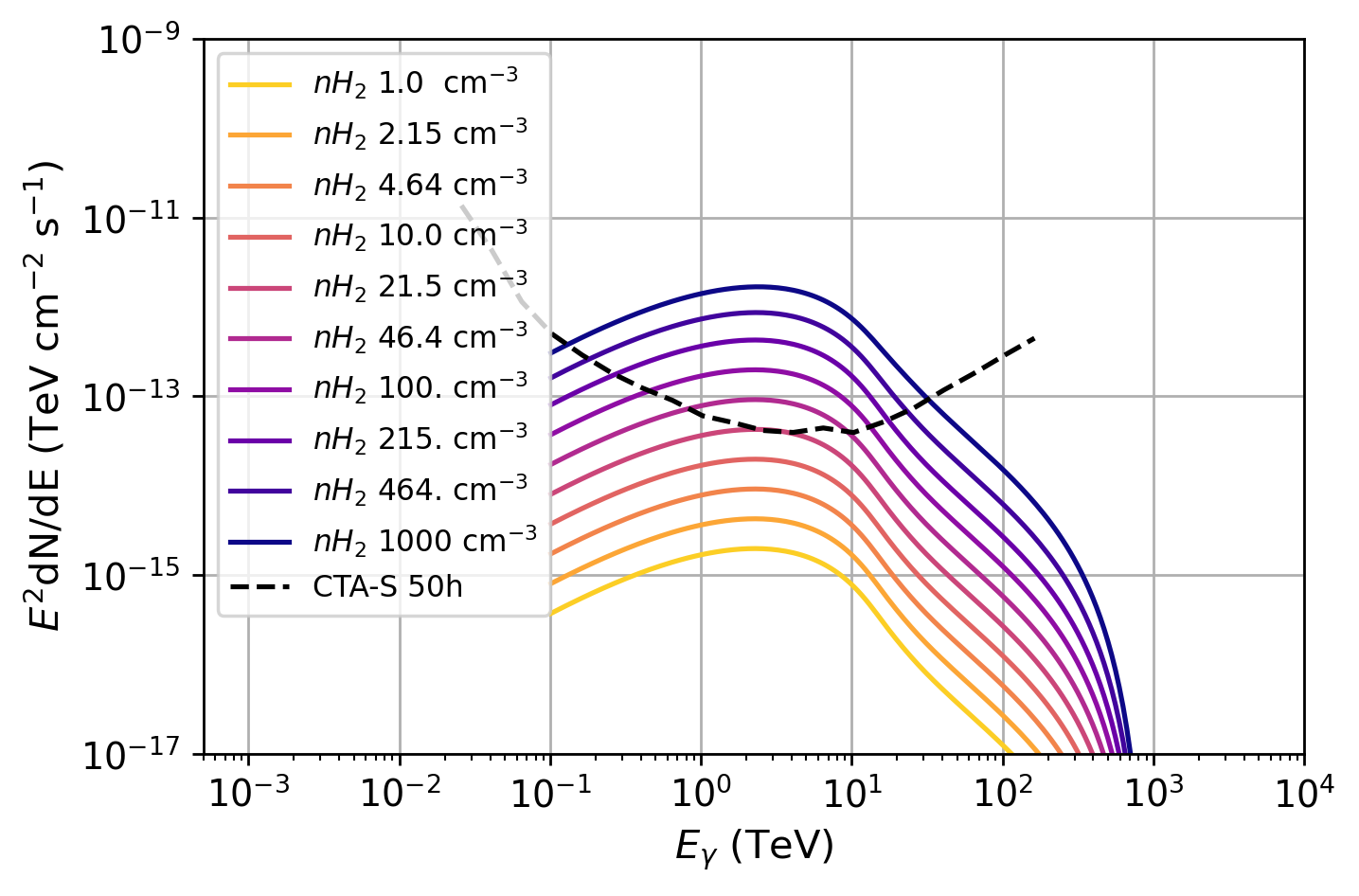}
\includegraphics[width=\columnwidth]{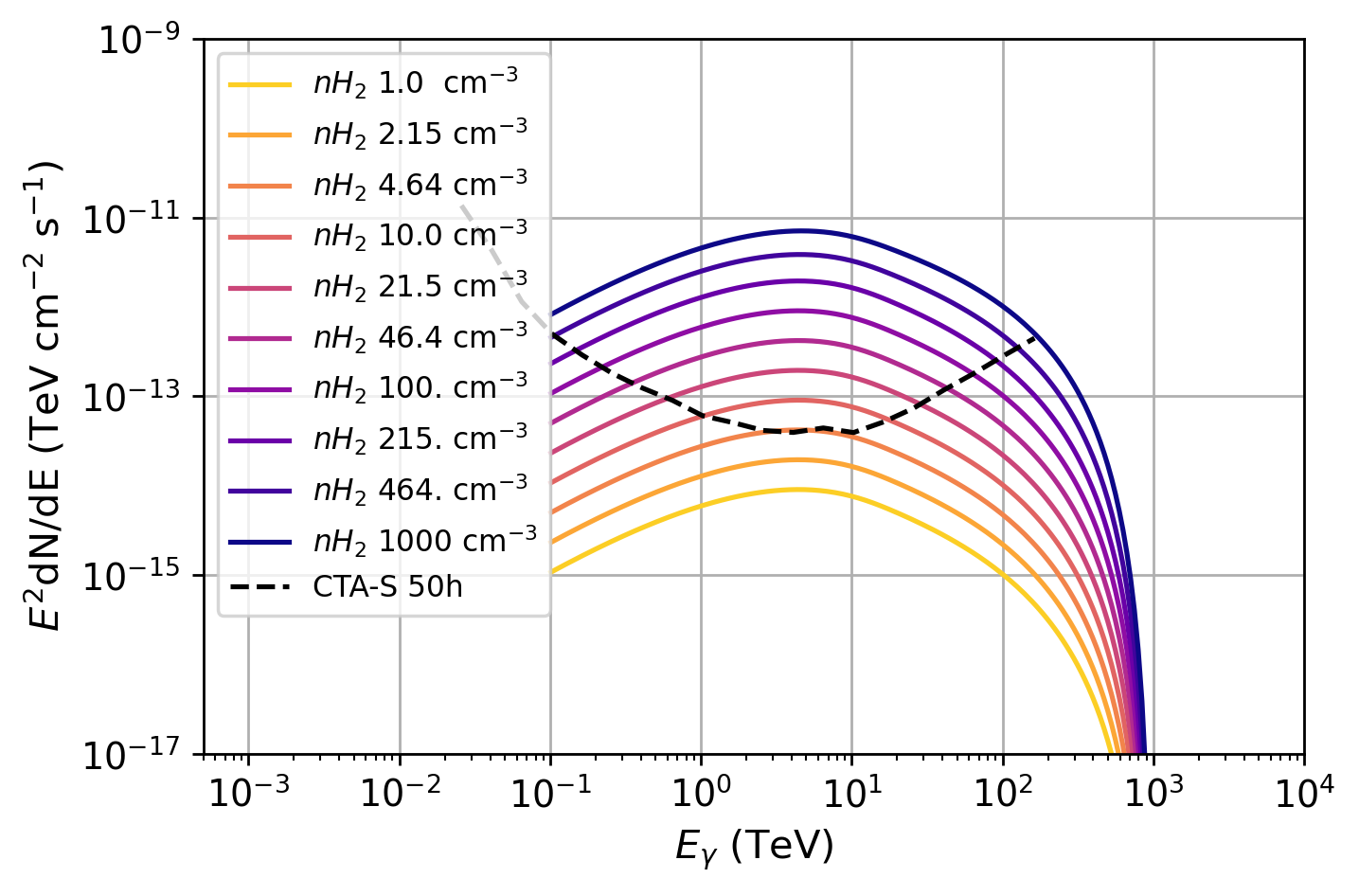}
\caption{
Variation in gamma-ray flux with density of the target material. The accelerator age $t$ is fixed to 8\,kyr and the separation distance $d$ to 32\,pc. 
{\bf Left:} using a fast diffusion coefficient $D_0$. {\bf Right:} using a slow value for $D_0$ (see table \ref{tab:constants}).
The black dashed line corresponds to the CTA-South 50 hour sensitivity to point-like sources \citep{2019scta.book.....C}.  }
\label{fig:gspectra2}
\end{figure*}

\begin{figure*}
    \centering
    \includegraphics[width=\columnwidth]{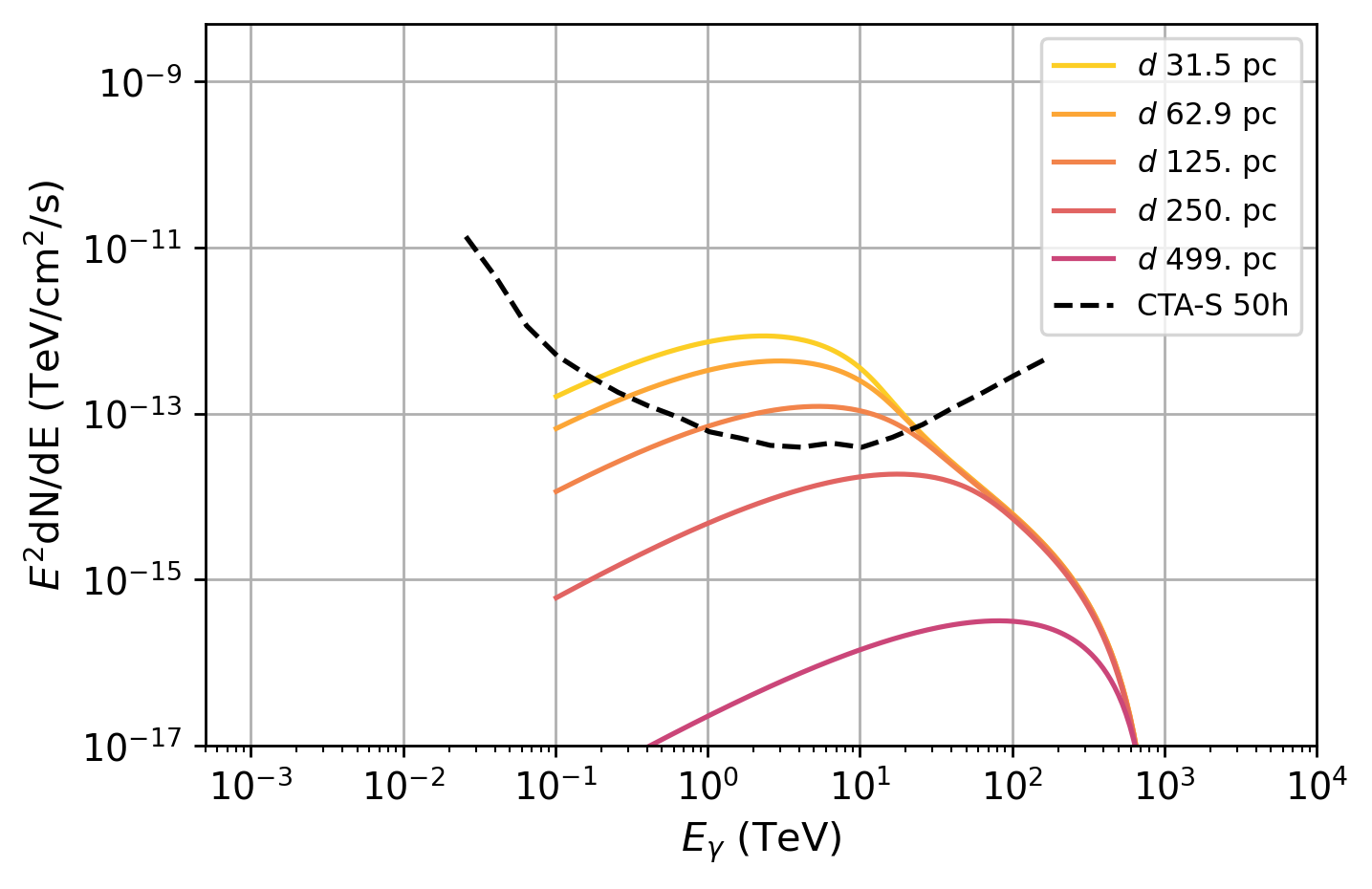}
    \includegraphics[width=\columnwidth]{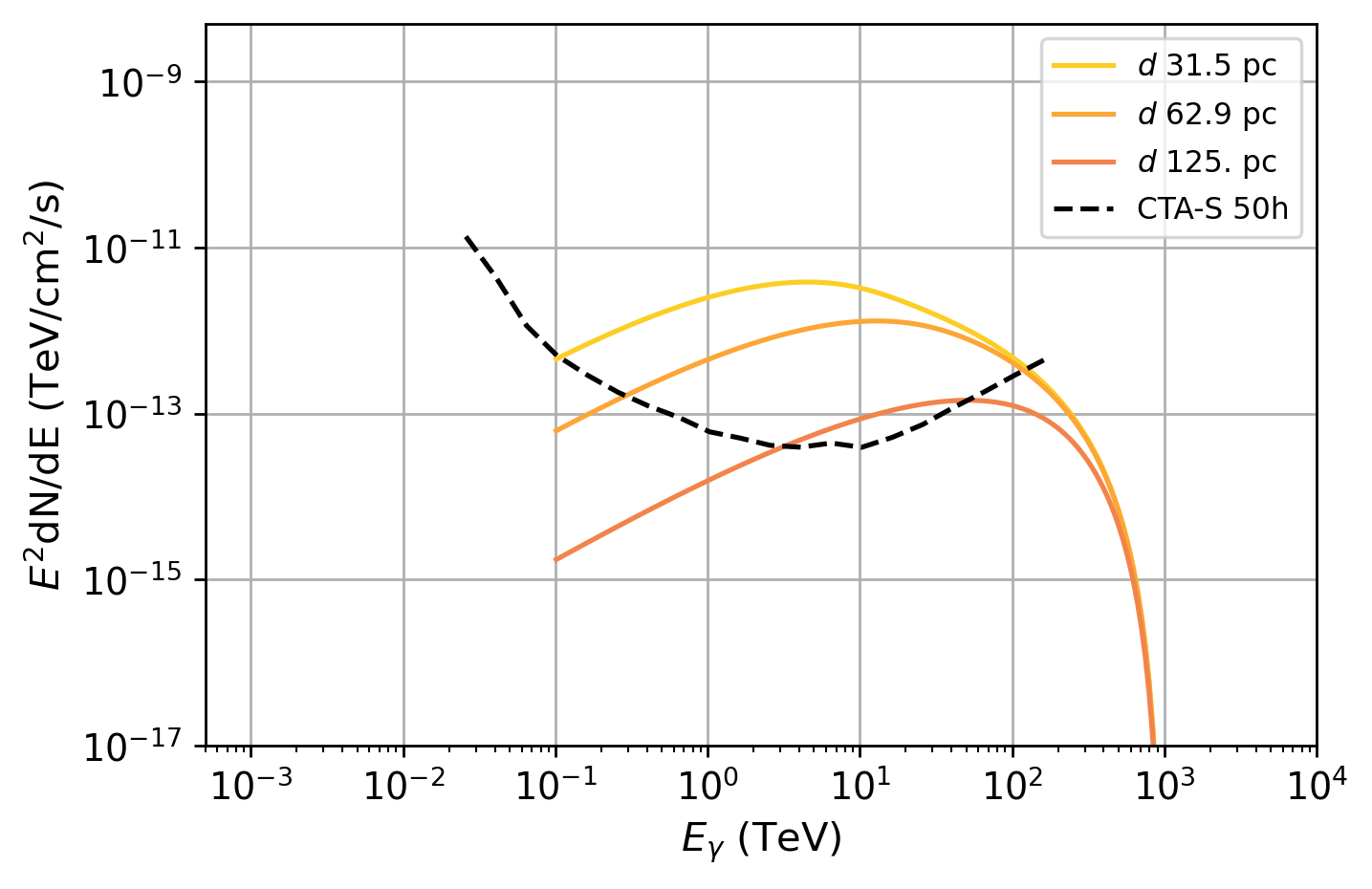}
    \caption{As for figure \ref{fig:gspectra2}, except with varying separation distance $d$. {\bf Left:} fast value for $D_0$. {\bf Right:} slow value for $D_0$. The age $t$ and density $n$ are kept fixed to $8$\,kyr and $460$\,cm$^{-3}$ respectively.}
    \label{fig:gspectrad}
\end{figure*}

\begin{figure*}
    \centering
    \includegraphics[width=\columnwidth]{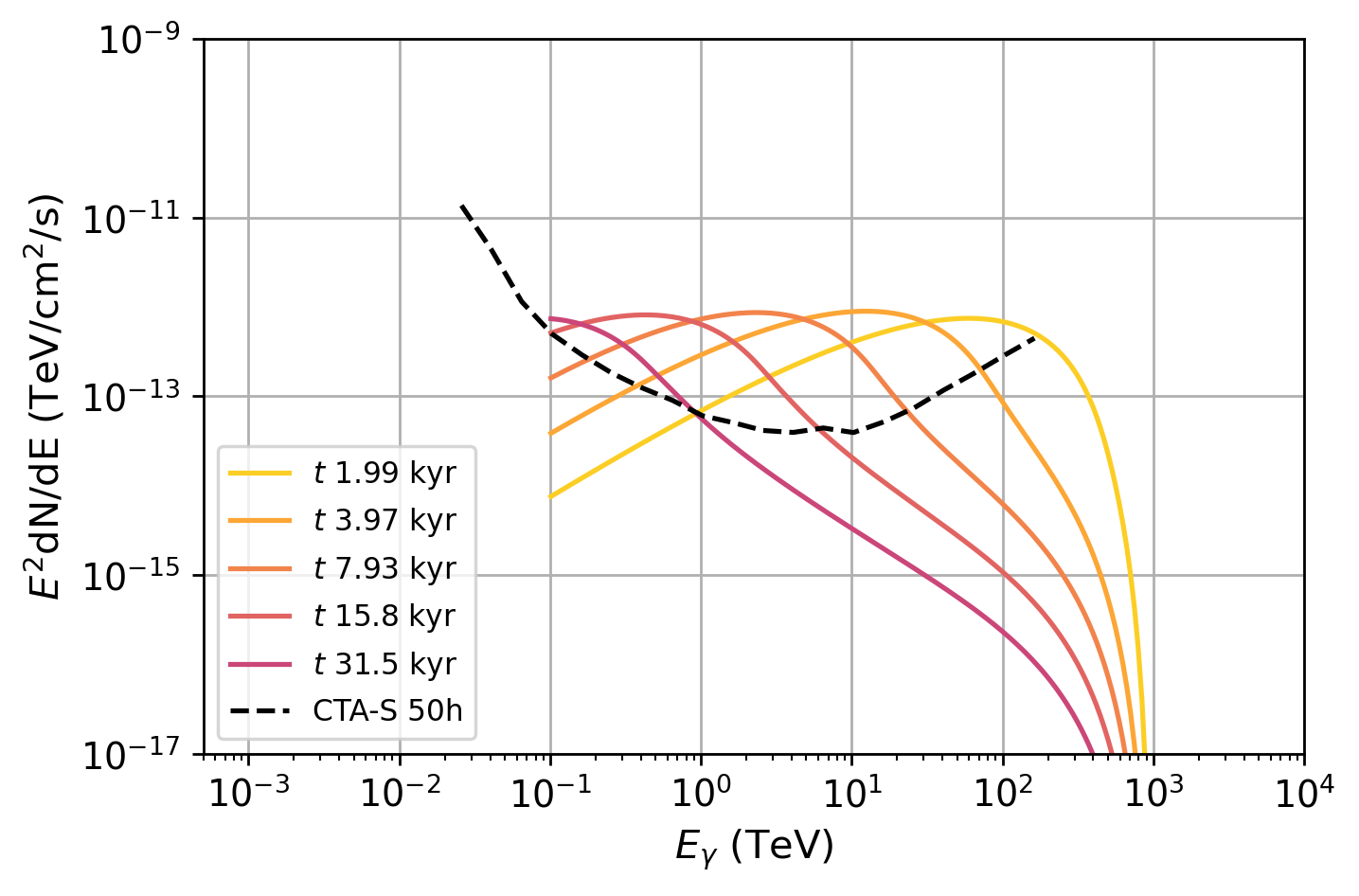}
    \includegraphics[width=\columnwidth]{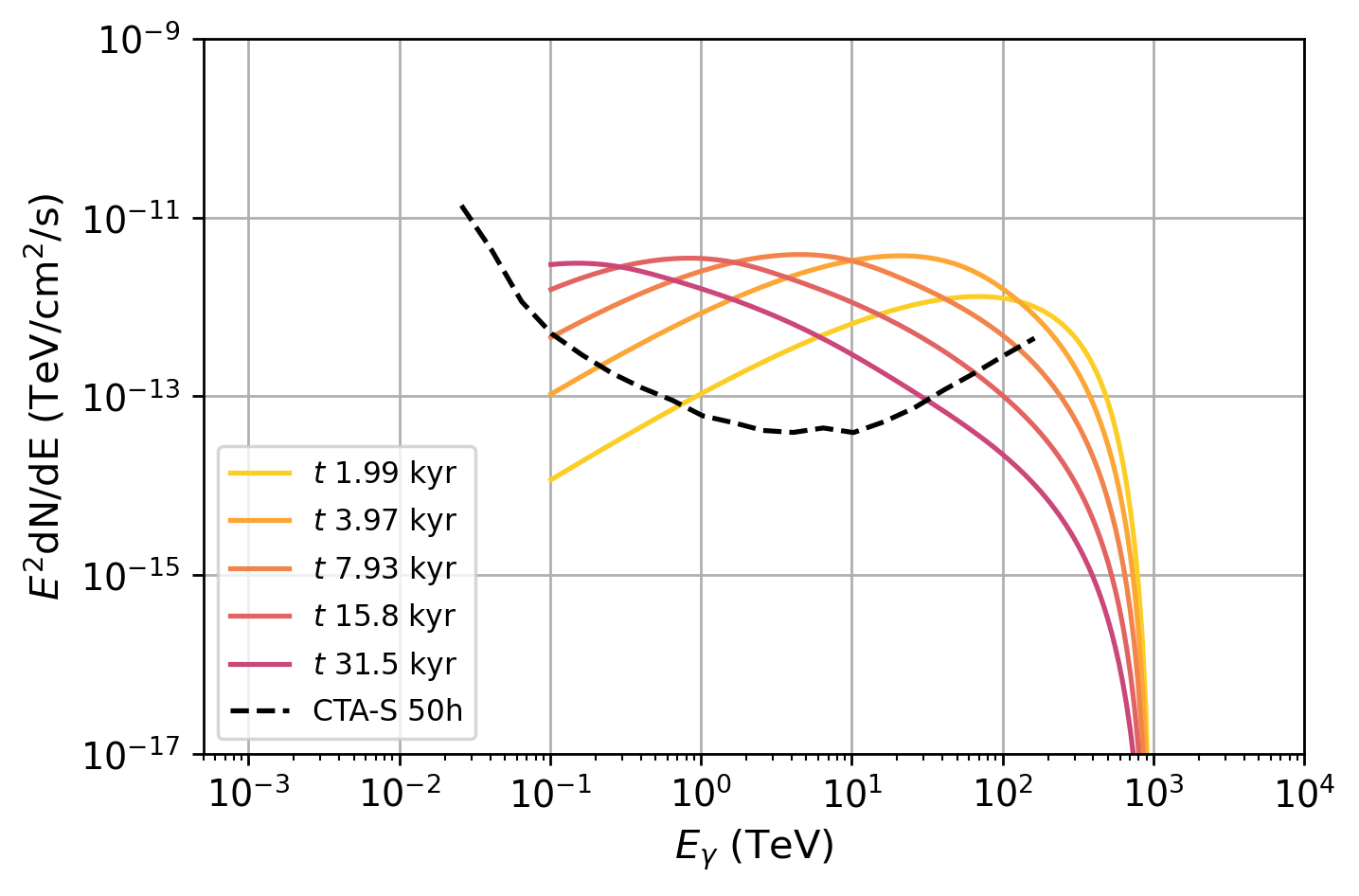}
    \caption{As for figure \ref{fig:gspectra2}, except with varying accelerator age $t$. {\bf Left:} fast value for $D_0$. {\bf Right:} slow value for $D_0$. The separation distance $d$ and density $n$ are kept fixed to $32$\,pc and $460$\,cm$^{-3}$ respectively.}
    \label{fig:gspectrat}
\end{figure*}

\subsubsection{Step 2: Interstellar Cloud }

To calculate the subsequent gamma-ray flux produced by particles reaching the target cloud, we use the treatment of \cite{Kelner06}, with the particle spectrum obtained in the previous section; 
$f(E,R',t')$ from equation \eqref{eq:pflux}.

The produced gamma-ray emissivity at fixed space and time position $\Phi_\gamma(E_\gamma,R',t')$, in ph~cm$^{-3}$~s$^{-1}$~TeV$^{-1}$, is given by:

\begin{equation}
\Phi_\gamma (E_\gamma,R',t') = cn \int_{E_\gamma}^{\infty} \sigma_{\mathrm{inel}}(E)f(E,R',t')F_\gamma \left(\frac{E_\gamma}{E},E\right)\frac{dE}{E}~,
\label{eq:gflux}
\end{equation}

\noindent where $f(E,R',t')$ is the particle density arriving at the cloud, as given by equation \eqref{eq:pflux}. Clouds are assumed to have a constant density profile, although a spatially varying profile could be considered in future work. 

For this second step, we wish to calculate the time for which the particles interact with the target material. If the particles have sufficient time to reach the cloud, i.e. $\tau_{\mathrm{ISM}} (p) < t_{\mathrm{SNR}}-t_{\mathrm{esc}}(p)$, then the remaining time spent in the cloud is  $\Delta t = t_{\mathrm{SNR}} - [t_{\mathrm{esc}}(p) + \tau_{\mathrm{ISM}}(p)]$. 
The depth, $d_c$, to which particles penetrate the cloud within the available time, $\Delta t$, is an energy dependent quantity given by equation \eqref{eq:rdiff}. If the cloud can be fully traversed, then the flux observed from the cloud $F(E_\gamma,t)$, in $\mathrm{ph~cm}^{-2}\mathrm{TeV}^{-1}\mathrm{s}^{-1}$, is:

\begin{equation}
    F(E_\gamma,t) = \Phi_\gamma (E_\gamma,t) V_c / (4\pi D^2)~,
    \label{eq:cellflux}
\end{equation}

\noindent where the gamma-ray emissivity was integrated over the volume $V_c$ of the cloud (assumed to be spherical) and $D$ is the distance from the cloud to the Earth.

However, in cases where the particles do not fully traverse the cloud, $d_c < 2 R_c$, we divide the cloud into a grid of cells and evaluate the total flux from the cloud by summing the contributions from all of the cells reached by the particles. Note that due to the energy dependent diffusion process, particles of different energy will reach different locations within the cloud. 
The volume $V_c$ in equation \eqref{eq:cellflux} is replaced by the sum of the cell volumes, $\sum_i A_i$ $dl_i$, where $A_i$ is the 2D projected area of the ith cell face and $dl_i$ is the depth of the ith cell along the line of sight. Figure \ref{fig:cloudcells} demonstrates this for a case where particles penetrate to a fixed depth. 

\subsubsection{Resulting gamma-ray spectrum}

Figures \ref{fig:gspectra2} to \ref{fig:gspectrat} show how the produced gamma-ray spectrum varies with properties of the accelerator--cloud system; the accelerator age, interstellar cloud density and distance between accelerator and cloud. 
The cloud radius is fixed to 10\,pc and the distance of the system to Earth is fixed to 1\,kpc throughout this section. For each variable, ten logarithmically spaced values were tested between 1-1000\,cm$^{-3}$ for the density; 1-500\,pc for separation distance and 1-500\,kyr for accelerator age.

In figure \ref{fig:gspectra2}, the gamma-ray flux is higher for more dense target clouds as expected (see Eq.~\eqref{eq:gflux}). As the accelerator age $t$ and separation distance $d$ are fixed, the change in density affects mainly the flux normalisation with an otherwise consistent shape to the spectra. 
Lower energy particles take longer to reach the target material, such that the proton spectrum reaching the cloud shows a low-energy cut-off, which is seen in gamma rays in the form of a smoothed suppression of the flux. 
\replaced{For the slow diffusion case, the flux normalisation is higher and the curve smoother with a less significant drop towards higher energies, as the particles typically spend a longer amount of time within the cloud.}{The flux normalisation is lowered while the energy at which the flux peaks appears higher for slower diffusion, as expected because of the fixed distance and the total age of the system, implying that low-energy particles cannot reach the cloud when the diffusion coefficient is reduced.}

The trend in figure \ref{fig:gspectrad} also matches that expected, with higher flux for lower separation distances. Again, a drop towards lower energies can be seen. 
Larger separation distances can be reached by lower energy particles in the case of fast diffusion. Similarly, the energy at which the flux peaks also moves towards higher energies for larger separation distances, as the higher energy particles will reach the cloud earlier and interact for longer than less energetic particles. The largest separation distances reached are 500\,pc or 125\,pc for fast or slower diffusion respectively; although with a comparatively low predicted flux $\sim10^{-13} - 10^{-16}$\,TeV\,cm$^{-2}$\,s$^{-1}$. 
The shortest distance shown in figure \ref{fig:gspectrad} is 31.5\,pc due to the requirement that the particles must first traverse the ISM (i.e. that the cloud does not overlap the SNR). At the escape time for 1\,PeV particles, $R_{\mathrm{esc}}=6$\,pc from equation \eqref{eq:snrrad},
whilst the cloud radius is fixed to 10\,pc, such that the minimum separation distance must be greater than 16\,pc.\footnote{Out of the ten values tested, 31.5\,pc is the next smallest value.}

Figure \ref{fig:gspectrat} shows how the age of the accelerator affects the predicted gamma-ray flux; 
the peak flux shifts towards lower energies with increasing age. 
Note that in the case of slower diffusion, a higher normalisation is achieved \added{as particles spend a longer time interacting within the cloud.} \deleted{with more particles reaching the cloud at earlier times.} 
This also emphasises how the shape of gamma-ray spectra can be used to infer accelerator ages, due to the energy losses of the underlying particle spectrum as well as the time needed for particles to travel through the ISM to reach the target cloud. 

For fixed $d$ and $t$, the gamma-ray flux is governed by the amount of target material available for interactions - i.e. by the density or size of the cloud.  Once the particles have had sufficient time to reach and fully traverse the cloud, the gamma-ray flux produced is fixed by properties of the cloud. The finite size of the target cloud therefore imposes a limit on the flux.

These figures \ref{fig:gspectra2} to \ref{fig:gspectrat} therefore demonstrate that properties of both the SNR and target cloud need to be favourable in order to generate a detectable gamma-ray flux. 

\subsection{Accelerator Target Distribution}

In principle, particles released from an accelerator may either travel through the ISM before reaching an interstellar cloud; or could be injected directly into a neighbouring cloud. The latter case may occur where the cloud is in close proximity and the SNR expands such that the forward shock encounters the cloud. Indeed, there are prominent examples where this is thought to have occurred, indicated by high levels of turbulence within a cloud and significant offset of leptonic gamma-ray emission away from the cloud direction, such as in the case of the pulsar wind nebula HESS\,J1825-137 and its progenitor supernova \citep{Voisin16}. However, we do not consider the case of particle injection into a directly adjacent cloud for two reasons. 

Firstly, the result of the impact of a shock on a cloud strongly depends on the system properties, namely shock speed and density contrast between the cloud and the surrounding medium. Diffuse clouds are typically destroyed by the shock, which means that the system would lose the spherical symmetry assumed here. In fact, for this model we consider evolution of the SNR radius with time and corresponding particle release using equations \eqref{eq:tesc} and \eqref{eq:snrrad} with spherically symmetric geometry; these approximations are no longer a valid description for the non-symmetric cases. In turn, more dense clumps might be able to survive the shock passage, however in this case the interaction between the shock and the clump might generate additional magnetic turbulence that would modify the transport of particles all around the clump, thus affecting their ability to interact with target gas there \citep{Inoue2012ApJ...744...71I,celli2019clumps}. The extent to which the particle transport is modified by magnetic turbulence depends on the length scale of the fluctuations, and may be limited for high energy particles \citep{Roh2016APh....73....1R}.

Secondly, the likelihood of such a situation occurring is low in comparison to the case of an intervening step through the ISM first. 
The volume of the Milky Way is approximately $10^{66}\mathrm{cm}^3$, assuming a disk with radius 15\,kpc and height 0.3\,kpc \citep{Rix13MWsize}. To estimate the volume occupied by SNRs in the Sedov-Taylor phase, we first assume an approximately uniform distribution of SNRs throughout our Galaxy, with a uniform distribution of SNR ages. On average, a supernova explosion occurs approximately three times per century. As the Sedov-Taylor phase typically lasts about 40\,kyr, this suggests a total of 600 hypothetical SNRs younger than 40\,kyr for this calculation, with a radius corresponding to their age as determined by the Sedov-Taylor self-similar expansion given in equation \eqref{eq:snrrad}. Assuming that each SNR forms a perfect sphere, the total volume occupied by SNRs with an age $t < 40\,\mathrm{kyr}$ is $6.7\times 10^{62}\mathrm{cm}^3$. 

The total volume of interstellar clouds within our Galaxy can be simply obtained from radii reported in the corresponding catalogues, yet assuming a spherical volume. From the \cite{Rice16} catalogue, as will be used in section \ref{sec:targets}, the total volume is estimated at $1.6\times 10^{64}\mathrm{cm}^3$. The catalogue of \cite{MDsurvey17} contains 8107 clouds, compared to the 1063 of \cite{Rice16}, although both use the CO survey of \cite{Dame01}. This difference is mostly due to inclusion of many smaller clouds by the methodology of \cite{MDsurvey17}.
The total volume occupied by the clouds in the catalogue of \cite{MDsurvey17} is $2.1\times 10^{65}\mathrm{cm}^3$.

Therefore, SNRs typically occupy just $0.01 \%$ 
of the Milky Way volume, whilst clouds occupy 0.25\% (or 3.3\%, using the more complete catalogue of \cite{MDsurvey17}). Hence, SNRs occupy a factor $\sim 100$ less Galactic volume than the most prominent clouds, implying that direct injection into a neighbouring cloud is unlikely. 

However, within 40\,kyr particles escaping from an SNR could reasonably travel to $\sim100$\,pc beyond the SNR radius (corresponding to particle energies of 100\,GeV and 10\,TeV 
for the fast and slow diffusion cases respectively). Let us consider instead the reachable volume with a radius of 100\,pc, attainable by more energetic particles within a shorter timescale. This increases the volume reachable by SNRs by a factor $\sim 350$ and is roughly 3.7\% of the total Milky Way volume. This latter estimate is comparable to the volume occupied by clouds. 

Considering that the true distribution is not uniform and both clouds and SNRs are more likely to be found in denser regions of the Galaxy, such as along spiral arms, this improves the likelihood of cloud-SNR encounters. Nevertheless, we can conclude that the two step approach with intervening propagation within the ISM is a much more likely scenario than that of direction injection of energetic particles by an SNR into an immediately adjacent cloud.

To further support this conclusion, we note that OH masers are an indicator of SNR shocks interacting directly with interstellar clouds. Of the over 800 maser sites recently catalogued, only $\sim$20  were associated with SNRs, which even after considering selection effects, may indicate that direct injection is a less common scenario \citep{BeutherOHmasers}. 

Lastly, the duration of $>$100 TeV gamma-ray emission from an adjacent cloud would be quite short-lived, as the highest energy particles will fully traverse the cloud early in the evolution of the SNR. 

\begin{figure}
\begin{overpic}[width=\columnwidth]{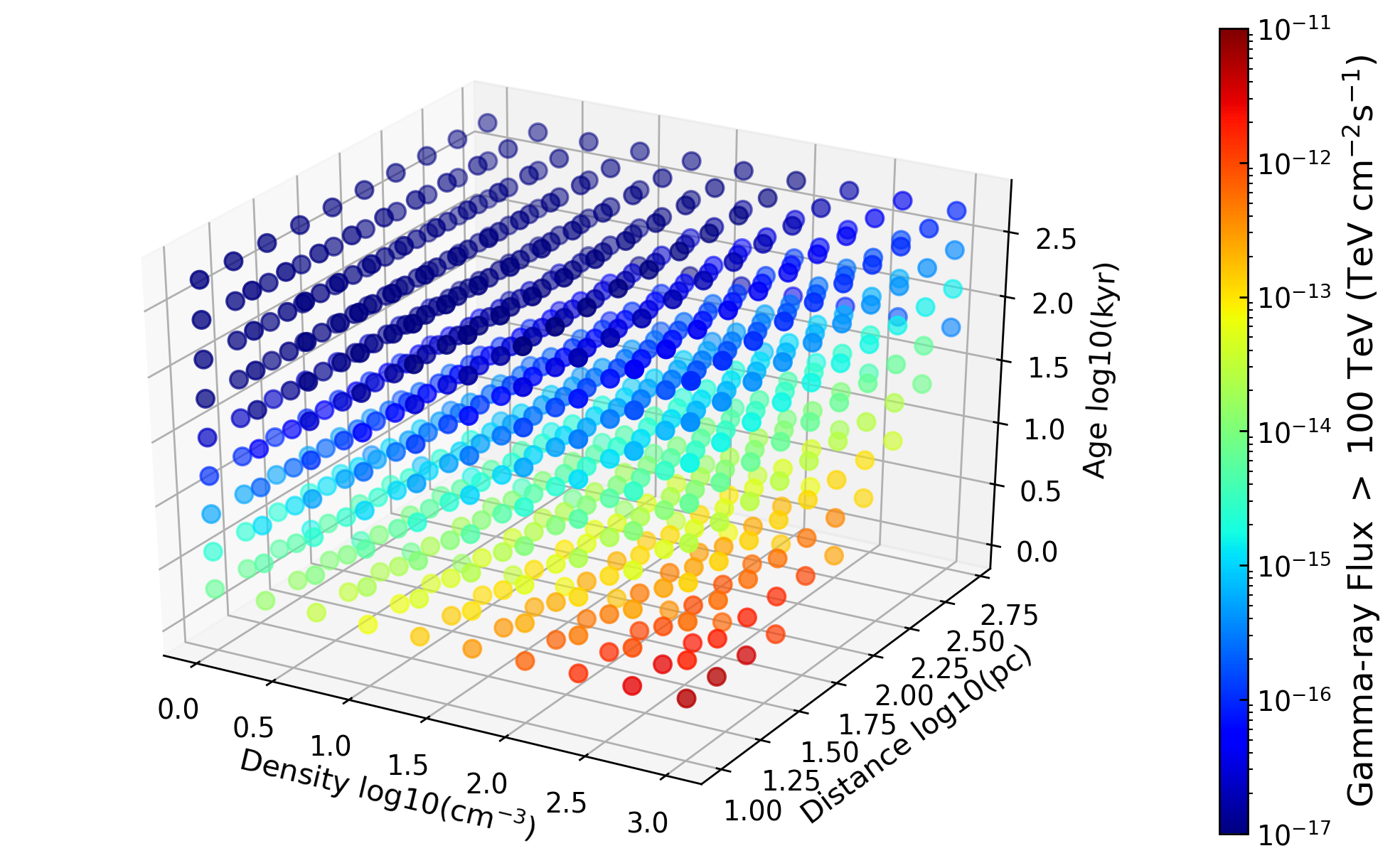}
\put(5,57){\Large{ $E_\gamma >$ 100 TeV}}
\end{overpic}
\caption{Phase space showing the integrated gamma-ray flux above 100\,TeV as a function of target material density, accelerator age and separation distance. A fast value for $D_0$ was used. }
\label{fig:phasespace100tevgamma2}
\end{figure}

\section{Dependence of Gamma-ray Flux above 100\,TeV on System Properties}
\label{sec:properties}

In this section, we explore how different system properties affect the gamma-ray flux produced. This is shown for our two-step model with initial propagation in the ISM to reach the target cloud.

Again, we assume a power law acceleration spectrum with slope $\alpha = 2$, as in table \ref{tab:constants}. 
The following properties are fixed throughout this section: fast value for $D_0$, SNR radius at 30\,pc, cloud radius at 10\,pc and distance from Earth to the SNR at 1\,kpc. 

\subsection{Variation in System Properties}

To explore the phase space, densities from 1 to 1000\,cm$^{-3}$ were chosen as test values, along with separation distances from 1 to 500\,pc and ages from 1 to 500\,kyr. All possible combinations of these three variables were evaluated and are shown in figure \ref{fig:phasespace100tevgamma2}. 

The largest flux is seen for low separation distances and high densities as expected. 
Provided the time is sufficient for particles to fully traverse the cloud, then the resulting flux is subsequently independent of the age; this is seen particularly at low separation distances.
Regions of the phase space at large distances and young ages do not have a detectable gamma-ray flux above 100\,TeV, because the time elapsed is insufficient for particles of the corresponding energy ($\sim$PeV) to travel the distance to the target material. Additionally, the flux is reduced towards lower target material densities, as the probability for p-p interactions is reduced.

\begin{figure}
\includegraphics[width=\columnwidth]{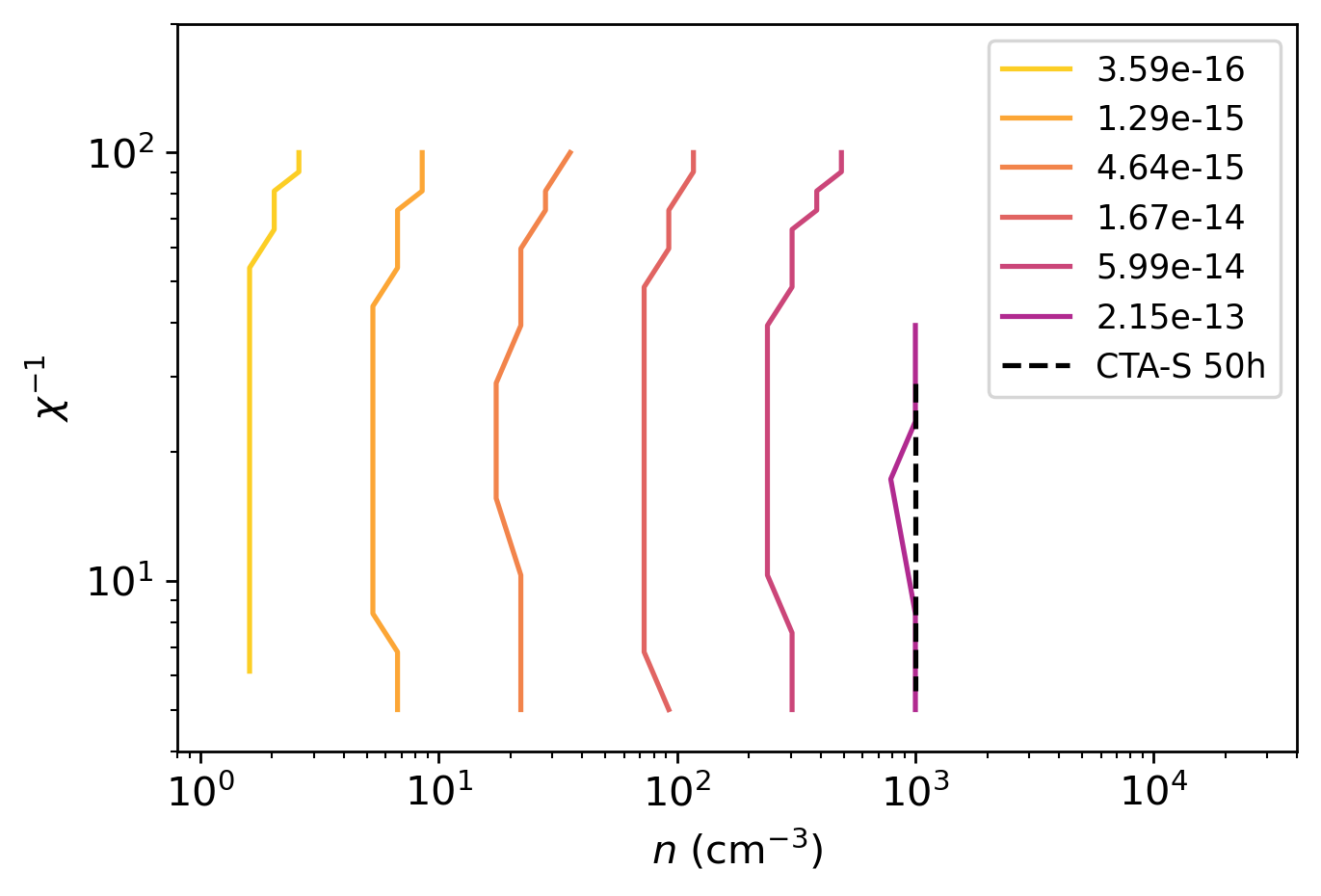}
\caption{Lines of constant integrated gamma-ray flux above 10\,TeV for different cloud densities and values of the diffusion suppression coefficient $\chi$, for fixed accelerator properties; $d=30$\,pc, $t=10$\,kyr.  Flux values are integral flux above 10 TeV in units of $\mathrm{TeV} \mathrm{cm}^{-2}\mathrm{s}^{-1}$.}
\label{fig:chin}
\end{figure}

\subsection{Diffusion Coefficient Variation}
\label{sec:Dcoeff}

Equation \eqref{eq:diffcoeff} describes the diffusion coefficient dependence on the particle energy $E$, the ambient magnetic field $B$ and the suppression factor $\chi$, which is related to the level of magnetic field turbulence \citep{Gabici07}. It should be noted that $\chi$ is assumed constant within the ISM and in this section we consider variation in $\chi$ within the cloud only, with particles fully traversing the cloud. 

Figure \ref{fig:chin} shows how the same gamma-ray flux level is reached by variation in the interstellar cloud properties, keeping the accelerator age and distance to the cloud fixed. In this case, the diffusion suppression coefficient, $\chi$, is plotted against the cloud density $n$, which relates to the amount of target material. With smaller values of $\chi$, the diffusion coefficient is more strongly suppressed, potentially corresponding to higher levels of magnetic field turbulence. 
This shows that the same gamma-ray flux can be achieved for a range of interstellar clouds with different properties (roughly analogous to the phase space for particle accelerators \citep{Hillas84}). 
The lines of constant flux show that the same flux level can be achieved with lower cloud densities if diffusion there is more efficient, namely particles remain trapped for longer times. 
For fixed $\chi$, higher densities lead to higher gamma-ray fluxes, as expected from equation \eqref{eq:gflux} and also seen in figure \ref{fig:gspectra2}.
For fixed density, lower $\chi$ values lead to higher gamma-ray fluxes. Suppression of diffusion coefficient might be achieved in the presence of a high ion-to-neutral ratio within the cloud \citep{nava}. Conversely, a high presence of neutral particles would act in the opposite direction, namely by suppressing magnetic turbulence in the cloud because of ion-neutral damping \citep{kulsrud}, hence increasing the diffusion coefficient.

As the density of an interstellar cloud can be inferred from catalogues and the separation distance and accelerator age determined for known systems, this relation between $\chi$ and $n$ could be used to constrain the diffusion properties within the cloud, given a gamma-ray flux measurement. For this purpose,  gamma-ray upper limits could also be used to constrain the diffusion properties of the system.

\section{Candidate target selection}
\label{sec:targets}

\subsection{Predicting gamma rays above 100\,TeV from SNR-cloud pairs}
\label{sec:snr}
To find suitable target clouds for gamma-ray detection, we use the Green's catalogue of SNRs \citep{Green2019} and the interstellar cloud catalogue of \cite{Rice16} using $^{12}$CO data from the DAME survey \citep{Dame01}, which includes a distance estimator for the interstellar clouds. 
\cite{Rice16} covers Galactic latitudes ranging from $|b| \leq 1^\circ$ to $|b| \leq 5^\circ$, depending on the Galactic quadrant. Longitudes $l$ within $\sim13^\circ$ of the Galactic Centre ($l=0^\circ$) are omitted by the survey. 

We scan the catalogues for pairings of interstellar clouds and SNRs that either overlap along the line of sight or are within 100\,pc distance of each other, as evaluated from the angular separation at a given distance. 

Only a limited number of SNRs have age and distance estimates from either \cite{Green2019}, \cite{snrcat} or \cite{2019SerAJ.199...23S}; where this information is available, we impose a distance cut of the cloud distance being within 20\% of the SNR distance along the line of sight for a plausible pair. In the majority of cases, however, no distance estimate is provided for the SNR and we assume for the purposes of the flux calculation that the SNR and interstellar cloud are located at the same distance along the line of sight (i.e. that of the interstellar cloud). For several SNRs, this results in calculated fluxes for multiple interstellar clouds at different assumed distances. 
For SNRs without an age estimate, the SNR age is evaluated by inverting equation \eqref{eq:snrrad}, given the SNR radius in pc (at the assumed distance). 

An acceleration particle spectrum of $f(E,r,t) \propto E^{-2}$ is assumed at the source location, as in section \ref{sec:model}. Here we take the true size of the interstellar cloud, again with spherical symmetry assumed. 

A total of 550 potential SNR-cloud pairs are found; however, for many of these pairs the age of the system is either not sufficient to allow particles to reach the cloud, or is long enough that the highest energy particles have fully traversed the cloud. Less than 10\% of these pairings ($\sim 40$) result in a detectable very high energy gamma-ray flux above 10\,TeV produced through particle interactions with the cloud under the slow diffusion scenario. 

Figure \ref{fig:chin} shows that \replaced{density dominates over diffusion in determining the gamma-ray flux. }{for fixed density, a higher flux is achieved under faster diffusion.} The gamma-ray flux was predicted using both fast and slow values for $D_0$ and fixed $\chi$ (see table \ref{tab:constants}). In both cases, few combinations result in a gamma-ray flux above 100\,TeV detectable by CTA-S within 50 hours\footnote{corresponding to a value of $2.5\times10^{-13}\,\mathrm{TeV}\mathrm{cm}^{-2}\mathrm{s}^{-1}$}; \replaced{with none achieving this in the case of fast diffusion and only $\sim$4 under slow diffusion}{13 in the case of fast diffusion and 36 under slow diffusion} conditions. \added{For both cases the four brightest clouds are given} in tables \ref{tab:targets} \& \ref{tab:slowtargets}). 

Results presented in the remainder of this section adopt the slow diffusion case in all figures. 
As the spectral shape predicted by this model \replaced{starts to drop at the highest energies at early times,}{is flatter than the CTA sensitivity curve,} a larger number of clouds are predicted to result in a detectable gamma-ray flux above 10\,TeV. We therefore provide results both for the integral flux above 100\,TeV, corresponding to the most promising candidates, and for the integral flux above 10\,TeV as providing a larger selection of candidates to which current instruments, including H.E.S.S. and HAWC, are sensitive. This will enable us to compare our model results to existing data in Appendix~\ref{sec:hgpsclouds}. 

\begin{figure}
\centering
\begin{overpic}[width=\columnwidth]{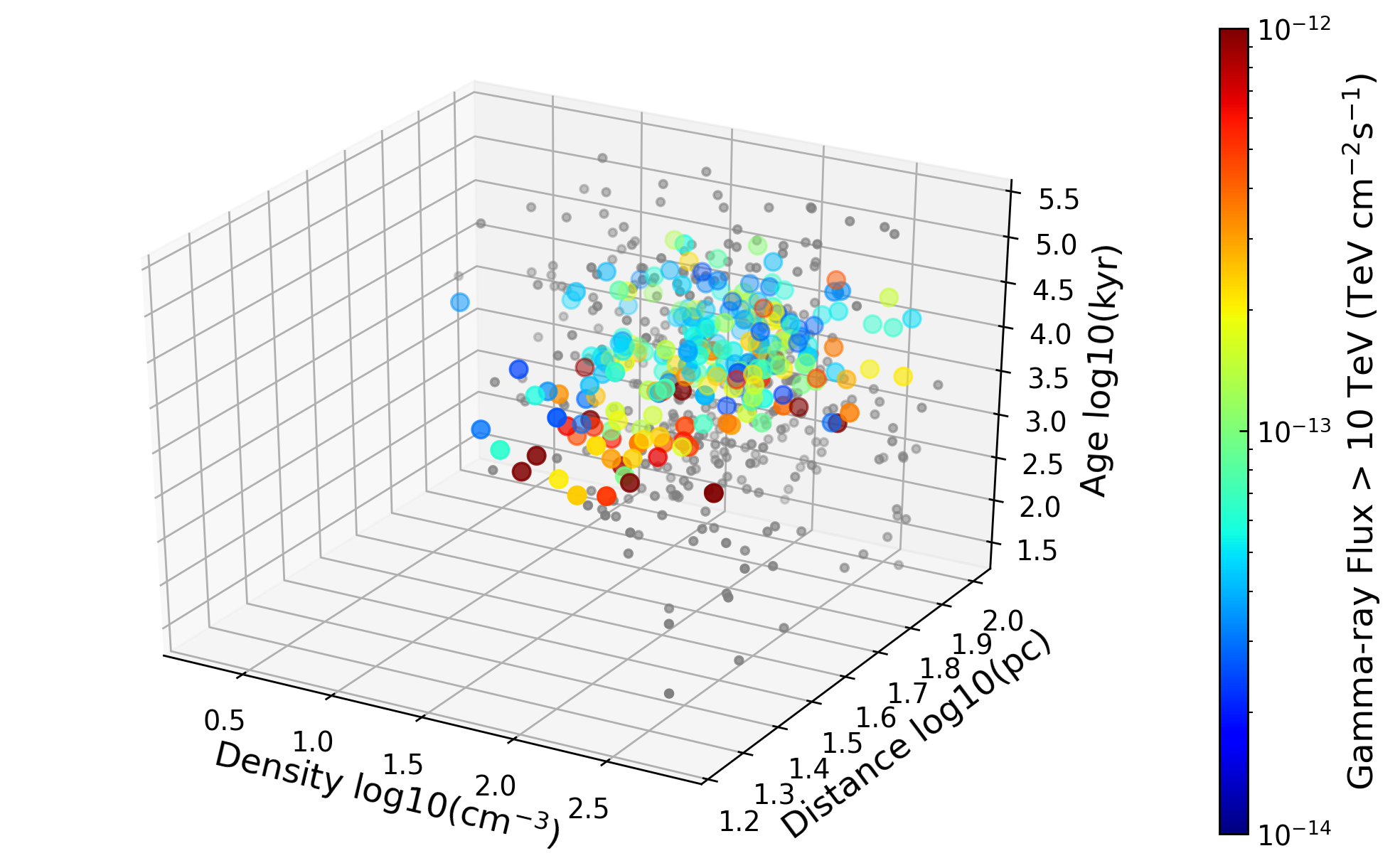}
\put(5,57){\Large{ $E_\gamma >$ 10 TeV}}
\end{overpic}
\caption{ The predicted gamma-ray flux above 10\,TeV from pairs of interstellar clouds and nearby SNRs, shown as a function of system properties. Cloud-SNR combinations that would be detectable by CTA-S under the slow diffusion scenario are shown as coloured dots, scaled according to the predicted flux. Non-detectable systems are shown as smaller grey dots. }  
\label{fig:pairphasespace}
\end{figure}

The predicted gamma-ray flux above 10\,TeV for SNR-cloud pairs detectable by CTA-S within 50 hours\footnote{corresponding to a value of $2.4\times10^{-14}\,\mathrm{TeV}\mathrm{cm}^{-2}\mathrm{s}^{-1}$} is shown in colour scale in the phase space of system properties in figure \ref{fig:pairphasespace}. Non-detectable systems are also plotted in figure \ref{fig:pairphasespace} according to their properties as smaller circles in grey.

\begin{figure}
    \centering
    \includegraphics[width=\columnwidth]{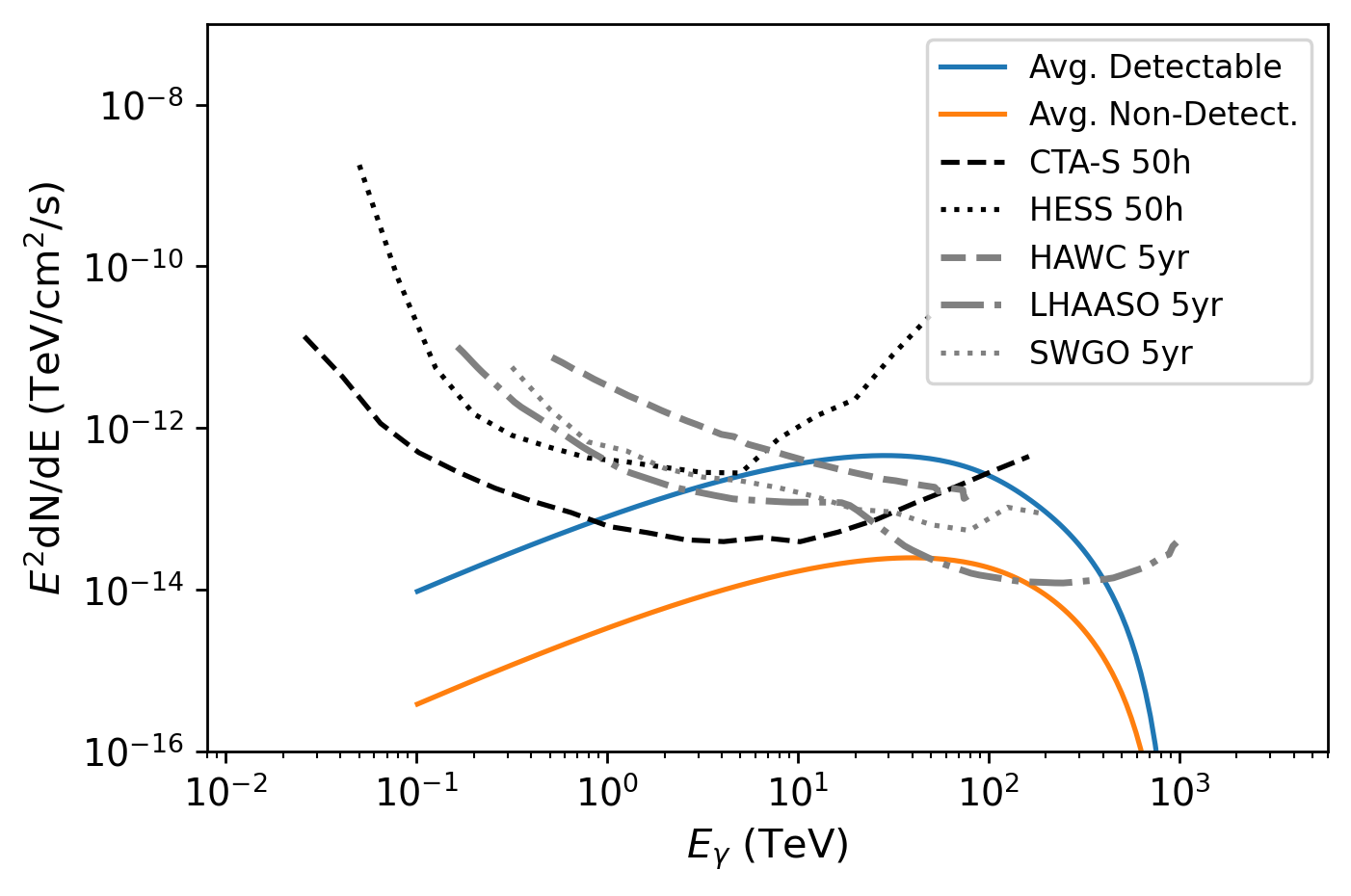}
    \caption{Gamma-ray spectrum for a system with the median properties of all SNR-cloud pairs predicted to have a gamma-ray flux detectable above 100~TeV by CTA-S, as in table \ref{tab:median}. The 50 hour sensitivities of CTA-S and of H.E.S.S. are shown for comparison, as well as the 5 year sensitivities of HAWC, LHAASO and SWGO.}
    \label{fig:medianspec}
\end{figure}

The median properties of systems detectable $>100$\,TeV are listed in table \ref{tab:median}. In general, the gamma-ray detection prospects are enhanced for systems that are closer to Earth, for younger SNRs and for larger amounts of target material. \replaced{The median cloud density for detectable systems is higher than that of non-detectable, as well as }{Although the median cloud density for detectable systems is lower than that of non-detectable; this is easily explained as compensated for by} the detectable clouds being on average more massive. Gamma-ray spectra for a system with the median properties of both detectable and non-detectable systems shown in figure \ref{fig:medianspec}. 
WCDs such as HAWC, LHAASO and SWGO have improved sensitivity over IACTs towards 100\,TeV, and may therefore be sensitive to a wider variety of systems.

\begin{table*}
    \centering
    \begin{tabular}{lccccc}
    Property & Median & Median & Median & Median \\
    & det. 100\,TeV & non-det. 100\,TeV & det. 10\,TeV & non-det. 10\,TeV \\
    \hline
    Distance from Earth (kpc) & 3.2 & 4.2 & 4.2 & 4.2 \\%
    SNR age (kyr) & 3.2 & 4.1 & 5.9 & 1.8 \\%
    SNR radius ($t=t_{\rm SNR}$, pc) & 4.9 & 8.2 & 10.0 & 6.02\\%
    SNR-cloud separation (pc) & 44 & 72 & 67 & 76 \\%
    Cloud density (cm$^{-3}$) & 86 & 55 & 55 & 56 \\%
    Cloud mass ($10^5$ M$_{\rm sun}$) & 3.01 & 0.67 & 0.87 & 0.62 \\%
    Cloud radius (pc) & 29 & 21 & 23 & 20 \\%
    \end{tabular}
    \caption{ Median properties of SNR-cloud pair systems predicted to have a gamma-ray flux above 10\,TeV or 100\,TeV that is detectable (det.) or non-detectable (non-det.) by CTA-S, under the slow diffusion scenario. }
    \label{tab:median}
\end{table*}

\begin{figure*}
    \centering
    \includegraphics[width=\columnwidth]{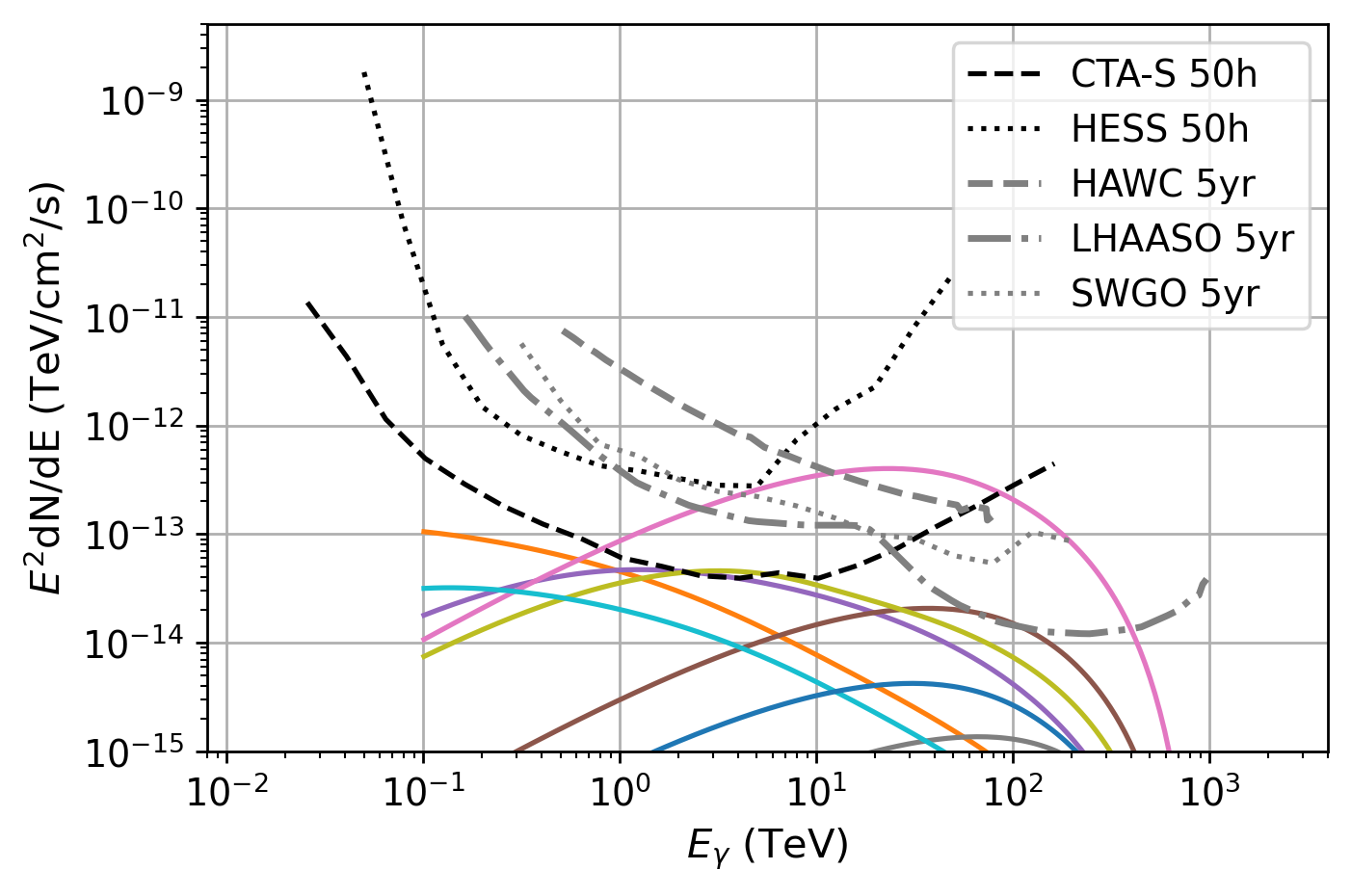}
    \includegraphics[width=\columnwidth]{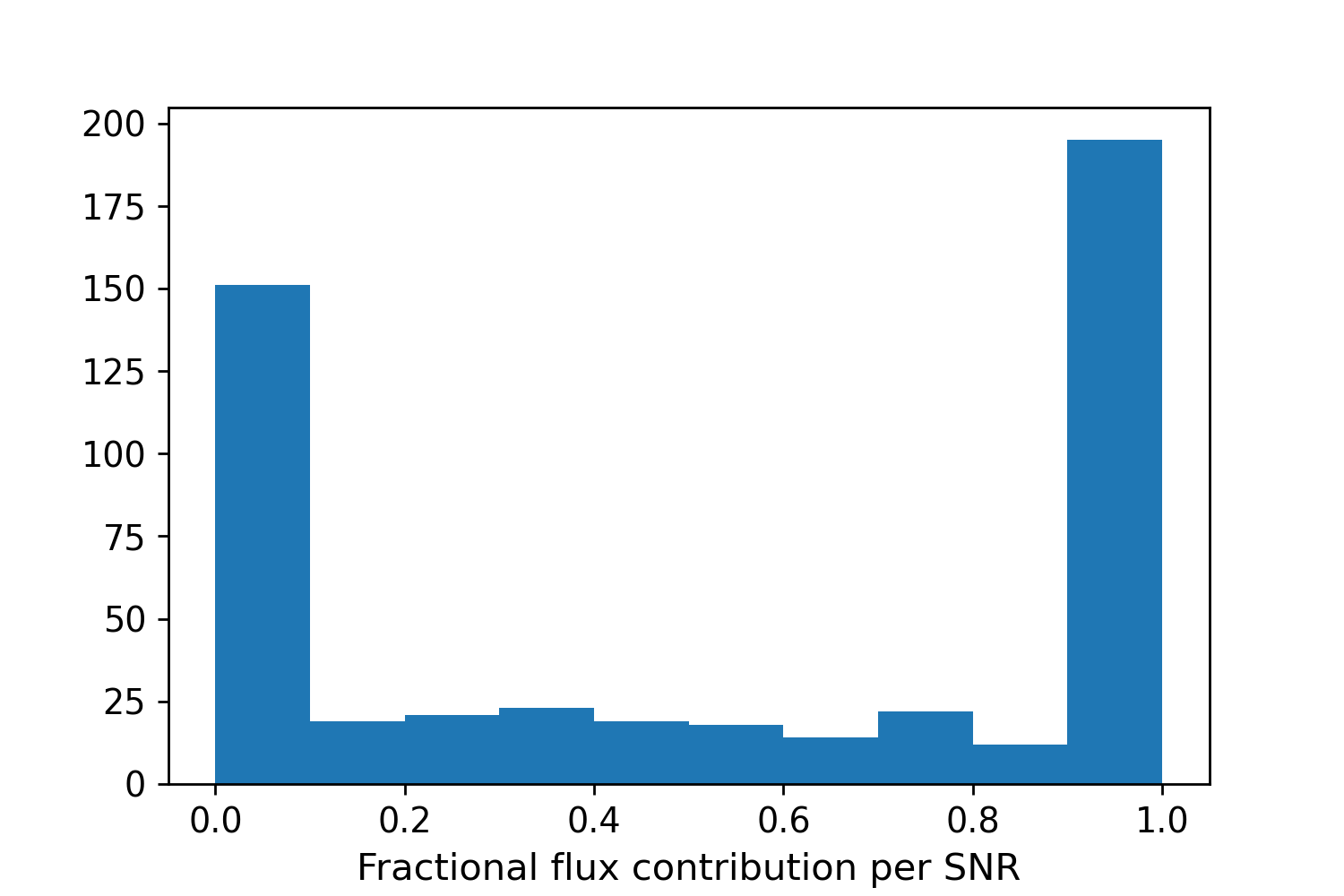}
    \caption{Contributions of different SNRs to the total predicted flux per cloud. {\bf Left:} an arbitrary sub-sample of example spectra are shown in different colours for different clouds (approx. 5\% of the total). 
    {\bf Right:} Distribution of fractional contributions of individual SNRs to the total flux from a given cloud. }
    \label{fig:totalcloudspec}
\end{figure*}

\subsection{Total gamma-ray spectrum per cloud}
\label{sec:totalgammaspec}

The total predicted gamma-ray flux from a given interstellar cloud is here overestimated, under the assumption that all SNRs contributing to the total flux are within a distance comparable to the cloud along the line of sight. This obviously cannot be true in reality, as SNRs without an intrinsic distance estimate have different assumed distances for various cloud pairings. Therefore, if one cloud is as bright as predicted, then another cloud associated with the same SNR yet at a different distance along the line of sight must be correspondingly dimmer in gamma-ray flux. 

In practise, when observing the sky in gamma rays, it will be challenging to disentangle the contributions of different SNRs (or other nearby accelerators) to the total gamma-ray flux emanating from an interstellar cloud. We therefore sum the predicted contributions of all SNRs to obtain a prediction for the maximum total gamma-ray flux from a given interstellar cloud.

Figure \ref{fig:totalcloudspec} shows the total gamma-ray spectrum from an arbitrary sub-sample of target clouds (approx. $5\%$), not all of which would be detectable by CTA. 
The right panel of figure \ref{fig:totalcloudspec} shows the fractional contribution of individual SNRs to the total flux per cloud; it is clear that in the majority of cases the emission is dominated by a single SNR, with other contributions only minor in comparison.

In table \ref{tab:targets}, the interstellar clouds are listed according to the predicted gamma-ray flux above 100\,TeV. 
The predicted gamma-ray fluxes are compared to flux upper limits taken from the H.E.S.S. Galactic Plane Survey (HGPS) with a $0.2^\circ$ correlation radius. As the HGPS coverage is from $l=250^\circ$ to $65^\circ$ and $|b| \leq 3^\circ$, no upper limits were evaluated for clouds located outside of this range.
These flux upper limits $F^{UL}$ evaluated using a $0.2^\circ$ correlation radius were scaled to the true angular size of the cloud $\sigma_c$ assuming a scaling behaviour as:

\begin{equation}
    F^{UL}_\gamma \propto \sqrt{\left(\sigma^2_{c}/\sigma^2_{\mathrm{psf}}\right) +1}\,,
    \label{eq:fscale}
\end{equation}

\noindent where $\sigma_{\mathrm{psf}}=0.2^\circ$ in this case. 
The flux upper limits from the HGPS assume a power law gamma-ray spectrum with an index of $-2.3$, enabling the integral flux upper limits and energy threshold to be converted into a comparable integral energy flux upper limit above 100\,TeV. 

Although the predicted flux from this model exceeds the upper limit from the HGPS in several cases, we note that for most of these clouds, multiple SNRs contribute to the total flux prediction for a given cloud. 
Figure \ref{fig:totalcloudspec} shows that in the majority of cases, the fractional contribution from a single SNR dominates the total flux ($>90\%$) whilst other SNRs only providing a minor contribution ($<10\%$).
In table \ref{tab:targets}, the fractional contributions from different SNRs are explicitly given; 
in many cases, one SNR dominates and the majority of the individual contributions to the total flux from different SNR lie below the upper limit from the HGPS. 

As noted above, in reality the corresponding SNRs will be located at different distances along the line of sight and cannot simultaneously contribute to the gamma-ray flux from a single cloud. 
In fact, flux upper limits corresponding to the locations of illuminated clouds could be used to constrain the line-of-sight distance to SNRs; in cases where a detectable flux is expected should the SNR be at a comparable distance to the cloud. 

For several cases, there is no known distance estimate to the SNR, which was assumed to be located at the same line-of-sight distance as the cloud for the purpose of this model. 
Therefore, if the predicted flux were to exceed the flux upper limit from H.E.S.S., this would imply that the SNR is not located at a distance comparable to the cloud.  
Those SNRs for which a distance estimate was available are indicated in bold in table \ref{tab:targets}.

Further uncertainties that could \replaced{contribute to the}{account for these discrepancies between the upper limits and} predicted gamma-ray flux are discussed in section \ref{sec:sysunc}.

\begin{table*}
	\centering
	\caption{The four brightest interstellar clouds in the vicinity of SNRs that result in a gamma-ray flux above 100\,TeV from our model. In cases where multiple SNRs contribute to the total predicted flux $F^{\mathrm{total}}_\gamma$, their fractional contributions are also given, ordered from highest to lowest flux. Contributions of $\ll 1\%$ to the total flux are omitted.
	SNRs with a distance estimate available are marked in bold. For those clouds covered by the HGPS, a measured upper limit on the gamma-ray flux $F^{UL}_\gamma$ is derived from the HGPS data \citep{HGPS}. } 
    \label{tab:targets}
	\begin{tabular}{c|l|l|l|l|l|c|l|l}
	\hline
	 Cloud ID & Cloud coordinates & Cloud size & Cloud distance  & $EF^{UL}_\gamma >$ 100\,TeV & $EF^{\mathrm{total}}_\gamma$ $>$ 100\,TeV & \# SNRs & SNR & flux contribution\\
	  & (l, b) deg & deg & kpc & TeV cm$^{-2}$ s$^{-1}$ & TeV cm$^{-2}$ s$^{-1}$ & & & TeV cm$^{-2}$ s$^{-1}$ \\
	 \hline
7 & (337.84,-0.40) & 0.465 & 2.910 & 3.76e-13 & 1.02e-13 & 2 & G337.2-00.7 & 5.57e-14 \\ 
& - & - & - & - & - & - & G338.1+00.4 & 4.61e-14 \\ 
5 & (336.73,-0.98) & 0.516 & 3.240 & 7.56e-13 & 9.94e-14 & 2 & G337.2-00.7 & 7.41e-14 \\ 
& - & - & - & - & - & - & G336.7+00.5 & 2.53e-14 \\ 
6 & (34.99,-0.96) & 0.727 & 2.620 & 3.45e-13 & 8.98e-14 & 2 & G033.2-00.6 & 6.77e-14 \\ 
& - & - & - & - & - & - & G036.6-00.7 & 2.21e-14 \\ 
1 & (333.46,-0.31) & 0.627 & 3.370 & 5.95e-13 & 7.93e-14 & 1 & G332.4-00.4 & 7.93e-14 \\ 
    \hline
	\end{tabular}
	
\end{table*}

\begin{table*}
	\centering
	\caption{As for table \ref{tab:targets} except for clouds with a CTA-S detectable gamma-ray flux above 100~TeV, as predicted in the case of a slow diffusion scenario.} 
    \label{tab:slowtargets}
	\begin{tabular}{c|l|l|l|l|l|c|l|l}
	\hline
	 Cloud ID & Cloud coordinates & Cloud size & Cloud distance  & $EF^{UL}_\gamma >$ 100\,TeV & $EF^{\mathrm{total}}_\gamma$ $>$ 100\,TeV & \# SNRs & SNR & flux contribution\\
	 & (l, b) deg & deg & kpc & TeV cm$^{-2}$ s$^{-1}$ & TeV cm$^{-2}$ s$^{-1}$ & & & TeV cm$^{-2}$ s$^{-1}$ \\
	 \hline
1 & (333.46,-0.31) & 0.627 & 3.370 & 5.95e-13 & 3.25e-13 & 1 & G332.4-00.4 & 3.25e-13 \\ 
3 & (110.43,1.89) & 1.303 & 0.600 & N/A & 2.79e-13 & 1 & G114.3+00.3 & 2.79e-13 \\ 
5 & (336.73,-0.98) & 0.516 & 3.240 & 7.56e-13 & 2.59e-13 & 1 & G337.2-00.7 & 2.59e-13 \\ 
9 & (341.34,0.21) & 0.284 & 2.410 & 3.56e-13 & 2.49e-13 & 2 & G341.2+00.9 & 2.08e-13 \\ 
& - & - & - & - & - & - & G343.1-00.7 & 4.14e-14 \\ 
	 \hline
	 \end{tabular}
\end{table*}

\subsection{Including contributions from `invisible' SNRs traced by pulsars}
\label{sec:pulsarsnr}

Given that on average three SNR are expected every century, the total number of known SNR according to Green's catalogue, 294, is rather low. Out of the 126 SNR with age estimates, 86 are estimated to be younger than 20\,kyr and therefore within the Sedov-Taylor expansion phase; compared with the 600 SNR expected from the average rate. 
This could be due to the fraction of the Galaxy that has been observed; surveys such as THOR and GLEAM are showing promising results with many candidate objects that could increase the number of known SNRs \citep{AndersonTHOR,GLEAM_new2019PASA,GLEAM_candidate2019PASA}.

However, there are many more pulsars known, with a total of \replaced{2360}{2294} detected within our Galaxy \citep{Manchester05}.\footnote{Millisecond pulsars are excluded from this estimate.} Every pulsar is born in a supernova explosion, due to the collapse of the stellar core as the outer layers are expelled outwards into the ISM. As such, the presence of pulsars can be used as an indicator for SNRs that are otherwise `invisible'; dimly radiating or no longer actively accelerating particles. 
The ATNF catalogue was used to select the most promising pulsar candidates, with \replaced{63}{230} listed as located within our Galaxy (distance $< 25$\,kpc); characteristic age younger than 30\,kyr and are not millisecond pulsars (rotation period $> 10$\,ms). 

The age of the progenitor SNR is then assumed to correspond to the pulsar characteristic age $\tau_c = P/2\dot{P}$; however this assumes magnetic dipole radiation (with braking index $n=3$) and is known to be a rough approximation to the true pulsar behaviour \citep{GaenslerSlane06}.
Additionally, several pulsars have a measurable proper motion; during the original SNR explosion, the pulsar may receive a kick velocity propelling it to travel away from the birth location \citep{GaenslerSlane06}. In these cases, the birth location of the pulsar was determined by extrapolating the proper motion of the pulsar in the opposite direction for a distance corresponding to the characteristic age $\tau_c$ -- this birth location was assumed to correspond to the central position of the corresponding `invisible' SNR. 
Similarly, the hypothetical SNR radius could be calculated based on the age $\tau_c$, and additional plausible pairs of hypothetical SNRs and nearby interstellar clouds were identified. The predicted gamma-ray fluxes arising from these clouds due to invisible SNRs as traced by pulsars were then added to the gamma-ray flux predictions calculated above. 
This then provided a revised estimate for the predicted gamma-ray flux from each cloud. Fractional contributions to the gamma-ray flux above 100\,TeV from hypothetical SNRs are given in table \ref{tab:psrsnrtargets} of Appendix \ref{sec:psrsnrgamma100}. 

Many more systems are predicted to result in a detectable flux above 10\,TeV, which nevertheless implies the presence of 0.1\,PeV protons. Up to $\sim40$ of these may be detectable by current IACT facilities. The full list is provided in table \ref{tab:targetspsrsnrhess} of Appendix \ref{sec:gamma10}. Maps of the predicted flux above 10\,TeV from clouds in the Galactic Plane in figure \ref{fig:GalPlane10TeV}; and above 100\,TeV in figure \ref{fig:GalPlane100TeV} show the total predicted gamma-ray flux, including contributions from both catalogued and hypothetical SNRs.

\subsection{Systematic Uncertainties due to model assumptions}
\label{sec:sysunc}
In order to estimate some of the systematic uncertainties inherent due to assumptions in the model, we tested the variation in predicted flux arising from varying the spectral slope of proton injection. 

As described in section \ref{sec:model}, we assumed a power law acceleration spectrum with slope $\alpha = 2$. This was found to have a strong influence on the predicted integral flux above 100\,TeV. Adopting a hard value of $\alpha = 1.8$ increased the predicted flux by up to a factor $\sim 30$, whilst a softer slope of $\alpha = 2.2$ decreased the flux \replaced{marginally}{by a factor $\sim 4$} with respect to $\alpha = 2$. Note that observationally most of the gamma-ray detected SNRs show steeper spectra than $E^{-2}$, especially middle-aged ones. An alternative spectral shape for the initial proton spectrum, such as a broken power law or power law with exponential cut-off, may also be expected to reduce the predicted flux at the highest energies. The resulting variation in cloud detectability is indicated in table \ref{tab:cloud_detectability} in the Appendix. 

Similarly, the flux upper limits from the HGPS were derived assuming a spectral index of 2.3, whilst the mean spectral index of sources listed in the HGPS is $2.4 \pm 0.3$. Therefore, we varied the spectral index for the derived upper limit and found that an index of 2.1 relaxes the upper limit by a factor 3, whilst an index of 2.5 tightens the upper limit by a factor 3 (with respect to the quoted value for an assumed index of 2.3).

Flux predictions provided by this model can hence be considered an order of magnitude estimate. In table \ref{tab:targets} \deleted{and \ref{tab:slowtargets}}, \replaced{the predicted fluxes that currently lie below the detectable limit may be enhanced by an order of magnitude}{cases where the predicted flux exceeds an upper limit obtained from the HGPS by less than an order of magnitude may be considered compatible} within the limitations of the model.

As mentioned in section \ref{sec:model}, the Sedov time occurs at characteristically different ages for different supernova types. Although a distinction between core collapse and type Ia events can be made for a small number of SNRs based on their immediate environment, for the majority of cases this is uncertain. The core collapse scenario was adopted as default, due to the progenitor massive stars being somewhat younger than the progenitor white dwarfs of type Ia events and therefore more likely to be in environments harbouring rich molecular clouds. Nevertheless, we explored the influence of a type Ia Sedov time on the results, finding that \replaced{only}{all except for} two of the clouds
\deleted{in table \ref{tab:targets}} predicted to be detectable by CTA remain so under the type Ia scenario. 

\replaced{The variation in predicted flux under type IIb and type Ia scenarios implies}{Interestingly, there were 3 clouds that only classed as detectable for type Ia events, implying} that flux constraints on clouds may also be used to investigate intrinsic properties of the SNR.

\section{Discussion}

As explained in section \ref{sec:snr}, although the integral gamma-ray flux above 100\,TeV indicates the most promising candidates for evidence of PeVatron activity, the integral gamma-ray flux above 10\,TeV provides a broader range of candidates to which current generation instruments are sensitive. Bright gamma-ray flux above 10\,TeV nevertheless indicates the presence of energetic particles within an order of magnitude of the CR knee. A summary of the clouds predicted by our model to be detectable by H.E.S.S. above 10\,TeV (under the slow diffusion scenario) is provided in Appendix \ref{sec:gamma10}. 

\subsection{Potentially bright clouds overlapping HGPS sources}

\begin{figure}
    \centering
    \includegraphics[width=22cm,angle=90]{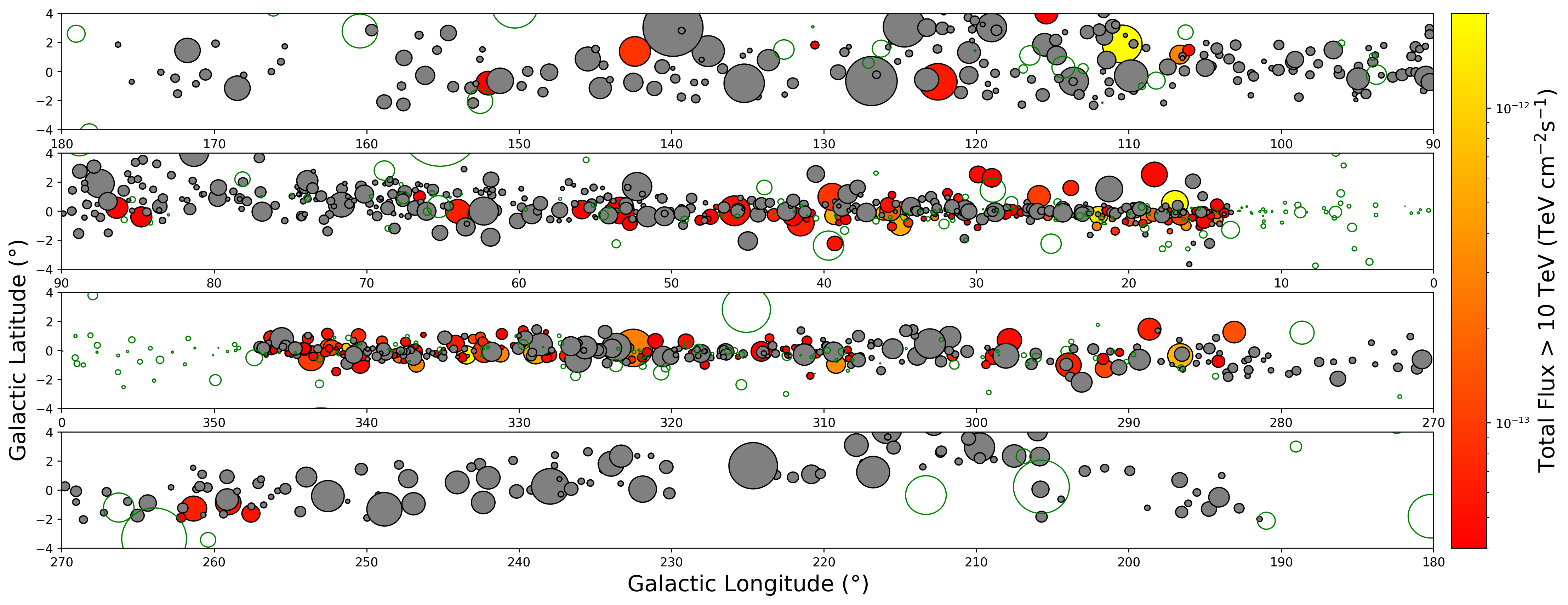}
    \caption{Predicted integral gamma-ray flux above 10\,TeV from interstellar clouds illuminated by escaped particles from nearby SNRs, using interstellar clouds identified in \citep{Rice16}. Clouds detectable by CTA are shown in colour scale, whilst non-detectable are shown in grey. 
    The total includes contributions from SNRs as listed in Green's catalogue (green circles) as well as contributions from hypothetical SNRs as traced by the energetic pulsar population \citep{Green2019}.}
    \label{fig:GalPlane10TeV}
\end{figure}

\begin{figure}
    \centering
    \includegraphics[width=22cm,angle=90]{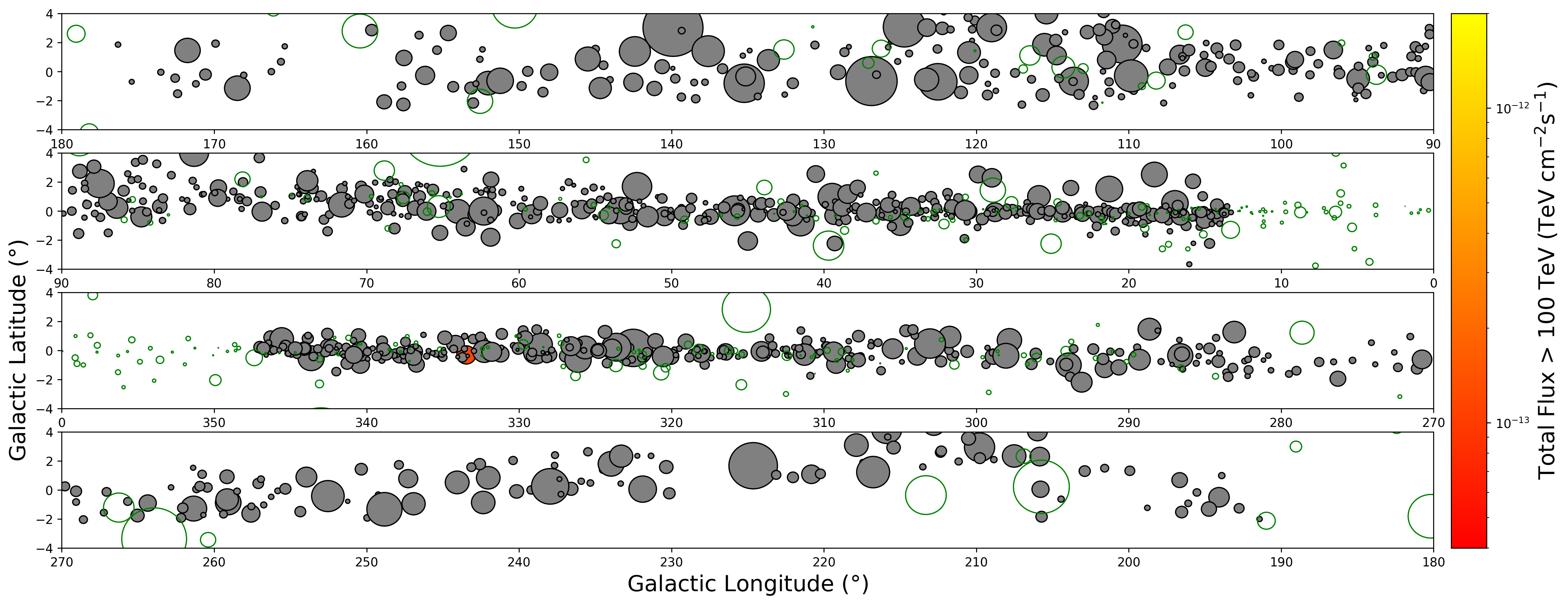}
    \caption{As for figure \ref{fig:GalPlane10TeV} except for gamma-ray flux above 100\,TeV.}
    \label{fig:GalPlane100TeV}
\end{figure}

Appendix \ref{sec:hgpsclouds} shows an enlarged view of the clouds in the Galactic plane predicted to harbour a gamma-ray flux above 10\,TeV detectable by H.E.S.S. (as in figure \ref{fig:GalPlane10TeV}). Figures \ref{fig:GPzoomab} to \ref{fig:GPzoomgh} show significance contours at 3, 5 and 15 $\sigma$ from the H.E.S.S. Galactic Plane Survey (HGPS) overlaid, together with the name and nominal location of each known source. Due to the limited coverage of the surveyed clouds by \cite{Rice16} and \cite{Dame01}, the Galactic latitude is restricted to $b \leq 2.5^\circ$. Similarly, due to the limited coverage of the HGPS, the longitude is restricted to between $l=64^\circ$ to $l=262^\circ$ running through the Galactic Centre. 

In several regions, potentially detectable clouds are seen to overlap with known H.E.S.S. sources, although caution should be exercised over the assumption of spherical symmetry. We discuss here a few selected regions in more detail in turn. 

\subsubsection{$49^\circ > l > 43^\circ$}

\deleted{The cloud situated at $l\sim49^\circ$ partially overlapping with HESS\,J1923+141, is illuminated by the SNR G049.2-0.7, with which the TeV source is associated and cloud interactions already expected, consistent with our prediction.}  

At $l\sim 46^\circ$, there are two clouds overlapping with a gamma-ray significance contour not currently associated with a H.E.S.S. source. This position is coincident with the unidentified TeV source discovered by HAWC, 2HWC J1914+117. The larger cloud at (45.86,0.01) is illuminated by the SNR \replaced{G043.9+1.6}{G045.7-0.04}, with which a shell-like Fermi source is associated, whilst the small cloud at this location is associated with the SNR \added{G046.8-0.3}. Two pulsars are located nearby, yet without $\dot{E}$ estimates: PSR J1915+1144 and PSR J1915+1149 at distances of 7.2\,kpc and 14\,kpc respectively. The predicted gamma-ray flux is compatible with the upper limit derived from the HGPS.
A further cloud coincident with HESS\,J1911+090 is illuminated by particles escaping from the SNR G042.8+0.6, which itself is tentatively associated with the nearby source HESS\,J1908+063. HESS\,J1911+090 is identified as W49B, corresponding to the SNR G043.3-0.2 already known to be interacting with molecular clouds. 

\subsubsection{$40^\circ > l > 37^\circ$}

In this region, the cloud with the highest predicted flux, at (39.33,-0.32), is well compatible with the upper limit from the HGPS, with contributions from the SNRs G040.5-0.5 and G038.7-1.3, with a distances constrained at 3.2\,kpc and 4\,kpc respectively. It is unlikely that both associated SNR contribute to the total flux; the prediction from either SNR separately is approximately half of the total predicted flux for this cloud.

\subsubsection{$37^\circ > l > 34^\circ$}

This region is shown in the second panel of figure \ref{fig:GPzoomab}, where a group of clouds overlap with HESS\,J1857+026 and HESS\,J1858+020. Both of these sources remain unidentified, although it is known that there is substantial molecular material in the vicinity \citep{Paredes14}. HESS\,J1857+026 was best described by two components that were merged to a single source in the HGPS, whereas the MAGIC collaboration favour a two source interpretation, associating MAGIC J1857.2+0263 with the pulsar PSR J1856+0245, whilst MAGIC J1857.6+0297 remains unidentified \citep{HGPS,MAGIC14_J1857}. A higher resolution view of the source reveals that the gamma-ray peak may coincide with a wind-blown cavity, suggesting a leptonic origin of the radiation; however there is considerable overlap with molecular gas and a hadronic origin is not excluded by \cite{MAGIC14_J1857}. 

The predicted gamma-ray flux from the brighter cloud at (36.100,-0.140) is provided by escaped particles from the SNR\,G035.6-00.4, for which the distance is constrained. MAGIC J1857.6+0297 is positionally coincident with this cloud, whilst G035.6-00.4 is itself associated with the TeV source HESS\,J1858+020. Detailed gamma-ray studies of this region with high angular resolution are required to disentangle the multiple components. 

Similarly, for HESS\,J1858+020, \citep{Paredes14} note the rich molecular environment and suggest that the non-thermal emission may arise from interactions between a SNR and nearby clouds. 
The predictions of our model that clouds should be detectable in gamma-rays in this region is therefore consistent with the data. 

The large and rather bright cloud at (34.99,-0.96) is illuminated predominantly by the SNR \replaced{G036.6-0.7}{G034.7-0.4}, for which no TeV detection has yet been made, despite a GeV detection by Fermi-LAT \deleted{and known cloud interactions} \citep{snrcat}. \replaced{However, as the distance to the SNR is known the flux prediction, which lies below the HGPS upper limit, is uncertain. }{As the distance is to the SNR known and the flux prediction comparable to the HGPS upper limit,lying approximately 10\% below it, further observations would soon be able to constrain our prediction.}

\subsubsection{$34^\circ > l > 31^\circ$}

The cloud overlapping with HESS\,J1852-000 - at (32.73,-0.42) - has a predicted flux with contributions from both the SNR \added{G033.2-0.6} and G033.6+0.1, \added{as well as lesser contributions from several other SNRs}. \deleted{and from hypothetical SNRs, and as such is inherently uncertain} 
The flux predicted by our model is consistent with the upper limit derived from the HGPS. However, it has been previously suggested that this unidentified TeV source is associated with a molecular cloud interaction, although a PWN associated with pulsars in the vicinity remains a plausible alternative scenario \citep{HGPS}.

A few clouds coincide with contours around HESS\,J\replaced{1848-018}{1852-000}; a complex source without a firm identification yet potentially corresponding to an SNR interacting with adjacent molecular clouds.

\subsubsection{$30^\circ > l > 27^\circ$}

HESS J1843-033 is a complex source, also detected by HAWC as 2HWC J1844-032, and is highly structured, showing multiple Gaussian components in the HGPS \citep{HGPS}. The location is \replaced{adjacent to}{coincident with} several clouds, \replaced{such as the cloud}{the largest} located at (28.770, -0.090) and illuminated predominantly by the SNR \replaced{G028.8+1.5}{G029.6+0.01}. 
This prediction is fairly reliable, as the SNR has a known distance of 2.8-4\,kpc. \replaced{Similarly, the adjacent cloud at (27.64,0.07) is}{Several clouds in the region are also} predicted to have a gamma-ray flux produced by particles escaping from the SNR G028.6-0.1, \added{of known distance and potentially associated to the source HESS\,J1843-033.} 
\deleted{The predictions of our model also indicate at least a contribution to the total flux from 
HESS J1843-033 is consistent with gamma-ray emission arising from escaped particles from the SNR G028.6-0.01 interacting with nearby clouds.} 

\subsubsection{$27^\circ > l > 25^\circ$}

\replaced{Several}{A cluster of} clouds overlap with the TeV gamma-ray sources HESS\,J1841-055 (unidentified) and HESS J1837-069, which is firmly identified as a pulsar wind nebula, powered by the energetic pulsar PSR J1838-0655 \citep{HGPS}.  
Nonetheless, the SNR associated with \deleted{several} clouds around $l=25^\circ-26^\circ$ in our model is G024.7+0.6, from which MAGIC has reported the detection of a source not listed in the HGPS; MAGIC J1835-069 located at (24.94, 0.37) \citep{MAGICJ1835}. 
The predicted fluxes are, however, compatible with derived upper limits from the HGPS. 
As G024.7+0.6 has a known distance of 3.2 - 3.7 kpc, this provide some confidence in the possibility that escaped particles may indeed be interacting with these clouds; yet the resulting flux is likely below that of the dominant gamma-ray source HESS\,J1837-069, and therefore not currently identifiable. 
Note that the adjacent cloud at \replaced{(23.80,1.58)}{(25.890, 1.000)} is also associated with SNR G024.7+0.6. 

\subsubsection{$23^\circ > l > 20^\circ$}

In this region, there is a large, bright cloud nestled between three known H.E.S.S. sources; the gamma-ray binary HESS\,J1832-093; the pulsar wind nebula HESS\,J1833-105 (associated with the pulsar PSR\,J1833-104 and the SNR G021.5-00.9); and the dark source HESS\,J1828-099, with no known associations. 
The cloud in question, 4 at (21.970, -0.290), derives its predicted flux above 10 TeV from a combination of the SNRs \deleted{G021.0-0.4 and} G021.6-00.8, about which little is known, and from G021.5-0.9.
The predicted integral flux above 10\,TeV is consistent with the derived upper limit, although as discussed in section \ref{sec:sysunc}, uncertainties of around an order of magnitude are inherent in our model assumptions. 

From figure \ref{fig:GPzoomab}, it can be seen that there appear to be significant contours at the location of this cloud. Indeed, there is a Gaussian component HGPSC\,072 reported here with a TS (test statistic) of 59, well above the detection threshold of TS$=30$, and a reported flux above 1\,TeV of $8.1\times 10^{-13}$cm$^{-2}$s$^{-1}$ \citep{HGPS}. This component was, however, not reported as a gamma-ray source due to the lack of a significant detection in the independent cross-check analysis (see section 4.9 of \cite{HGPS}). We conclude that the lack of a currently known TeV gamma-ray source at the location of this cloud does not rule out the existence of detectable gamma-ray emission emanating from this cloud due to interactions with  energetic particles originating from the SNRs \deleted{G021.0-0.4 and} G021.6-00.8 \added{and G021.5-0.9}.

\subsubsection{$20^\circ > l > 14^\circ$}
\label{sec:j1825}
This is a complex region in which the large, bright pulsar wind nebula HESS\,J1825-137 dominates, exhibiting strong energy dependent morphology with extended emission reaching as far as the gamma-ray binary LS\,5039 (HESS\,J1826-148) and rendering the source HESS\,J1826-130 identifiable only at the highest energies \citep{j1825}. The presence of molecular material in this region is, however, well established; the offset of the leptonic gamma-ray emission (i.e. the parent electron population) from the pulsar PSR\,J1826-1334 is thought to be due to shock interactions between the PWN and molecular material towards HESS\,J1826-130. 
Supporting this scenario is the identification of high turbulence within the nearby cloud at (18.16, -0.34) \citep{Voisin16}. 
It is worth noting that emission from this region has recently been confirmed to extend beyond 100\,TeV, as reported for the source eHWC\,J1825-134, which overlaps multiple H.E.S.S. sources due to the limited angular resolution of HAWC \citep{HAWC56tev}. 

Adjacent to HESS\,J1826-130, the cloud at (18.81,-0.46) is illuminated by the SNR \replaced{G018.1-0.1}{G018.6-0.2} which is already potentially associated to the HESS source itself; the distance to both cloud and SNR is estimated at around 5.6-6.8\,kpc. 

Many clouds are seen to overlap with HESS\,J1825-137; however, clouds with predicted fluxes lower than that measured from the PWN are likely dwarfed by its gamma-ray emission. 
The larger cloud at (16.970, 0.530) slightly offset from the PWN has \replaced{two}{four} SNRs contributing to the total predicted flux; \deleted{only one of which} \added{G019.1+0.2 and} G018.9-01.1, \added{whereby only the latter} has an estimated distance. The contribution from this more certain SNR-cloud association is at the level of $\sim 40\%$ of the total predicted flux (see  
table \ref{tab:targetspsrsnrhess}). 
For this system, the total predicted flux \deleted{from all four contributions} is compatible with the derived upper limits from the HGPS. 
\added{Note that the bright cloud at (16.97, 0.53) is additionally illuminated by a hypothetical SNR, corresponding to the pulsar J1826-1256; which is the pulsar potentially powering the HESS\,J1826-130 PWN (see table \ref{tab:psrsnrtargets}). }

\added{The brighter cloud at (16.24,-1.02) has a predicted flux that exceeds the HGPS upper limit and is almost evenly split between two SNRs, G016.4-0.5 and G016.0-0.5 (see tables \ref{tab:targets} and \ref{tab:slowtargets}), neither of which has an available distance estimate. }
As the region has been extensively observed in recent years, we can surmise that it is likely the primary contributing SNR, G016.4-00.5 (flux $>$10\,TeV of $5.81\times 10^{-13}$TeV cm$^{-2}$s$^{-1}$, table \ref{tab:targetspsrsnrhess}), is not located at $\sim 2$ kpc and hence not associated with this cloud. 

\added{The smaller cloud at (14.14,-0.59) has a predicted flux exceeding the HGPS upper limit by $\sim50\%$, arising solely from the SNR G015.1-1.6, which has a known distance of $\sim 2.1-2.2$\,kpc well compatible with the distance to the cloud of 2.09\,kpc. In this case, one or more of the model assumptions concerning propagation through the intervening medium, age of the SNR or spectral properties of the accelerated particles, must be incorrect. }

\subsubsection{$347^\circ > l > 341^\circ$}

Two of the known TeV sources in this region, HESS\,J1708-410 and HESS\,J1702-420, remain without any clear counterparts or multiwavelength associations. Several of the smaller clouds coincident with significant emission in the region are predictions for particles accelerated by the SNR G344.7-0.1, which is known to be interacting with clouds. 

\added{An extended cloud at (343.64,-0.54) is illuminated primarily by the SNR G343.1-2.3, for a total flux at a level of a quarter of the upper limit obtained from the HGPS; for this SNR again the distance estimate of 2\,kpc is based on an associated pulsar. }

\added{The small bright cloud at (341.34,0.21) is illuminated by the SNR G341.2+0.9 for a predicted flux slightly above the corresponding HGPS upper limit, although again the distance to this SNR is unknown, suggesting that it is plausibly not located at a distance comparable to the cloud at $\sim2.4$\,kpc }

\replaced{Several of the}{The three brightest} cloud predictions at Galactic longitudes slightly lower than HESS\,J1702-420, derive a significant fraction of their predicted flux from the SNR G342.0-0.2, for which no distance or age estimates are provided. Given the absence of significant gamma-ray emission, this may indicate that the SNR is not located at the same distance as the clouds (of $\sim2.5$\,kpc). 

\subsubsection{$341^\circ > l > 336^\circ$}
Westerlund 1, with the designation HESS\,J1646-458, is a massive stellar cluster with highly complex gamma-ray morphology comprising multiple components \citep{HESSWest1}. Massive stellar clusters are thought to be a suitable complementary (or even alternative) PeVatron candidate to SNRs, accelerating cosmic rays through stellar winds and multiple shock interactions from several supernovae \citep{AharonianMassiveStars19}. There is evidence for at least one young SNR $\lesssim 10$\,kyr in Westerlund 1, established by the presence of a magnetar associated with the cluster \citep{Wd1magnetar}. Although for this paper we focus on SNRs, a dedicated study would be required to model the complex Westerlund 1 region in detail. Here, we merely note that the predicted fluxes for clouds at Galactic longitudes $\gtrsim 338^\circ$ are in general compatible with derived upper limits from the HGPS, 
whilst the reported flux has been quoted with an uncertainty of a factor 2 \citep{HGPS}.

There are several clouds situated around HESS\,J1640-465 and HESS\,J1641-463, where the former is a composite SNR and the latter is (similarly to HESS\,J1826-130) a hard spectrum source of unknown origin that reveals its presence towards higher energies \citep{HESSJ1641}. Molecular material coincident with the emission `bridge' between the two H.E.S.S. sources has been identified \citep{Lau17}, however we note that the gamma-ray emission contours do not consistently correspond to all of these clouds. 
\added{For the cloud at (337.84,-0.40) located at $\sim2.91\,$kpc, the SNR G337.2-0.7 has an estimated distance spanning a wide range from $2-9.3$\,kpc such that the true distance may indeed be different from $\sim3$\,kpc. }
The other contributing SNRs do not have firm distance estimates, such that these model results are in any case intrinsically uncertain. 

Next, we turn to clouds overlapping at $l\sim337^\circ$ with HESS\,J1634-472 - an unidentified source known to overlap dense CO gas and with plausible SNR associations \citep{somJ1634}. In this case, it seems that our model may be verified and the emission is associated with escaped particles from nearby SNRs. 
At least some of the distance estimates are broadly consistent with the large 14\,kpc distance - for several clouds in the vicinity, the reported distance is $\sim10-14$ kpc. \deleted{For the bright cloud at (337.84,-0.40) however, the distance is estimated on the low side at 2.91\,kpc, for an association with SNR G337.2-0.2 which has a distance estimated in the range $2-9.3$\,kpc.} 

Lastly, we note the brighter cloud at (336.73,-0.98) \deleted{and (336.92,-0.49) both of} which also derives \replaced{its flux predominantly}{the largest fraction of their flux} from the \replaced{aforementioned}{same} SNR, G337.2-0.7, \added{which has an estimated distance in the wide range} 2 - 9.3\,kpc, \added{comptable with this cloud at $3.24$\,kpc}. 
\deleted{although in the case of the latter cloud the flux is almost evenly split with a second SNR G336.7+0.5, for which the distance is unconstrained.} 

\subsubsection{$334^\circ > l > 330^\circ$}

Previous studies have established the presence of diffuse gas overlapping with HESS\,J1616-508 (a likely PWN) and HESS\,J1614-518 (an SNR candidate with shell-like morphology) along the line of sight, yet without being a plausible association \citep{LauJ1616}. 
\deleted{Indeed, the estimated distance for the adjacent cloud at (332.58,0.6) is $\sim2.8$\,kpc, with the CR flux originating from SNR G332.0+0.2 which has an unconstrained distance. HESS\,J1616-508, meanwhile, is estimated to be situated at a distance of 6.5\,kpc.}

The neighbouring bright cloud at (333.46,-0.31) derives its predicted flux \added{almost evenly split between the SNR G332.4-0.4 and a hypothetical SNR -- whilst the contributions are separately comparable to the derived upper limit, the summation is in violation of it. As the SNR G332.4-0.4 has an estimated distance of } 2.7 - 3.3\,kpc and is already known to be interacting with molecular clouds, \added{it is the contribution from the hypothetical SNR which is considered less reliable and can be neglected. }

\subsubsection{$330^\circ > l > 325^\circ$}

All of the predictions for illuminated clouds in this region are compatible with upper limits obtained from the HGPS. The brightest prediction, for the cloud at \replaced{(328.91,-0.13)}{(326.60,0.29)} and a distance of \replaced{1.51}{2.84}\,kpc, originates from an association with the SNR \replaced{G329.7+0.4}{G327.4+1.0}, for which neither age nor distance are constrained.

\subsubsection{$324^\circ > l > 318^\circ$}

Firstly, we note that the most prominent cloud in this region at (322.51,0.17); as listed in Table \ref{tab:targetspsrsnrhess}, \added{remains nevertheless compatible with} the upper limit derived from the HGPS. This cloud is assumed to be illuminated by CRs from two SNRs, \replaced{G321.9-1.1 and G320.6-1.6}{G323.5+0.1 and G322.5-0.1}, about which little is known and no distance estimate is currently available. As the total flux is \replaced{dominated by the former SNR, tighter upper limits would constrain the distance to the former SNR more than the latter. }{almost evenly split between the two SNRs, we suggest that this implies the two SNRs cannot both be located at the same distance as the cloud of 2.2\,kpc. (The contribution from a single SNR would, however, be compatible with the upper limits.)}

Further clouds can be seen to overlap with HESS\,J1457-593, associated with the SNR G318.2+0.1 (distance estimated between 3.5\,kpc and 9.2\,kpc); a source that was first announced with the HGPS \citep{HGPS}. \deleted{Several of these clouds have compatible distance estimates and are illuminated by the SNR G318.2+0.1, which indicates a potential association with this TeV source.}
\added{This SNR is suggested by our model to illuminate the cloud at (319.05,0.61), adjacent to HESS\,J1503-582, whilst clouds overlapping with HESS\,J1457-593 are illuminated by G318.9+0.4, for which no distance is available.}

We note that a cloud located at (318.07, -0.21) was recently analysed by \cite{Peron18} using data from Fermi-LAT to probe the CR sea by testing for agreement between predicted and measured flux. This cloud was among those exhibiting a higher gamma-ray flux than expected due to the CR sea alone, 
\added{ whilst our model does not predict an additional contribution due to nearby SNRs. }
\deleted{Our model also suggests that a gamma-ray flux is expected from this cloud, assuming CR illumination by the SNR G318.2+0.1.}

\subsubsection{$316^\circ > l > 308^\circ$}

At $l \sim311^\circ$ two clouds overlap with significance contours not associated to any known TeV source. Both are predicted to be illuminated by CRs from the SNR \replaced{G309.8+0.0 about which little is known.}{G311.5-0.3, situated at $\sim6-7$\,kpc distance and suggested to be interacting with a molecular cloud.} 
\deleted{We therefore suggest that } A dedicated analysis to search for emission from these clouds is worthwhile to either confirm our model predictions, or place constraints on \replaced{properties of the corresponding SNR.}{the central hypothesis that this SNR accelerated particles to PeV energies.} 

The rather bright cloud at (309.2,-0.96) results from the sum of CR contributions from three SNRs; whilst the individual contributions are well compatible with upper limits, the summation likely over-predicts the total flux associated to this cloud\added{, resulting in a total flux incompatible with the HGPS upper limit in the slow diffusion case. Of the two contributing SNRs, only G308.8-0.1 has a constrained distance and can be considered the more reliable prediction, with the flux prediction due to this SNR alone lying approximately 20\% below the HGPS upper limit (the other is G309.8+0.0 mentioned above).}

\subsubsection{$287^\circ > l > 282^\circ$}

Westerlund 2 (HESS J1023-575) is another massive stellar cluster; as for Westerlund 1, in such a rich environment (of massive stars and molecular material) we can expect energetic particles illuminating nearby clouds to be a frequent occurrence \citep{DameJ1023}. The resulting morphology is complex and may be comprised of multiple CR accelerators; with the improved angular resolution of CTA in conjunction with multi-wavelength information, the origins of gamma-ray emission in this region may be revealed \citep{HESSWest2}.

The cloud at (284.12,-0.75) does, however, have a distance which is nevertheless compatible with the kinematic distance of molecular material in the region, $\sim 6$ kpc as determined in previous studies, although the distance to Westerlund 2 itself is thought to be closer to $\sim8$ kpc \citep{DameJ1023,West2clouds14}. This suggests that such a scenario may indeed contribute to, but not dominate the flux in the region. 

The environments of massive stellar clusters may be worth further dedicated modelling efforts, to establish if these environments are indeed suitable PeVatron candidates \citep{AharonianMassiveStars19}. 

We note that the larger cloud at (286.62,-0.36) which does not overlap with any known TeV source is predicted to be illuminated by the SNR G286.5-1.2, generating a gamma-ray flux at a level approximately half that of the corresponding HGPS upper limit, although neither age nor distance are constrained for this SNR. With a continued lack of detectable emission with increased exposure in this region, this could constrain the distance to this SNR as lying outside the range of 2-2.5\,kpc, where the cloud is situated. 

\subsection{Potentially bright clouds observable by HAWC}
Although H.E.S.S. is based in Namibia observing the Southern sky and HAWC, based in Mexico, observes the Northern sky, there is a region of the Galactic Plane from $60^\circ \gtrsim l \gtrsim 10^\circ$ which is observable by both. 
A dedicated study will compare the two instruments data in this overlapping region \citep{HAWC_HESS_GPS_2019ICRC,2021ApJ...917....6A_HAWCHESS}. 
For clouds within this common region, we do not repeat discussion of H.E.S.S. sources that have also been observed by HAWC, with the exception of sources for which gamma rays $>100$\,TeV have been detected \citep{HAWC56tev}. These include eHWC\,J1907+063 and eHWC\,J1825-134; the location of the former is consistent with the source HESS\,J1908+063, whilst the latter is associated with the complex region discussed in section \ref{sec:j1825}. 
From this study, there are no bright clouds $>100$\,TeV predicted that are coincident with eHWC\,J1907+063, which is consistent with the current scenario of the observed TeV sources above 100\,TeV being associated with energetic pulsars \citep{HAWC56tev}. 

The centroid of the highest energy emission $>$100\,TeV from eHWC\,J1825-134 is located between the two known sources HESS\,J1825-137 and HESS\,J1826-130. A recent detailed study attempts to disentangle contributions from multiple components in the region, with indications for an independent emission region emerging at the highest energies, potentially associated with molecular material \citep{HAWCj1825cloud2020}.

There are several new TeV sources discovered by HAWC in  the common region - with work ongoing to confirm their detection in the H.E.S.S. data \citep{HAWC_HESS_GPS_2019ICRC}. We now briefly discuss two of these regions. 

\subsubsection{$54^\circ > l > 52^\circ$}

Adjacent to HESS\,J1930+188 and nearby clouds lies the source HAWC\,J1928+178, newly discovered by HAWC and confirmed in subsequent HAWC catalogues, although it was not detected in an initial follow-up study by VERITAS and Fermi-LAT \citep{VERITAS_FERMI_HAWC_2018}. The dedicated comparison study of HAWC and H.E.S.S. analysis techniques does, however, appear to reveal HAWC\,J1928+178 in the HGPS dataset. 

A dedicated study of this source suggested an association with the energetic pulsar PSR\,J1928+1746, implying a leptonic, PWN origin for the gamma-ray emission \citep{J1928_2017ICRC...35..732L}. More recently, alternative scenarios including an illuminated molecular cloud or a gamma-ray binary were shown to remain plausible by \cite{MoriJ1928}. They also state that further X-ray observations are necessary to clarify a PWN association for the emission. 

\subsubsection{$47^\circ > l > 45^\circ$}

The TeV source HAWC\,J1914+118 is located at (46.00, 0.25) and overlaps with the clouds located at (45.92,-0.45) and (45.86,0.01). 
As yet, there is no firm identification for HAWC\,J1914+118, with several HII regions, pulsars and X-ray sources either coincident with or nearby to the gamma-ray emission \cite{2HWC_2017}. Our model predicts illumination of the former cloud by the SNR G046.8-0.3, and of the latter cloud by G043.9+1.6, both of which have constrained distance \citep{snrcat}. We suggest that further investigations are necessary to firmly establish the nature of this new TeV source. 

\subsubsection{Regions outside the HGPS: $260^\circ > l > 60^\circ$}

Although there is a large portion of the Galactic Plane that is not covered by the HGPS, there are only a few candidate clouds identified by our model as potential targets for CRs from nearby SNRs. The most prominent of these is the cloud at (110.43, 1.89)
illuminated by the potentially young (7.7kyr) SNR G114.3+0.3, at a known nearby distance of 0.7 kpc, compatible with the distance to the cloud of 0.6\,kpc.
The angular size of the cloud is quite large at $1.30^\circ$, which will make it more challenging to detect with point source searches and will likely require a dedicated analysis (for any pointing gamma-ray instrument). Although this region is observable by MAGIC and VERITAS, observations of this region may have not yet taken place and/or dedicated searches not yet performed. 
An additional challenge comes from the sensitivity of HAWC degrading towards this region of the sky, with declination $>+60^\circ$ \cite{2HWC_2017}. 
According to our model, this could be a promising region to study for evidence of PeVatron activity from SNRs, suggesting that dedicated observations and analysis searching for extended sources in this region would be worthwhile. 

\subsection{Limitations of the model}

Despite our efforts to construct a realistic model for particle acceleration, transport and interaction, there are nevertheless several additional known limitations of the modelling. Firstly, we have assumed that all SNRs are in the Sedov-Taylor phase, when evaluating the SNR age based on its current radius or equivalently in calculating the spatial evolution of the SNR to establish the radius at the escape time for particles of a given energy. This is typically valid for SNRs with ages between $\sim300$\,yr and $\sim20$\,kyr, and is therefore valid for the majority of the systems we consider. However, SNRs generally undergo three phases during their evolution; an initial free-expansion phase, prior to adiabatic expansion during the Sedov-Taylor phase; followed by a period of radiative expansion up to approximately $\sim500$\,kyr. An extension to the modelling could therefore be to incorporate the radiative phase into the evolution. In the radiative phase, particles that remain attached to the shock (have not yet escaped) become less energetic with time \citep{1999ApJ...511..798C}. 

However, the transition time between different phases strongly depends on the system properties, including details of the supernova explosion and pre-existing conditions in the circumstellar medium, which are often unknown.
Similarly, throughout the modelling of both the SNRs and the interstellar clouds we have assumed the morphology to be spherically symmetric, whereas in practise the morphologies can be significantly more complex. 

The largest uncertainty of the model resides in the diffusion properties of both the ISM and the interstellar clouds, which remain so far elusive. We have assumed a suppression of the diffusion coefficient within the clouds of $\chi = 0.05$ with respect to the ISM. As figure \ref{fig:chin} shows, for the same cloud density $n$, different $\chi$ values can lead to variations in gamma-ray flux. Also, the average diffusion coefficient within the ISM is uncertain and unlikely to be uniform. 

The normalisation of $D_0$ assumed in table \ref{tab:constants} is a typical Galactic value as derived from CR diffusion time in our Galaxy \citep{Gabici07}. Recent measurements have shown, however, that the diffusion coefficient may be suppressed within regions of the ISM with turbulent activity due to nearby accelerators \citep{HAWCgeminga17}. Transport of energetic particles within the intervening ISM between an accelerator and a nearby cloud may therefore proceed somewhat slower than our assumed transport here. 
Comparing the results using two different assumed values of $D_0$ provides an indication of this effect. 

To estimate the magnetic field strength within interstellar clouds, we adopted in equation \eqref{eq:bfield} a parameterisation based on the cloud density from \cite{Crutcher10}, using estimates from Zeeman splitting which is significant only for dense clouds. Although a fixed value comparable to the ISM field strength has been used for clouds with $n < 300\,\mathrm{cm}^{-3}$, the magnetic field strength is likely to vary further. Alternative methods for estimating the magnetic field must then be used, such as Faraday tomography, which could revise the flux prediction \citep{VanEckFaradayTomo17}. 

To describe the energy dependence of the p--p interaction cross-section, we adopted the most recent parameterisation from \cite{Kafexhiu14} in equation \eqref{eq:ppcross}. This differs quite substantially from the parameterisation used in \cite{Kelner06} at around 1\,GeV, however the agreement improves and the two expressions are quite compatible from a few tens of GeV towards higher energies. 
A minimum energy of 10\,GeV was used in the modelling, above which these two parameterisations are in reasonable agreement. 

Additionally, Green's catalogue \citep{Green2019} has limited availability of distance and age estimates for the SNRs; where these are provided (for SNRs indicated in bold in tables \ref{tab:targets}, \ref{tab:slowtargets} and \ref{tab:targetspsrsnrhess}), uncertainties are typically large with a range of possible values given. Distance estimates to interstellar clouds may suffer from a distance ambiguity between `near' and `far' estimates based on the observed velocites, although \cite{Rice16} removed this effect as far as possible, using the kinematic distance model of \cite{Reid09}.

\subsection{Applications of the results} 

Despite the uncertainties surrounding several input parameters of the model, the consequence is that observations can in turn be used to constrain the system parameters. For example, in many cases there was no distance estimate available for a particular SNR, such that it was assumed to be located at a comparable distance to the nearby cloud. However, if the resulting flux prediction exceeds a measured upper limit, then the distance to the SNR can be constrained (as unrelated to the cloud). Lack of detectable gamma-ray emission at the highest energies may, however, also be due to gamma-ray absorption in surrounding radiation fields. For clouds at distances $\gtrsim10$\,kpc, this effect should be corrected for \citep{Porter2018PhRvD..98d1302P}.

Similarly, as shown in section \ref{sec:Dcoeff}, the flux produced depends on the highly uncertain diffusion properties of the cloud. For a given density, the measured flux (or flux upper limit) can be used to constrain the diffusion properties of the cloud, provided the SNR - cloud association is secure. 

As discussed in section \ref{sec:intro}, particles accelerated to PeV energies and beyond will escape the source environment, such that it is considerably more likely that evidence for PeVatron activity will be found from target clouds located nearby an accelerator, than at the accelerators themselves \citep{Celli20}. We consider the predictions made here for the clouds listed in tables \ref{tab:targets}, \ref{tab:slowtargets} and \ref{tab:targetspsrsnrhess} to be suitable targets for future observing programs with CTA, searching for evidence of PeVatron activity from SNRs.

\subsection{Further work and next steps} 

This study can be extended to other source classes, such as stellar clusters and pulsars, considering not only impulsive but also  continuous acceleration scenarios. Indeed, it has been shown that pulsars are capable of accelerating hadronic particles to energies beyond 1\,PeV and injecting these into the pulsar wind; albeit at a rate dwarfed by the production of electron - positron pairs \citep{Lemoine15,Kotera15}. 
Massive stellar clusters have also been established as prime PeVatron candidates, that warrant dedicated modelling efforts \cite{AharonianMassiveStars19}. Additionally, detailed studies of individual sources and  specific regions can be made using the modelling framework established here.

\section{Conclusions}

Interstellar clouds are a suitable tool that can be exploited to search for evidence of PeVatron activity within our Galaxy. The case of target material for hadronic CR interactions being located near to but separated from an astrophysical accelerator is reasonably common. Under the hypothesis that energetic CRs escaping an SNR and illuminating nearby clouds could provide an alternative signature for PeV particle acceleration to the gamma-ray spectrum at the accelerator itself, we constrain the necessary phase space of system properties. 

In general, a higher target cloud density and lower separation distances are preferred for a detectable gamma-ray flux at very high energies. The predicted flux is dominated by the amount of target material, however there is a slight preference for older systems, to allow sufficient time for particles to reach the cloud. Faster diffusion, consistent with measurements from the Boron to Carbon ratio in the ISM and corresponding to a lower magnetic field within the cloud, leads to lower flux levels due to fewer interactions within the cloud. Slower diffusion within the ISM, however, may prevent particles from reaching the cloud at all for larger separation distances.  

The four brightest clouds, for which a detectable flux above 10\,TeV is predicted under a range of model scenarios, are located at (l,b) =  (333.46,-0.31), (16.97,0.53), (110.43,1.89) and (336.73,-0.98).
We remark that these clouds are spatially extended with respect to the angular resolution that CTA will achieve. However, no significant degradation of the minimum flux detectable by the instrument is expected above 100~TeV for sources whose extension is less than 1~deg, which applies to the majority of the clouds in this study \citep{Ambrogi18}.

On average, a detectable gamma-ray flux is more likely for more massive clouds; systems with lower separation distance between the SNR and cloud; and for slightly older SNRs.
We provide a further key target list for observations with future facilities, and suggest that even gamma-ray upper limits can provide useful information, constraining the distances to SNRs or diffusion properties of the ISM. 

Currently operational IACT facilities such as H.E.S.S., MAGIC and VERITAS have already demonstrated the capabilities of IACTs to perform detailed studies of specific regions, untangling the origins of the gamma-ray emission thanks to their good angular and energy resolution. The forthcoming CTA will further improve on these; this work provides a suitable list of target areas for further investigation. Facilities such as HAWC, LHASSO and the proposed SWGO employ WCD technology; these are survey instruments with a large effective area particularly towards the highest gamma-ray energies, ideally complementing the IACT observations \citep{HAWCcrab17,LHAASOwhitepaper,sgsowhitepaper}. Continuous observations by WCD facilities will rapidly establish locations of the highest energy gamma-ray emission, suitable for dedicated follow-up studies with IACTs. We encourage detailed studies of the promising regions highlighted by our model and look forward to further developments in the ongoing search for PeVatron candidates. 

\section*{Acknowledgements}

We thank T. Collins for useful input to the modelling. 
This research has made use of the CTA instrument response functions provided by the CTA Consortium and Observatory, see \url{http://www.cta-observatory.org/science/cta-performance/} (version prod3b-v2) for more details.
The authors gratefully acknowledge the assistance of R. Burley and A. Specovius in preparing this revision. 
As of Oct 2021, AM is supported by the Deutsche Forschungsgemeinschaft (DFG, German Research Foundation) - Project Number 452934793.

\section*{Data Availability}
The data underlying this article are available in the article and in its online supplementary material.




\bibliographystyle{mnras}
\bibliography{MCpev.bib} 

\begin{thebibliography}{}
\makeatletter
\relax
\def\mn@urlcharsother{\let\do\@makeother \do\$\do\&\do\#\do\^\do\_\do\%\do\~}
\def\mn@doi{\begingroup\mn@urlcharsother \@ifnextchar [ {\mn@doi@}
  {\mn@doi@[]}}
\def\mn@doi@[#1]#2{\def\@tempa{#1}\ifx\@tempa\@empty \href
  {http://dx.doi.org/#2} {doi:#2}\else \href {http://dx.doi.org/#2} {#1}\fi
  \endgroup}
\def\mn@eprint#1#2{\mn@eprint@#1:#2::\@nil}
\def\mn@eprint@arXiv#1{\href {http://arxiv.org/abs/#1} {{\tt arXiv:#1}}}
\def\mn@eprint@dblp#1{\href {http://dblp.uni-trier.de/rec/bibtex/#1.xml}
  {dblp:#1}}
\def\mn@eprint@#1:#2:#3:#4\@nil{\def\@tempa {#1}\def\@tempb {#2}\def\@tempc
  {#3}\ifx \@tempc \@empty \let \@tempc \@tempb \let \@tempb \@tempa \fi \ifx
  \@tempb \@empty \def\@tempb {arXiv}\fi \@ifundefined
  {mn@eprint@\@tempb}{\@tempb:\@tempc}{\expandafter \expandafter \csname
  mn@eprint@\@tempb\endcsname \expandafter{\@tempc}}}

\bibitem[\protect\citeauthoryear{{Abdalla} et~al.,}{{Abdalla}
  et~al.}{2021}]{2021ApJ...917....6A_HAWCHESS}
{Abdalla} H.,  et~al., 2021, \mn@doi [\apj] {10.3847/1538-4357/abf64b}, \href
  {https://ui.adsabs.harvard.edu/abs/2021ApJ...917....6A} {917, 6}

\bibitem[\protect\citeauthoryear{{Abeysekara} et~al.,}{{Abeysekara}
  et~al.}{2017a}]{HAWCgeminga17}
{Abeysekara} A.~U.,  et~al., 2017a, \mn@doi [Science]
  {10.1126/science.aan4880}, \href
  {https://ui.adsabs.harvard.edu/abs/2017Sci...358..911A} {358, 911}

\bibitem[\protect\citeauthoryear{{Abeysekara} et~al.,}{{Abeysekara}
  et~al.}{2017b}]{HAWCcrab17}
{Abeysekara} A.~U.,  et~al., 2017b, \mn@doi [\apj] {10.3847/1538-4357/aa7555},
  \href {https://ui.adsabs.harvard.edu/abs/2017ApJ...843...39A} {843, 39}

\bibitem[\protect\citeauthoryear{{Abeysekara} et~al.,}{{Abeysekara}
  et~al.}{2017c}]{2HWC_2017}
{Abeysekara} A.~U.,  et~al., 2017c, \mn@doi [\apj] {10.3847/1538-4357/aa7556},
  \href {https://ui.adsabs.harvard.edu/abs/2017ApJ...843...40A} {843, 40}

\bibitem[\protect\citeauthoryear{{Abeysekara} et~al.,}{{Abeysekara}
  et~al.}{2018}]{VERITAS_FERMI_HAWC_2018}
{Abeysekara} A.~U.,  et~al., 2018, \mn@doi [\apj] {10.3847/1538-4357/aade4e},
  \href {https://ui.adsabs.harvard.edu/abs/2018ApJ...866...24A} {866, 24}

\bibitem[\protect\citeauthoryear{Abeysekara et~al.,}{Abeysekara
  et~al.}{2020}]{HAWC56tev}
Abeysekara A.~U.,  et~al., 2020, \mn@doi [Phys. Rev. Lett.]
  {10.1103/PhysRevLett.124.021102}, 124, 021102

\bibitem[\protect\citeauthoryear{{Abramowski} et~al.,}{{Abramowski}
  et~al.}{2012}]{HESSWest1}
{Abramowski} A.,  et~al., 2012, \mn@doi [\aap] {10.1051/0004-6361/201117928},
  \href {https://ui.adsabs.harvard.edu/abs/2012A&A...537A.114A} {537, A114}

\bibitem[\protect\citeauthoryear{{Abramowski} et~al.,}{{Abramowski}
  et~al.}{2014}]{HESSJ1641}
{Abramowski} A.,  et~al., 2014, \mn@doi [\apjl] {10.1088/2041-8205/794/1/L1},
  \href {https://ui.adsabs.harvard.edu/abs/2014ApJ...794L...1A} {794, L1}

\bibitem[\protect\citeauthoryear{{Ackermann} et~al.,}{{Ackermann}
  et~al.}{2013}]{FermiPionSNR13}
{Ackermann} M.,  et~al., 2013, \mn@doi [Science] {10.1126/science.1231160},
  \href {https://ui.adsabs.harvard.edu/abs/2013Sci...339..807A} {339, 807}

\bibitem[\protect\citeauthoryear{{Aharonian} \& {Atoyan}}{{Aharonian} \&
  {Atoyan}}{1996}]{AA96}
{Aharonian} F.~A.,  {Atoyan} A.~M.,  1996, \aap, \href
  {https://ui.adsabs.harvard.edu/abs/1996A&A...309..917A} {309, 917}

\bibitem[\protect\citeauthoryear{{Aharonian} et~al.,}{{Aharonian}
  et~al.}{2006}]{HESSCrab2006}
{Aharonian} F.,  et~al., 2006, \mn@doi [\aap] {10.1051/0004-6361:20065351},
  \href {https://ui.adsabs.harvard.edu/abs/2006A&A...457..899A} {457, 899}

\bibitem[\protect\citeauthoryear{{Aharonian}, {Yang}  \& {de O{\~n}a
  Wilhelmi}}{{Aharonian} et~al.}{2019}]{AharonianMassiveStars19}
{Aharonian} F.,  {Yang} R.,   {de O{\~n}a Wilhelmi} E.,  2019, \mn@doi [Nature
  Astronomy] {10.1038/s41550-019-0724-0}, \href
  {https://ui.adsabs.harvard.edu/abs/2019NatAs...3..561A} {3, 561}

\bibitem[\protect\citeauthoryear{Aharonian, Peron, Yang, Casanova  \&
  Zanin}{Aharonian et~al.}{2020}]{Peron18}
Aharonian F.,  Peron G.,  Yang R.,  Casanova S.,   Zanin R.,  2020, \mn@doi
  [Phys. Rev. D] {10.1103/PhysRevD.101.083018}, 101, 083018

\bibitem[\protect\citeauthoryear{{Albert} et~al.,}{{Albert}
  et~al.}{2019}]{sgsowhitepaper}
{Albert} A.,  et~al., 2019, arXiv e-prints, \href
  {https://ui.adsabs.harvard.edu/abs/2019arXiv190208429A} {p. arXiv:1902.08429}

\bibitem[\protect\citeauthoryear{{Albert} et~al.,}{{Albert}
  et~al.}{2021}]{HAWCj1825cloud2020}
{Albert} A.,  et~al., 2021, \mn@doi [\apjl] {10.3847/2041-8213/abd77b}, \href
  {https://ui.adsabs.harvard.edu/abs/2021ApJ...907L..30A} {907, L30}

\bibitem[\protect\citeauthoryear{{Aleksi{\'c}} et~al.,}{{Aleksi{\'c}}
  et~al.}{2015}]{MagicCrab2014}
{Aleksi{\'c}} J.,  et~al., 2015, \mn@doi [Journal of High Energy Astrophysics]
  {10.1016/j.jheap.2015.01.002}, \href
  {https://ui.adsabs.harvard.edu/abs/2015JHEAp...5...30A} {5, 30}

\bibitem[\protect\citeauthoryear{{Ambrogi}, {Celli}  \& {Aharonian}}{{Ambrogi}
  et~al.}{2018}]{Ambrogi18}
{Ambrogi} L.,  {Celli} S.,   {Aharonian} F.,  2018, \mn@doi [Astroparticle
  Physics] {10.1016/j.astropartphys.2018.03.001}, \href
  {https://ui.adsabs.harvard.edu/abs/2018APh...100...69A} {100, 69}

\bibitem[\protect\citeauthoryear{{Anderson} et~al.,}{{Anderson}
  et~al.}{2017}]{AndersonTHOR}
{Anderson} L.~D.,  et~al., 2017, \mn@doi [\aap] {10.1051/0004-6361/201731019},
  \href {https://ui.adsabs.harvard.edu/abs/2017A&A...605A..58A} {605, A58}

\bibitem[\protect\citeauthoryear{{Bai} et~al.,}{{Bai}
  et~al.}{2019}]{LHAASOwhitepaper}
{Bai} X.,  et~al., 2019, arXiv e-prints, \href
  {https://ui.adsabs.harvard.edu/abs/2019arXiv190502773B} {p. arXiv:1905.02773}

\bibitem[\protect\citeauthoryear{{Beuther} et~al.,}{{Beuther}
  et~al.}{2019}]{BeutherOHmasers}
{Beuther} H.,  et~al., 2019, \mn@doi [\aap] {10.1051/0004-6361/201935936},
  \href {https://ui.adsabs.harvard.edu/abs/2019A&A...628A..90B} {628, A90}

\bibitem[\protect\citeauthoryear{{Blasi}}{{Blasi}}{2013}]{BlasiCRreview}
{Blasi} P.,  2013, \mn@doi [\aapr] {10.1007/s00159-013-0070-7}, \href
  {https://ui.adsabs.harvard.edu/abs/2013A&ARv..21...70B} {21, 70}

\bibitem[\protect\citeauthoryear{{Celli}, {Palladino}  \& {Vissani}}{{Celli}
  et~al.}{2017}]{celli17}
{Celli} S.,  {Palladino} A.,   {Vissani} F.,  2017, \mn@doi [European Physical
  Journal C] {10.1140/epjc/s10052-017-4635-x}, \href
  {https://ui.adsabs.harvard.edu/abs/2017EPJC...77...66C} {77, 66}

\bibitem[\protect\citeauthoryear{{Celli}, {Morlino}, {Gabici}  \&
  {Aharonian}}{{Celli} et~al.}{2019a}]{celli2019clumps}
{Celli} S.,  {Morlino} G.,  {Gabici} S.,   {Aharonian} F.~A.,  2019a, \mn@doi
  [\mnras] {10.1093/mnras/stz1425}, \href
  {https://ui.adsabs.harvard.edu/abs/2019MNRAS.487.3199C} {487, 3199}

\bibitem[\protect\citeauthoryear{{Celli}, {Morlino}, {Gabici}  \&
  {Aharonian}}{{Celli} et~al.}{2019b}]{Celli19}
{Celli} S.,  {Morlino} G.,  {Gabici} S.,   {Aharonian} F.~A.,  2019b, \mn@doi
  [\mnras] {10.1093/mnras/stz2897}, \href
  {https://ui.adsabs.harvard.edu/abs/2019MNRAS.490.4317C} {490, 4317}

\bibitem[\protect\citeauthoryear{{Celli}, {Aharonian}  \& {Gabici}}{{Celli}
  et~al.}{2020}]{Celli20}
{Celli} S.,  {Aharonian} F.,   {Gabici} S.,  2020, \mn@doi [\apj]
  {10.3847/1538-4357/abb805}, \href
  {https://ui.adsabs.harvard.edu/abs/2020ApJ...903...61C} {903, 61}

\bibitem[\protect\citeauthoryear{{Cherenkov Telescope Array Consortium}
  et~al.,}{{Cherenkov Telescope Array Consortium}
  et~al.}{2019}]{2019scta.book.....C}
{Cherenkov Telescope Array Consortium} et~al., 2019, {Science with the
  Cherenkov Telescope Array}, \mn@doi{10.1142/10986.
}

\bibitem[\protect\citeauthoryear{{Chevalier}}{{Chevalier}}{1999}]{1999ApJ...511..798C}
{Chevalier} R.~A.,  1999, \mn@doi [\apj] {10.1086/306710}, \href
  {https://ui.adsabs.harvard.edu/abs/1999ApJ...511..798C} {511, 798}

\bibitem[\protect\citeauthoryear{Crutcher}{Crutcher}{2012}]{CrutcherReview}
Crutcher R.~M.,  2012, \mn@doi [Annual Review of Astronomy and Astrophysics]
  {10.1146/annurev-astro-081811-125514}, 50, 29

\bibitem[\protect\citeauthoryear{{Crutcher}, {Wandelt}, {Heiles}, {Falgarone}
  \& {Troland}}{{Crutcher} et~al.}{2010}]{Crutcher10}
{Crutcher} R.~M.,  {Wandelt} B.,  {Heiles} C.,  {Falgarone} E.,   {Troland}
  T.~H.,  2010, \mn@doi [\apj] {10.1088/0004-637X/725/1/466}, \href
  {https://ui.adsabs.harvard.edu/abs/2010ApJ...725..466C} {725, 466}

\bibitem[\protect\citeauthoryear{{Cui}, {Yeung}, {Tam}  \&
  {P{\"u}hlhofer}}{{Cui} et~al.}{2018}]{W2818}
{Cui} Y.,  {Yeung} P. K.~H.,  {Tam} P.~H.~T.,   {P{\"u}hlhofer} G.,  2018,
  \mn@doi [\apj] {10.3847/1538-4357/aac37b}, \href
  {https://ui.adsabs.harvard.edu/abs/2018ApJ...860...69C} {860, 69}

\bibitem[\protect\citeauthoryear{{D'Angelo}, {Morlino}, {Amato}  \&
  {Blasi}}{{D'Angelo} et~al.}{2018}]{dangelo}
{D'Angelo} M.,  {Morlino} G.,  {Amato} E.,   {Blasi} P.,  2018, \mn@doi
  [\mnras] {10.1093/mnras/stx2828}, \href
  {https://ui.adsabs.harvard.edu/abs/2018MNRAS.474.1944D} {474, 1944}

\bibitem[\protect\citeauthoryear{{Dame}}{{Dame}}{2007}]{DameJ1023}
{Dame} T.~M.,  2007, \mn@doi [\apjl] {10.1086/521363}, \href
  {https://ui.adsabs.harvard.edu/abs/2007ApJ...665L.163D} {665, L163}

\bibitem[\protect\citeauthoryear{{Dame}, {Hartmann}  \& {Thaddeus}}{{Dame}
  et~al.}{2001}]{Dame01}
{Dame} T.~M.,  {Hartmann} D.,   {Thaddeus} P.,  2001, \mn@doi [\apj]
  {10.1086/318388}, \href
  {https://ui.adsabs.harvard.edu/abs/2001ApJ...547..792D} {547, 792}

\bibitem[\protect\citeauthoryear{{Ferrand} \& {Safi-Harb}}{{Ferrand} \&
  {Safi-Harb}}{2012}]{snrcat}
{Ferrand} G.,  {Safi-Harb} S.,  2012, \mn@doi [Advances in Space Research]
  {10.1016/j.asr.2012.02.004}, \href
  {https://ui.adsabs.harvard.edu/abs/2012AdSpR..49.1313F} {49, 1313}

\bibitem[\protect\citeauthoryear{{Furukawa} et~al.,}{{Furukawa}
  et~al.}{2014}]{West2clouds14}
{Furukawa} N.,  et~al., 2014, \mn@doi [\apj] {10.1088/0004-637X/781/2/70},
  \href {https://ui.adsabs.harvard.edu/abs/2014ApJ...781...70F} {781, 70}

\bibitem[\protect\citeauthoryear{{Gabici} \& {Aharonian}}{{Gabici} \&
  {Aharonian}}{2007}]{GabiciAharonianPeVatron07}
{Gabici} S.,  {Aharonian} F.~A.,  2007, \mn@doi [\apjl] {10.1086/521047}, \href
  {https://ui.adsabs.harvard.edu/abs/2007ApJ...665L.131G} {665, L131}

\bibitem[\protect\citeauthoryear{{Gabici}, {Aharonian}  \& {Blasi}}{{Gabici}
  et~al.}{2007}]{Gabici07}
{Gabici} S.,  {Aharonian} F.~A.,   {Blasi} P.,  2007, \mn@doi [\apss]
  {10.1007/s10509-007-9427-6}, \href
  {https://ui.adsabs.harvard.edu/abs/2007Ap&SS.309..365G} {309, 365}

\bibitem[\protect\citeauthoryear{{Gaensler} \& {Slane}}{{Gaensler} \&
  {Slane}}{2006}]{GaenslerSlane06}
{Gaensler} B.~M.,  {Slane} P.~O.,  2006, \mn@doi [\araa]
  {10.1146/annurev.astro.44.051905.092528}, \href
  {https://ui.adsabs.harvard.edu/abs/2006ARA&A..44...17G} {44, 17}

\bibitem[\protect\citeauthoryear{{Green}}{{Green}}{2019}]{Green2019}
{Green} D.~A.,  2019, \mn@doi [Journal of Astrophysics and Astronomy]
  {10.1007/s12036-019-9601-6}, \href
  {https://ui.adsabs.harvard.edu/abs/2019JApA...40...36G} {40, 36}

\bibitem[\protect\citeauthoryear{{Grenier}, {Black}  \& {Strong}}{{Grenier}
  et~al.}{2015}]{GrenierCRreview}
{Grenier} I.~A.,  {Black} J.~H.,   {Strong} A.~W.,  2015, \mn@doi [\araa]
  {10.1146/annurev-astro-082214-122457}, \href
  {https://ui.adsabs.harvard.edu/abs/2015ARA&A..53..199G} {53, 199}

\bibitem[\protect\citeauthoryear{{H.E.S.S. Collaboration}}{{H.E.S.S.
  Collaboration}}{2010}]{somJ1634}
{H.E.S.S. Collaboration} 2010, Forgotten sources? HESS J1634-472 and HESS
  J1632-478, \url {https://www.mpi-hd.mpg.de/hfm/HESS/pages/home/som/2010/02/}

\bibitem[\protect\citeauthoryear{{H.E.S.S. Collaboration} et~al.,}{{H.E.S.S.
  Collaboration} et~al.}{2011}]{HESSWest2}
{H.E.S.S. Collaboration} et~al., 2011, \mn@doi [\aap]
  {10.1051/0004-6361/201015290}, \href
  {https://ui.adsabs.harvard.edu/abs/2011A&A...525A..46H} {525, A46}

\bibitem[\protect\citeauthoryear{{H.E.S.S. Collaboration} et~al.,}{{H.E.S.S.
  Collaboration} et~al.}{2018}]{HGPS}
{H.E.S.S. Collaboration} et~al., 2018, \mn@doi [\aap]
  {10.1051/0004-6361/201732098}, \href
  {https://ui.adsabs.harvard.edu/abs/2018A&A...612A...1H} {612, A1}

\bibitem[\protect\citeauthoryear{{H.E.S.S. Collaboration} et~al.,}{{H.E.S.S.
  Collaboration} et~al.}{2019}]{j1825}
{H.E.S.S. Collaboration} et~al., 2019, \mn@doi [\aap]
  {10.1051/0004-6361/201834335}, \href
  {https://ui.adsabs.harvard.edu/abs/2019A&A...621A.116H} {621, A116}

\bibitem[\protect\citeauthoryear{{Hillas}}{{Hillas}}{1984}]{Hillas84}
{Hillas} A.~M.,  1984, \mn@doi [\araa] {10.1146/annurev.aa.22.090184.002233},
  \href {https://ui.adsabs.harvard.edu/abs/1984ARA&A..22..425H} {22, 425}

\bibitem[\protect\citeauthoryear{{Hurley-Walker} et~al.,}{{Hurley-Walker}
  et~al.}{2019a}]{GLEAM_new2019PASA}
{Hurley-Walker} N.,  et~al., 2019a, \mn@doi [\pasa] {10.1017/pasa.2019.34},
  \href {https://ui.adsabs.harvard.edu/abs/2019PASA...36...45H} {36, e045}

\bibitem[\protect\citeauthoryear{{Hurley-Walker} et~al.,}{{Hurley-Walker}
  et~al.}{2019b}]{GLEAM_candidate2019PASA}
{Hurley-Walker} N.,  et~al., 2019b, \mn@doi [\pasa] {10.1017/pasa.2019.33},
  \href {https://ui.adsabs.harvard.edu/abs/2019PASA...36...48H} {36, e048}

\bibitem[\protect\citeauthoryear{{Inoue}}{{Inoue}}{2019}]{Inoue2019ApJ...872...46I}
{Inoue} T.,  2019, \mn@doi [\apj] {10.3847/1538-4357/aafb70}, \href
  {https://ui.adsabs.harvard.edu/abs/2019ApJ...872...46I} {872, 46}

\bibitem[\protect\citeauthoryear{{Inoue}, {Yamazaki}, {Inutsuka}  \&
  {Fukui}}{{Inoue} et~al.}{2012}]{Inoue2012ApJ...744...71I}
{Inoue} T.,  {Yamazaki} R.,  {Inutsuka} S.-i.,   {Fukui} Y.,  2012, \mn@doi
  [\apj] {10.1088/0004-637X/744/1/71}, \href
  {https://ui.adsabs.harvard.edu/abs/2012ApJ...744...71I} {744, 71}

\bibitem[\protect\citeauthoryear{{Jansson} \& {Farrar}}{{Jansson} \&
  {Farrar}}{2012}]{Jansson_ISMB}
{Jansson} R.,  {Farrar} G.~R.,  2012, \mn@doi [\apj]
  {10.1088/0004-637X/757/1/14}, \href
  {https://ui.adsabs.harvard.edu/abs/2012ApJ...757...14J} {757, 14}

\bibitem[\protect\citeauthoryear{{Jardin-Blicq}, {Marandon}  \&
  {Brun}}{{Jardin-Blicq} et~al.}{2019}]{HAWC_HESS_GPS_2019ICRC}
{Jardin-Blicq} A.,  {Marandon} V.,   {Brun} F.,  2019, in 36th International
  Cosmic Ray Conference (ICRC2019). p.~706 (\mn@eprint {arXiv} {1908.06658})

\bibitem[\protect\citeauthoryear{{Jogler} \& {Funk}}{{Jogler} \&
  {Funk}}{2016}]{W51C16}
{Jogler} T.,  {Funk} S.,  2016, \mn@doi [\apj] {10.3847/0004-637X/816/2/100},
  \href {https://ui.adsabs.harvard.edu/abs/2016ApJ...816..100J} {816, 100}

\bibitem[\protect\citeauthoryear{{Kafexhiu}, {Aharonian}, {Taylor}  \&
  {Vila}}{{Kafexhiu} et~al.}{2014}]{Kafexhiu14}
{Kafexhiu} E.,  {Aharonian} F.,  {Taylor} A.~M.,   {Vila} G.~S.,  2014, \mn@doi
  [\prd] {10.1103/PhysRevD.90.123014}, \href
  {https://ui.adsabs.harvard.edu/abs/2014PhRvD..90l3014K} {90, 123014}

\bibitem[\protect\citeauthoryear{{Kelner}, {Aharonian}  \& {Bugayov}}{{Kelner}
  et~al.}{2006}]{Kelner06}
{Kelner} S.~R.,  {Aharonian} F.~A.,   {Bugayov} V.~V.,  2006, \mn@doi [\prd]
  {10.1103/PhysRevD.74.034018}, \href
  {https://ui.adsabs.harvard.edu/abs/2006PhRvD..74c4018K} {74, 034018}

\bibitem[\protect\citeauthoryear{{Kotera}, {Amato}  \& {Blasi}}{{Kotera}
  et~al.}{2015}]{Kotera15}
{Kotera} K.,  {Amato} E.,   {Blasi} P.,  2015, \mn@doi [\jcap]
  {10.1088/1475-7516/2015/08/026}, \href
  {https://ui.adsabs.harvard.edu/abs/2015JCAP...08..026K} {2015, 026}

\bibitem[\protect\citeauthoryear{{Kulsrud} \& {Pearce}}{{Kulsrud} \&
  {Pearce}}{1969}]{kulsrud}
{Kulsrud} R.,  {Pearce} W.~P.,  1969, \mn@doi [\apj] {10.1086/149981}, \href
  {https://ui.adsabs.harvard.edu/abs/1969ApJ...156..445K} {156, 445}

\bibitem[\protect\citeauthoryear{{Lau} et~al.,}{{Lau} et~al.}{2017a}]{LauJ1616}
{Lau} J.~C.,  et~al., 2017a, \mn@doi [\pasa] {10.1017/pasa.2017.59}, \href
  {https://ui.adsabs.harvard.edu/abs/2017PASA...34...64L} {34, e064}

\bibitem[\protect\citeauthoryear{{Lau} et~al.,}{{Lau} et~al.}{2017b}]{Lau17}
{Lau} J.~C.,  et~al., 2017b, \mn@doi [\mnras] {10.1093/mnras/stw2692}, \href
  {https://ui.adsabs.harvard.edu/abs/2017MNRAS.464.3757L} {464, 3757}

\bibitem[\protect\citeauthoryear{{Lemoine}, {Kotera}  \& {P{\'e}tri}}{{Lemoine}
  et~al.}{2015}]{Lemoine15}
{Lemoine} M.,  {Kotera} K.,   {P{\'e}tri} J.,  2015, \mn@doi [\jcap]
  {10.1088/1475-7516/2015/07/016}, \href
  {https://ui.adsabs.harvard.edu/abs/2015JCAP...07..016L} {2015, 016}

\bibitem[\protect\citeauthoryear{{Lopez-Coto}, {Marandon}, {Brun}, {HAWC
  Collaboration}  \& {H.E.S.S. Collaboration}}{{Lopez-Coto}
  et~al.}{2017}]{J1928_2017ICRC...35..732L}
{Lopez-Coto} R.,  {Marandon} V.,  {Brun} F.,  {HAWC Collaboration}  {H.E.S.S.
  Collaboration} 2017, in 35th International Cosmic Ray Conference (ICRC2017).
  p.~732 (\mn@eprint {arXiv} {1708.03137})

\bibitem[\protect\citeauthoryear{{MAGIC Collaboration} et~al.,}{{MAGIC
  Collaboration} et~al.}{2014}]{MAGIC14_J1857}
{MAGIC Collaboration} et~al., 2014, \mn@doi [\aap]
  {10.1051/0004-6361/201423517}, \href
  {https://ui.adsabs.harvard.edu/abs/2014A&A...571A..96M} {571, A96}

\bibitem[\protect\citeauthoryear{{MAGIC Collaboration} et~al.,}{{MAGIC
  Collaboration} et~al.}{2019}]{MAGICJ1835}
{MAGIC Collaboration} et~al., 2019, \mn@doi [\mnras] {10.1093/mnras/sty3387},
  \href {https://ui.adsabs.harvard.edu/abs/2019MNRAS.483.4578M} {483, 4578}

\bibitem[\protect\citeauthoryear{{Manchester}, {Hobbs}, {Teoh}  \&
  {Hobbs}}{{Manchester} et~al.}{2005}]{Manchester05}
{Manchester} R.~N.,  {Hobbs} G.~B.,  {Teoh} A.,   {Hobbs} M.,  2005, \mn@doi
  [\aj] {10.1086/428488}, \href
  {https://ui.adsabs.harvard.edu/abs/2005AJ....129.1993M} {129, 1993}

\bibitem[\protect\citeauthoryear{{Meagher} \& {VERITAS
  Collaboration}}{{Meagher} \& {VERITAS
  Collaboration}}{2015}]{VeritasCrab_2015ICRC}
{Meagher} K.,  {VERITAS Collaboration} 2015, in 34th International Cosmic Ray
  Conference (ICRC2015). p.~792 (\mn@eprint {arXiv} {1508.06442})

\bibitem[\protect\citeauthoryear{{Miville-Desch{\^e}nes}, {Murray}  \&
  {Lee}}{{Miville-Desch{\^e}nes} et~al.}{2017}]{MDsurvey17}
{Miville-Desch{\^e}nes} M.-A.,  {Murray} N.,   {Lee} E.~J.,  2017, \mn@doi
  [\apj] {10.3847/1538-4357/834/1/57}, \href
  {https://ui.adsabs.harvard.edu/abs/2017ApJ...834...57M} {834, 57}

\bibitem[\protect\citeauthoryear{{Mori} et~al.,}{{Mori}
  et~al.}{2020}]{MoriJ1928}
{Mori} K.,  et~al., 2020, \mn@doi [\apj] {10.3847/1538-4357/ab9631}, \href
  {https://ui.adsabs.harvard.edu/abs/2020ApJ...897..129M} {897, 129}

\bibitem[\protect\citeauthoryear{{Muno} et~al.,}{{Muno}
  et~al.}{2006}]{Wd1magnetar}
{Muno} M.~P.,  et~al., 2006, \mn@doi [\apjl] {10.1086/499776}, \href
  {https://ui.adsabs.harvard.edu/abs/2006ApJ...636L..41M} {636, L41}

\bibitem[\protect\citeauthoryear{{Nava}, {Gabici}, {Marcowith}, {Morlino}  \&
  {Ptuskin}}{{Nava} et~al.}{2016}]{nava}
{Nava} L.,  {Gabici} S.,  {Marcowith} A.,  {Morlino} G.,   {Ptuskin} V.~S.,
  2016, \mn@doi [\mnras] {10.1093/mnras/stw1592}, \href
  {https://ui.adsabs.harvard.edu/abs/2016MNRAS.461.3552N} {461, 3552}

\bibitem[\protect\citeauthoryear{{Paredes}, {Ishwara-Chandra}, {Bosch-Ramon},
  {Zabalza}, {Iwasawa}  \& {Rib{\'o}}}{{Paredes} et~al.}{2014}]{Paredes14}
{Paredes} J.~M.,  {Ishwara-Chandra} C.~H.,  {Bosch-Ramon} V.,  {Zabalza} V.,
  {Iwasawa} K.,   {Rib{\'o}} M.,  2014, \mn@doi [\aap]
  {10.1051/0004-6361/201322306}, \href
  {https://ui.adsabs.harvard.edu/abs/2014A&A...561A..56P} {561, A56}

\bibitem[\protect\citeauthoryear{{Porter}, {Rowell}, {J{\'o}hannesson}  \&
  {Moskalenko}}{{Porter} et~al.}{2018}]{Porter2018PhRvD..98d1302P}
{Porter} T.~A.,  {Rowell} G.~P.,  {J{\'o}hannesson} G.,   {Moskalenko} I.~V.,
  2018, \mn@doi [\prd] {10.1103/PhysRevD.98.041302}, \href
  {https://ui.adsabs.harvard.edu/abs/2018PhRvD..98d1302P} {98, 041302}

\bibitem[\protect\citeauthoryear{{Reid} et~al.,}{{Reid} et~al.}{2009}]{Reid09}
{Reid} M.~J.,  et~al., 2009, \mn@doi [\apj] {10.1088/0004-637X/700/1/137},
  \href {https://ui.adsabs.harvard.edu/abs/2009ApJ...700..137R} {700, 137}

\bibitem[\protect\citeauthoryear{{Reynolds}}{{Reynolds}}{2008}]{2008ARA&A..46...89Reynolds}
{Reynolds} S.~P.,  2008, \mn@doi [\araa]
  {10.1146/annurev.astro.46.060407.145237}, \href
  {https://ui.adsabs.harvard.edu/abs/2008ARA&A..46...89R} {46, 89}

\bibitem[\protect\citeauthoryear{{Rice}, {Goodman}, {Bergin}, {Beaumont}  \&
  {Dame}}{{Rice} et~al.}{2016}]{Rice16}
{Rice} T.~S.,  {Goodman} A.~A.,  {Bergin} E.~A.,  {Beaumont} C.,   {Dame}
  T.~M.,  2016, \mn@doi [\apj] {10.3847/0004-637X/822/1/52}, \href
  {https://ui.adsabs.harvard.edu/abs/2016ApJ...822...52R} {822, 52}

\bibitem[\protect\citeauthoryear{{Rix} \& {Bovy}}{{Rix} \&
  {Bovy}}{2013}]{Rix13MWsize}
{Rix} H.-W.,  {Bovy} J.,  2013, \mn@doi [\aapr] {10.1007/s00159-013-0061-8},
  \href {https://ui.adsabs.harvard.edu/abs/2013A&ARv..21...61R} {21, 61}

\bibitem[\protect\citeauthoryear{{Roh}, {Inutsuka}  \& {Inoue}}{{Roh}
  et~al.}{2016}]{Roh2016APh....73....1R}
{Roh} S.,  {Inutsuka} S.-i.,   {Inoue} T.,  2016, \mn@doi [Astroparticle
  Physics] {10.1016/j.astropartphys.2015.06.001}, \href
  {https://ui.adsabs.harvard.edu/abs/2016APh....73....1R} {73, 1}

\bibitem[\protect\citeauthoryear{{Truelove} \& {McKee}}{{Truelove} \&
  {McKee}}{1999}]{TrueloveMcKee}
{Truelove} J.~K.,  {McKee} C.~F.,  1999, \mn@doi [\apjs] {10.1086/313176},
  \href {https://ui.adsabs.harvard.edu/abs/1999ApJS..120..299T} {120, 299}

\bibitem[\protect\citeauthoryear{{Van Eck} et~al.,}{{Van Eck}
  et~al.}{2017}]{VanEckFaradayTomo17}
{Van Eck} C.~L.,  et~al., 2017, \mn@doi [\aap] {10.1051/0004-6361/201629707},
  \href {https://ui.adsabs.harvard.edu/abs/2017A&A...597A..98V} {597, A98}

\bibitem[\protect\citeauthoryear{{Voisin}, {Rowell}, {Burton}, {Walsh}, {Fukui}
   \& {Aharonian}}{{Voisin} et~al.}{2016}]{Voisin16}
{Voisin} F.,  {Rowell} G.,  {Burton} M.~G.,  {Walsh} A.,  {Fukui} Y.,
  {Aharonian} F.,  2016, \mn@doi [\mnras] {10.1093/mnras/stw473}, \href
  {https://ui.adsabs.harvard.edu/abs/2016MNRAS.458.2813V} {458, 2813}

\bibitem[\protect\citeauthoryear{{Vukoti{\'c}}, {{\'C}iprijanovi{\'c}},
  {Vu{\v{c}}eti{\'c}}, {Oni{\'c}}  \& {Uro{\v{s}}evi{\'c}}}{{Vukoti{\'c}}
  et~al.}{2019}]{2019SerAJ.199...23S}
{Vukoti{\'c}} B.,  {{\'C}iprijanovi{\'c}} A.,  {Vu{\v{c}}eti{\'c}} M.~M.,
  {Oni{\'c}} D.,   {Uro{\v{s}}evi{\'c}} D.,  2019, \mn@doi [Serbian
  Astronomical Journal] {10.2298/SAJ1999023V}, \href
  {https://ui.adsabs.harvard.edu/abs/2019SerAJ.199...23S} {199, 23}

\makeatother
\end{thebibliography}



\newpage
\onecolumn


\twocolumn
\newpage
\appendix
\onecolumn

\section{Predicted gamma-ray flux from clouds above 10\,TeV}
\label{sec:gamma10}

\begin{table}
	\centering
	\caption{As for Table~\ref{tab:targets}, for the predicted flux and upper limits above 10 TeV. All clouds in this table are predicted to have a flux above the H.E.S.S. 50 hour sensitivity threshold in the slow diffusion case. } 
	\label{tab:targetspsrsnrhess}
	\begin{tabular}{c|l|l|l|l|l|c|l|l} 
		\hline 
	Cloud ID & Cloud coordinates & Cloud size & Cloud distance  & $EF^{UL}_\gamma >$ 10\,TeV & $EF^{\mathrm{total}}_\gamma$ $>$ 10\,TeV & \# SNRs & SNR & flux contribution\\
	& (l, b) deg & deg & kpc & TeV cm$^{-2}$ s$^{-1}$ & TeV cm$^{-2}$ s$^{-1}$ & & & TeV cm$^{-2}$ s$^{-1}$ \\
	 \hline
2 & (110.43,1.89) & 1.303 & 0.600 & N/A & 3.55e-12 & 1 & G114.3+00.3 & 3.55e-12 \\ 
1 & (333.46,-0.31) & 0.627 & 3.370 & 1.78e-12 & 2.02e-12 & 1 & G332.4-00.4 & 2.02e-12 \\ 
9 & (341.34,0.21) & 0.284 & 2.410 & 1.07e-12 & 1.59e-12 & 2 & G341.2+00.9 & 1.39e-12 \\ 
& - & - & - & - & - & - & G343.1-00.7 & 2.04e-13 \\ 
10 & (36.10,-0.14) & 0.492 & 3.400 & 5.26e-12 & 1.33e-12 & 3 & G035.6-00.4 & 8.13e-13 \\ 
& - & - & - & - & - & - & G036.6-00.7 & 3.11e-13 \\ 
& - & - & - & - & - & - & G034.7-00.4 & 2.11e-13 \\ 
13 & (328.91,-0.13) & 0.797 & 1.510 & 2.16e-12 & 1.28e-12 & 1 & G329.7+00.4 & 1.28e-12 \\ 
14 & (14.14,-0.59) & 0.333 & 2.090 & 8.30e-13 & 1.21e-12 & 1 & G015.1-01.6 & 1.21e-12 \\ 
16 & (309.20,-0.96) & 0.619 & 2.020 & 6.70e-13 & 1.15e-12 & 2 & G308.8-00.1 & 6.37e-13 \\ 
& - & - & - & - & - & - & G309.8+00.0 & 5.16e-13 \\ 
15 & (16.24,-1.02) & 0.328 & 4.330 & 4.93e-13 & 1.06e-12 & 2 & G016.4-00.5 & 5.81e-13 \\ 
& - & - & - & - & - & - & G016.0-00.5 & 4.80e-13 \\ 
5 & (336.73,-0.98) & 0.516 & 3.240 & 2.26e-12 & 1.05e-12 & 1 & G337.2-00.7 & 1.05e-12 \\ 
19 & (106.66,1.16) & 0.648 & 0.880 & N/A & 1.04e-12 & 1 & G106.3+02.7 & 1.04e-12 \\ 
11 & (286.62,-0.36) & 0.827 & 2.390 & 2.26e-12 & 1.04e-12 & 1 & G286.5-01.2 & 1.04e-12 \\ 
20 & (322.51,0.17) & 1.287 & 2.210 & 1.30e-12 & 1.02e-12 & 2 & G321.9-01.1 & 8.90e-13 \\ 
& - & - & - & - & - & - & G320.6-01.6 & 1.35e-13 \\ 
6 & (34.99,-0.96) & 0.727 & 2.620 & 1.03e-12 & 9.37e-13 & 2 & G036.6-00.7 & 9.27e-13 \\ 
& - & - & - & - & - & - & G033.2-00.6 & 9.96e-15 \\ 
2 & (16.97,0.53) & 0.870 & 2.080 & 2.34e-12 & 8.66e-13 & 2 & G019.1+00.2 & 5.12e-13 \\ 
& - & - & - & - & - & - & G018.9-01.1 & 3.54e-13 \\ 
22 & (32.73,-0.42) & 0.257 & 2.910 & 9.56e-13 & 8.50e-13 & 5 & G033.2-00.6 & 5.12e-13 \\ 
& - & - & - & - & - & - & G033.6+00.1 & 1.81e-13 \\ 
& - & - & - & - & - & - & G031.5-00.6 & 6.66e-14 \\ 
& - & - & - & - & - & - & G034.7-00.4 & 5.28e-14 \\ 
& - & - & - & - & - & - & G032.1-00.9 & 3.73e-14 \\ 
7 & (337.84,-0.40) & 0.465 & 2.910 & 1.13e-12 & 8.35e-13 & 2 & G337.2-00.7 & 6.58e-13 \\ 
& - & - & - & - & - & - & G338.1+00.4 & 1.77e-13 \\ 
25 & (35.64,0.01) & 0.691 & 1.860 & 3.42e-12 & 7.79e-13 & 2 & G036.6-00.7 & 4.56e-13 \\ 
& - & - & - & - & - & - & G034.7-00.4 & 3.24e-13 \\ 
28 & (342.52,0.26) & 0.518 & 2.580 & 7.29e-13 & 7.36e-13 & 2 & G343.1-00.7 & 4.92e-13 \\ 
& - & - & - & - & - & - & G341.2+00.9 & 2.44e-13 \\ 
4 & (21.97,-0.29) & 0.608 & 3.570 & 2.07e-12 & 7.15e-13 & 2 & G021.6-00.8 & 5.76e-13 \\ 
& - & - & - & - & - & - & G021.5-00.9 & 1.39e-13 \\ 
17 & (18.81,-0.46) & 0.422 & 4.510 & 1.07e-12 & 7.13e-13 & 2 & G018.1-00.1 & 5.44e-13 \\ 
& - & - & - & - & - & - & G019.1+00.2 & 1.68e-13 \\ 
23 & (34.87,0.38) & 0.248 & 3.210 & 5.78e-13 & 6.89e-13 & 3 & G034.7-00.4 & 3.87e-13 \\ 
& - & - & - & - & - & - & G033.6+00.1 & 2.74e-13 \\ 
& - & - & - & - & - & - & G035.6-00.4 & 2.87e-14 \\ 
32 & (25.31,-0.37) & 0.294 & 3.660 & 4.36e-12 & 6.42e-13 & 2 & G024.7-00.6 & 4.93e-13 \\ 
& - & - & - & - & - & - & G024.7+00.6 & 1.49e-13 \\ 
34 & (342.30,0.19) & 0.537 & 5.420 & 1.66e-12 & 5.83e-13 & 1 & G342.1+00.9 & 5.83e-13 \\ 
35 & (291.58,-1.25) & 0.622 & 3.070 & 1.36e-12 & 5.78e-13 & 2 & G291.0-00.1 & 3.90e-13 \\ 
& - & - & - & - & - & - & G290.1-00.8 & 1.88e-13 \\ 
30 & (17.30,-1.40) & 0.179 & 3.140 & 9.32e-13 & 5.64e-13 & 2 & G017.4-02.3 & 3.52e-13 \\ 
& - & - & - & - & - & - & G017.8-02.6 & 2.11e-13 \\ 
38 & (343.64,-0.54) & 0.855 & 2.730 & 2.33e-12 & 5.14e-13 & 1 & G343.1-02.3 & 5.14e-13 \\ 
39 & (25.89,1.00) & 0.744 & 2.970 & 1.25e-12 & 5.13e-13 & 2 & G024.7+00.6 & 4.06e-13 \\ 
& - & - & - & - & - & - & G027.8+00.6 & 1.07e-13 \\ 
12 & (39.33,-0.32) & 0.611 & 3.840 & 1.45e-12 & 4.97e-13 & 2 & G040.5-00.5 & 2.88e-13 \\ 
& - & - & - & - & - & - & G038.7-01.3 & 2.09e-13 \\ 
21 & (22.10,-1.06) & 0.297 & 4.210 & 8.32e-13 & 4.66e-13 & 2 & G021.6-00.8 & 3.95e-13 \\ 
& - & - & - & - & - & - & G022.7-00.2 & 6.76e-14 \\ 
44 & (338.99,0.63) & 0.359 & 4.320 & 2.53e-13 & 4.58e-13 & 1 & G338.1+00.4 & 4.58e-13 \\ 
47 & (142.40,1.38) & 1.018 & 0.660 & N/A & 4.31e-13 & 1 & G150.3+04.5 & 4.31e-13 \\ 
51 & (322.60,-0.63) & 0.242 & 3.450 & 3.13e-13 & 3.69e-13 & 4 & G322.5-00.1 & 2.35e-13 \\ 
& - & - & - & - & - & - & G321.9-01.1 & 5.64e-14 \\ 
& - & - & - & - & - & - & G323.7-01.0 & 3.92e-14 \\ 
& - & - & - & - & - & - & G321.9-00.3 & 3.91e-14 \\ 
49 & (15.60,-0.39) & 0.293 & 4.470 & 8.65e-13 & 3.66e-13 & 3 & G016.0-00.5 & 1.59e-13 \\ 
& - & - & - & - & - & - & G015.4+00.1 & 1.35e-13 \\ 
& - & - & - & - & - & - & G016.4-00.5 & 7.16e-14 \\ 
\end{tabular}
\end{table}

\begin{table}
    \centering
    \caption{Table \ref{tab:targetspsrsnrhess} continued.}
\begin{tabular}{c|l|l|l|l|l|c|l|l} 
\hline
Cloud ID & Cloud coordinates & Cloud size & Cloud distance  & $EF^{UL}_\gamma >$ 10\,TeV & $EF^{\mathrm{total}}_\gamma$ $>$ 10\,TeV & \# SNRs & SNR & flux contribution\\
& (l, b) deg & deg & kpc & TeV cm$^{-2}$ s$^{-1}$ & TeV cm$^{-2}$ s$^{-1}$ & & & TeV cm$^{-2}$ s$^{-1}$ \\
\hline
27 & (19.96,-0.69) & 0.347 & 3.260 & 8.08e-13 & 3.64e-13 & 1 & G019.1+00.2 & 3.64e-13 \\ 
52 & (340.53,1.02) & 0.469 & 4.300 & 5.41e-13 & 3.63e-13 & 2 & G340.4+00.4 & 2.20e-13 \\ 
& - & - & - & - & - & - & G341.2+00.9 & 1.43e-13 \\ 
53 & (332.38,0.60) & 0.322 & 2.840 & 7.97e-13 & 3.60e-13 & 1 & G332.4-00.4 & 3.60e-13 \\ 
54 & (301.56,-0.37) & 0.361 & 2.330 & 6.28e-13 & 3.56e-13 & 1 & G301.4-01.0 & 3.56e-13 \\ 
55 & (310.48,-0.24) & 0.279 & 4.430 & 8.14e-13 & 3.50e-13 & 2 & G310.8-00.4 & 3.04e-13 \\ 
& - & - & - & - & - & - & G309.8+00.0 & 4.64e-14 \\ 
24 & (21.78,-0.40) & 0.216 & 5.110 & 7.87e-13 & 3.37e-13 & 2 & G021.6-00.8 & 3.01e-13 \\ 
& - & - & - & - & - & - & G022.7-00.2 & 3.44e-14 \\ 
57 & (344.09,0.12) & 0.457 & 5.090 & 2.90e-12 & 3.31e-13 & 1 & G344.7-00.1 & 3.31e-13 \\ 
58 & (43.28,-0.28) & 0.435 & 2.690 & 7.80e-13 & 3.28e-13 & 2 & G042.8+00.6 & 3.09e-13 \\ 
 & - & - & - & - & - & - & G043.9+01.6 & 1.96e-14 \\ 
26 & (18.40,-0.27) & 0.451 & 3.600 & 3.29e-12 & 3.26e-13 & 1 & G019.1+00.2 & 3.26e-13 \\ 
59 & (23.80,1.58) & 0.517 & 2.580 & 6.28e-13 & 3.23e-13 & 1 & G024.7+00.6 & 3.23e-13 \\ 
\hline
\end{tabular}
\end{table}

\newpage
\section{Clouds in the Galactic Plane compared to H.E.S.S.}
\label{sec:hgpsclouds}

\begin{figure}
    \centering
    \includegraphics[width=23cm,angle=90]{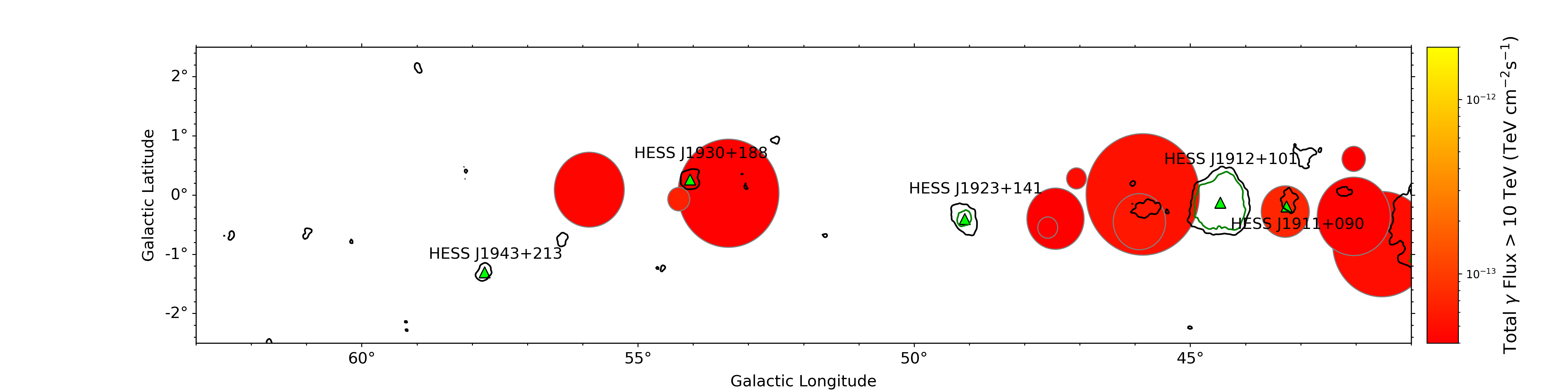}
    \hspace{2cm}
    \includegraphics[width=23cm,angle=90]{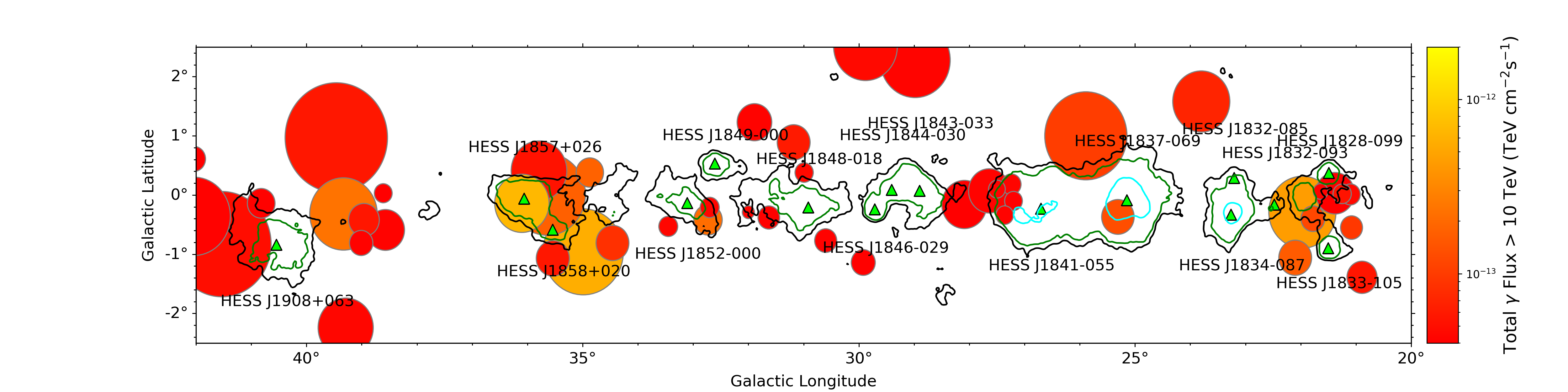}
    \vspace{-15mm}
    \caption{Predicted integral gamma-ray flux above 10\,TeV from interstellar clouds that could (under our model assumptions) be detectable by H.E.S.S.. Significance contours at 3, 5 and 15 sigma (black, green and cyan respectively) from the HGPS are shown together with the best fit positions of known H.E.S.S. sources as green triangles \citep{HGPS}. (1 of 4)  }
    \label{fig:GPzoomab}
\end{figure}

\begin{figure}
    \centering
    \includegraphics[width=23cm,angle=90]{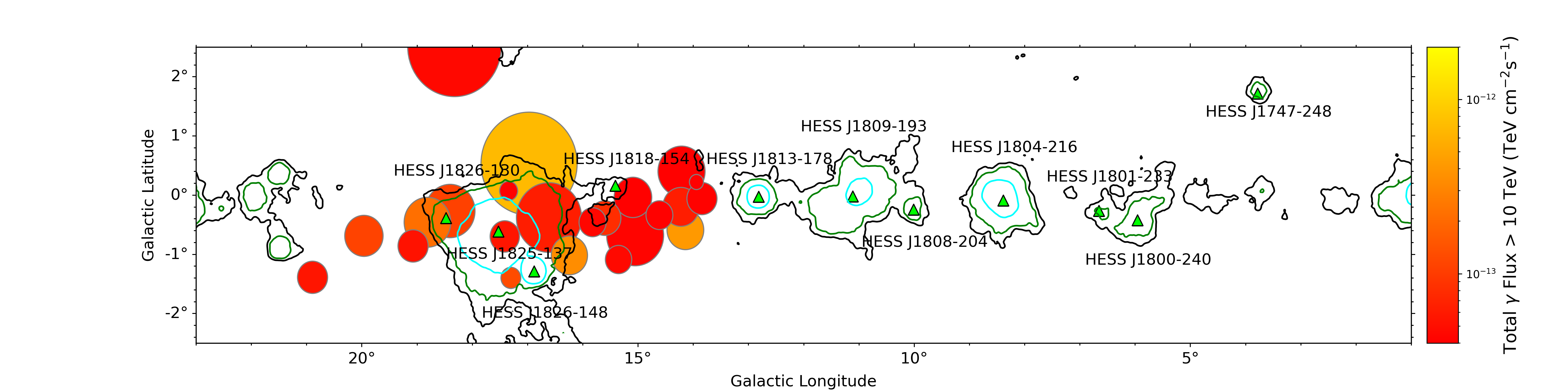}
    \hspace{2cm}
    \includegraphics[width=23cm,angle=90]{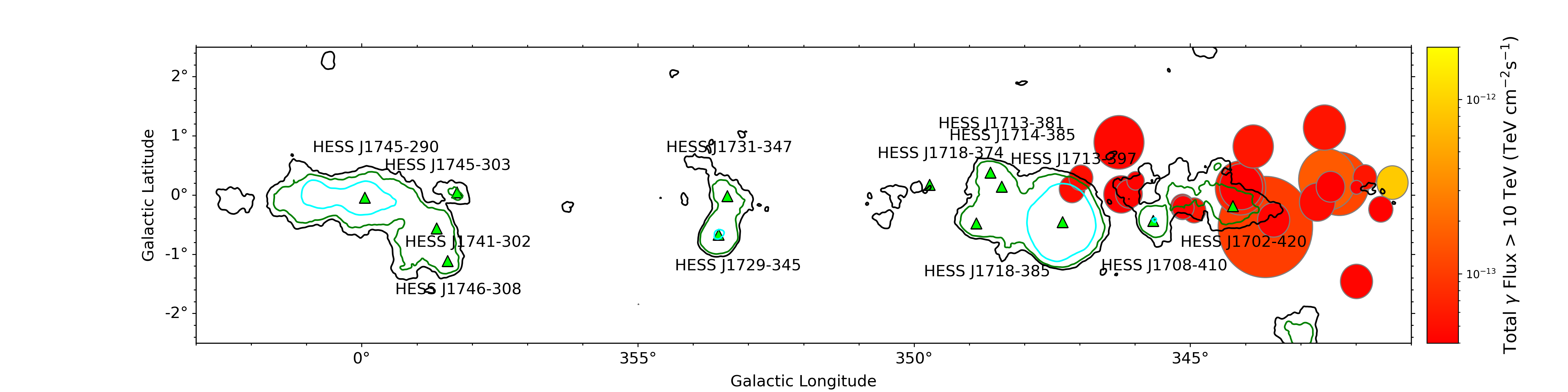}
    \vspace{-15mm}
    \caption{Galactic Plane continued from figure \ref{fig:GPzoomab}. (2 of 4)}
    \label{fig:GPzoomcd}
\end{figure}

\begin{figure}
    \centering
    \includegraphics[width=23cm,angle=90]{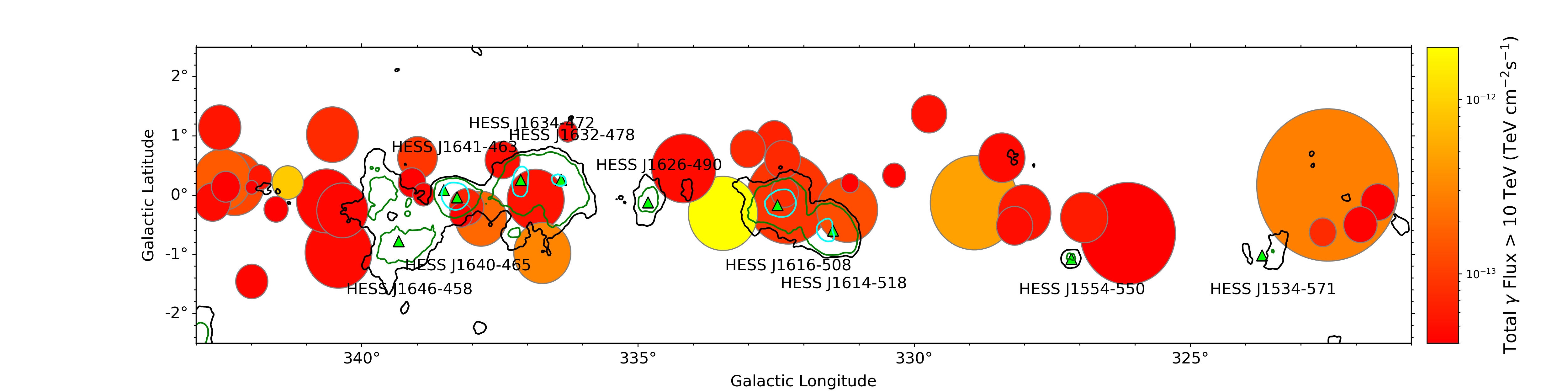}
    \hspace{2cm}
    \includegraphics[width=23cm,angle=90]{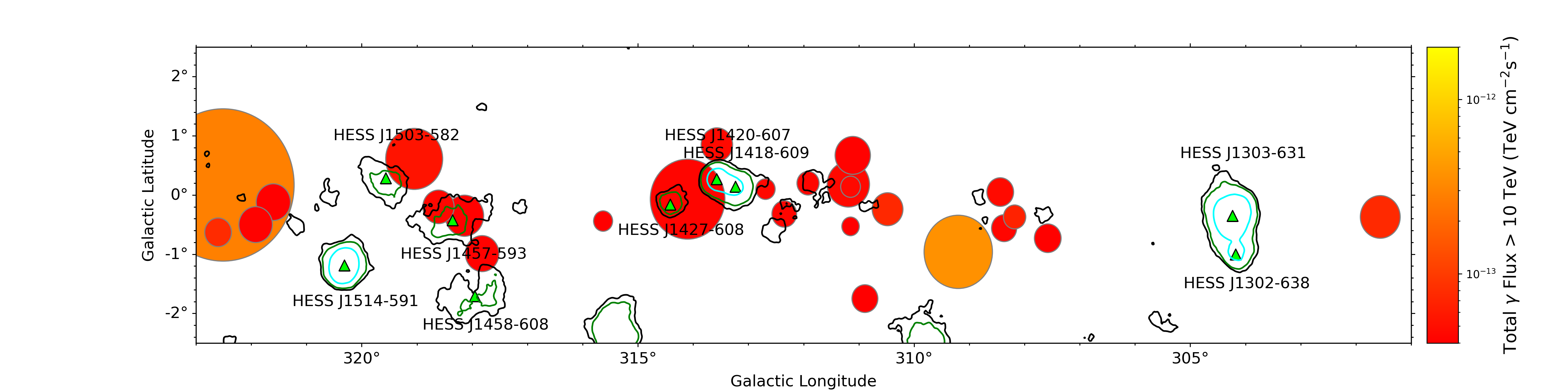}
    \vspace{-15mm}
    \caption{Galactic Plane continued. (3 of 4)}
    \label{fig:GPzoomef}
\end{figure}

\begin{figure}
    \centering
    \includegraphics[width=23cm,angle=90]{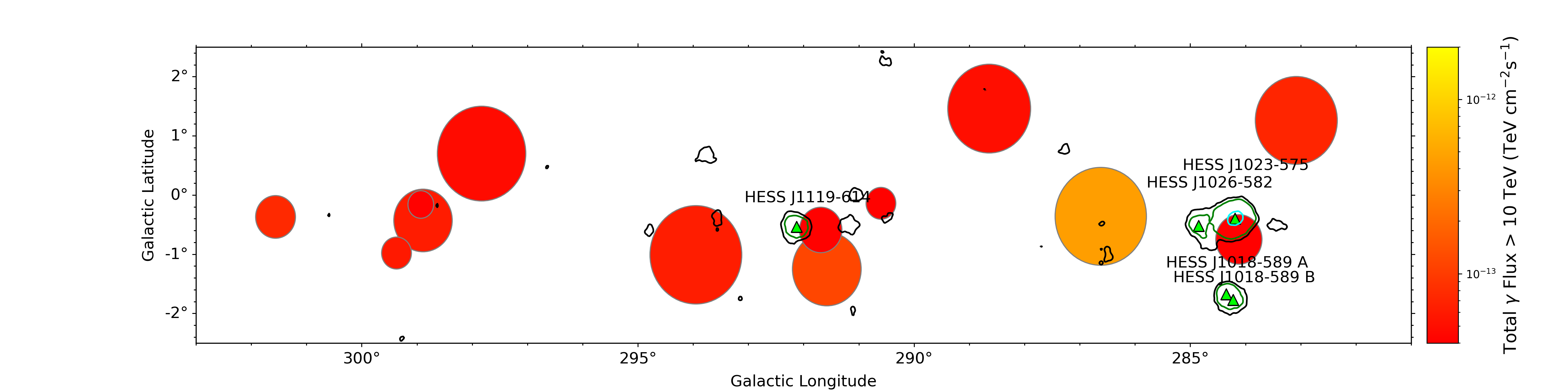}
    \hspace{2cm}
    \includegraphics[width=23cm,angle=90]{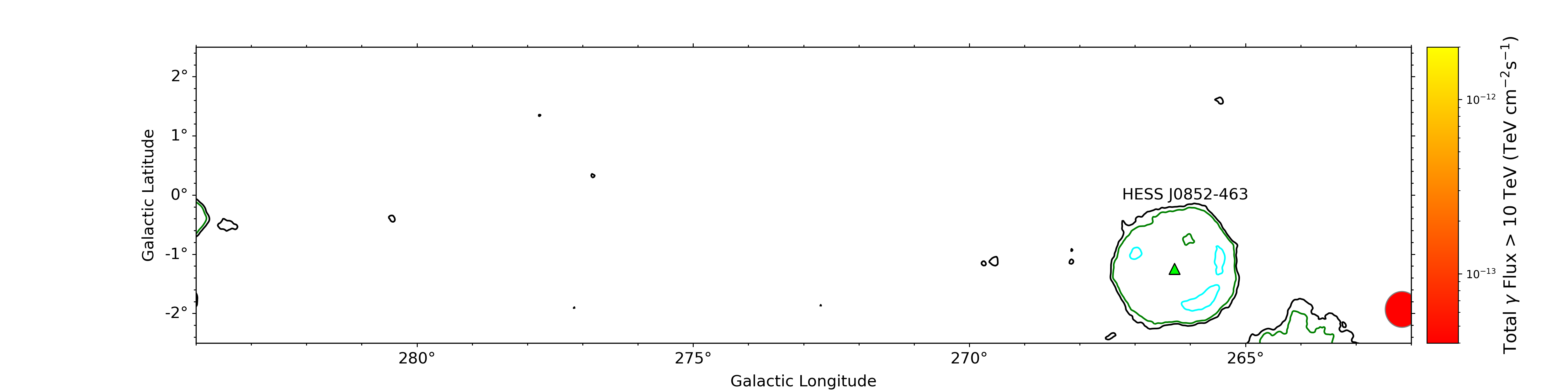}
    \vspace{-15mm}
    \caption{Galactic Plane continued. (4 of 4)}
    \label{fig:GPzoomgh}
\end{figure}

\newpage
\section{Predicted gamma-ray flux from clouds above 100\,TeV due to hypothetical SNRs traced by pulsars}
\label{sec:psrsnrgamma100}

\begin{table}
    \centering
    \caption{As for Table~\ref{tab:targets}, for the predicted flux and upper limits above 100 TeV due to cosmic rays escaped from hypothetical SNRs, as traced by pulsars. In this case, the J2000 name of the pulsar is given, and all pulsars have an associated distance estimate. Note that only fractional contributions from hypothetical SNRs are shown; fractional contributions to the total flux from known SNRs are not given here (see tables \ref{tab:targets} and \ref{tab:slowtargets}). All clouds in this table are predicted to have a flux above the CTA 50 hour sensitivity threshold in the slow diffusion case. }
    \begin{tabular}{c|l|l|l|l|l|c|l|l} 
		\hline 
	Cloud ID & Cloud coordinates & Cloud size & Cloud distance  & $EF^{UL}_\gamma >$ 100\,TeV & $EF^{\mathrm{total}}_\gamma$ $>$ 100\,TeV & \# PSRs & PSR & flux contribution\\
	& (l, b) deg & deg & kpc & TeV cm$^{-2}$ s$^{-1}$ & TeV cm$^{-2}$ s$^{-1}$ & & & TeV cm$^{-2}$ s$^{-1}$ \\
	 \hline
1 & (333.46,-0.31) & 0.627 & 3.370 & 1.78e-12 & 4.92e-12 & 1 & J1617-5055 & 1.45e-12 \\ 
2 & (16.97,0.53) & 0.870 & 2.080 & 2.34e-12 & 2.06e-12 & 1 & J1826-1256 & 5.98e-13 \\ 
4 & (21.97,-0.29) & 0.608 & 3.570 & 2.07e-12 & 1.77e-12 & 1 & J1833-1034 & 5.29e-13 \\ 
10 & (36.10,-0.14) & 0.492 & 3.400 & 5.26e-12 & 1.58e-12 & 1 & J1856+0113 & 1.21e-13 \\ 
11 & (286.62,-0.36) & 0.827 & 2.390 & 2.26e-12 & 1.47e-12 & 1 & J1048-5832 & 2.16e-13 \\ 
12 & (39.33,-0.32) & 0.611 & 3.840 & 1.45e-12 & 1.35e-12 & 2 & J1907+0631 & 2.56e-13 \\ 
& - & - & - & - & - & - & J1907+0602 & 1.73e-13 \\ 
15 & (16.24,-1.02) & 0.328 & 4.330 & 4.93e-13 & 1.18e-12 & 1 & J1821-1419 & 6.00e-14 \\ 
17 & (18.81,-0.46) & 0.422 & 4.510 & 1.07e-12 & 1.08e-12 & 1 & J1826-1334 & 1.84e-13 \\ 
18 & (331.21,-0.25) & 0.550 & 4.060 & 2.75e-12 & 1.06e-12 & 2 & J1614-5048 & 2.47e-13 \\ 
& - & - & - & - & - & - & J1617-5055 & 1.82e-13 \\ 
21 & (22.10,-1.06) & 0.297 & 4.210 & 8.32e-13 & 1.02e-12 & 1 & J1833-1034 & 2.78e-13 \\ 
 & (32.73,-0.42) & 0.257 & 2.910 & 9.56e-13 & 9.14e-13 & 1 & J1856+0113 & 3.19e-14 \\ 
23 & (34.87,0.38) & 0.248 & 3.210 & 5.78e-13 & 8.93e-13 & 1 & J1856+0113 & 1.02e-13 \\ 
24 & (21.78,-0.40) & 0.216 & 5.110 & 7.87e-13 & 8.32e-13 & 1 & J1833-1034 & 2.47e-13 \\ 
26 & (18.40,-0.27) & 0.451 & 3.600 & 3.29e-12 & 7.54e-13 & 1 & J1826-1334 & 2.14e-13 \\ 
27 & (19.96,-0.69) & 0.347 & 3.260 & 8.08e-13 & 7.53e-13 & 1 & J1833-1034 & 1.95e-13 \\ 
29 & (21.08,-0.55) & 0.199 & 4.530 & 4.00e-13 & 6.84e-13 & 1 & J1833-1034 & 2.09e-13 \\ 
30 & (17.30,-1.40) & 0.179 & 3.140 & 9.32e-12 & 6.77e-13 & 1 & J1826-1334 & 5.67e-14 \\ 
31 & (283.08,1.26) & 0.742 & 1.890 & 3.08e-12 & 6.48e-13 & 1 & J1023-5746 & 3.24e-13 \\ 
33 & (333.01,0.78) & 0.318 & 3.260 & 3.61e-13 & 6.00e-13 & 1 & J1617-5055 & 2.52e-13 \\ 
36 & (16.61,-0.38) & 0.589 & 3.650 & 2.45e-12 & 5.39e-13 & 1 & J1826-1334 & 2.69e-13 \\ 
37 & (14.22,-0.20) & 0.329 & 3.590 & 6.44e-13 & 5.18e-13 & 1 & J1813-1749 & 2.56e-13 \\ 
40 & (34.46,-0.81) & 0.298 & 3.460 & 3.01e-13 & 4.99e-13 & 1 & J1856+0113 & 9.07e-14 \\ %
41 & (332.53,0.93) & 0.327 & 4.730 & 4.48e-13 & 4.91e-13 & 2 & J1614-5048 & 1.33e-13 \\ 
& - & - & - & - & - & - & J1617-5055 & 6.50e-14 \\ 
42 & (332.34,0.04) & 0.248 & 4.250 & 5.85e-12 & 4.68e-13 & 1 & J1617-5055 & 8.35e-14 \\ 
43 & (54.26,-0.07) & 0.199 & 4.870 & 3.39e-13 & 4.58e-13 & 1 & J1930+1852 & 1.68e-13 \\ 
46 & (39.46,0.97) & 0.927 & 1.980 & 1.66e-12 & 4.41e-13 & 1 & J1907+0602 & 2.21e-13 \\ 
48 & (38.96,-0.42) & 0.278 & 2.640 & 7.32e-13 & 4.18e-13 & 2 & J1907+0631 & 1.39e-13 \\ 
& - & - & - & - & - & - & J1907+0602 & 7.00e-14 \\ 
49 & (15.60,-0.39) & 0.293 & 4.470 & 8.65e-13 & 4.04e-13 & 1 & J1821-1419 & 1.89e-14 \\ 
50 & (20.89,-1.39) & 0.273 & 3.680 & 2.28e-13 & 3.77e-13 & 1 & J1833-1034 & 1.76e-13 \\ 
56 & (344.93,-0.26) & 0.212 & 3.000 & 1.21e-12 & 3.39e-13 & 2 & J1708-4008 & 9.27e-14 \\ 
& - & - & - & - & - & - & J1702-4310 & 4.54e-14 \\ 
    \hline
    \end{tabular}
    \label{tab:psrsnrtargets}
\end{table}

\newpage
\section{Summary of Predicted Cloud Detectability}

\begin{table}
    \centering
    \caption{Summary of cloud detection prospects based on the gamma-ray flux predicted by this model under different assumptions. These are evaluated for the integral gamma-ray flux above 100\,TeV (compared to CTA), or 10\,TeV (compared to H.E.S.S.) respectively. Unless otherwise specified, the faster value for the diffusion coefficient $D_0$ from table \ref{tab:constants}, a particle spectrum index $\alpha=2$ and core collapse supernovae (as opposed to type Ia, see section \ref{sec:model}) are assumed. 
    } 
    \begin{tabular}{c|l|c|c|c|c|c|c|c}
    Cloud ID & Location (l, b) deg & $>100$ & $>100$ (slow $D_0$) & $> 10$ & $>10$ (slow $D_0$) & $>100$ (type Ia) & $>100$ ($\alpha=1.8$) & $>100$ ($\alpha=2.2$) \\
         \hline
1 & (333.46, -0.31) &   & \checkmark & \checkmark & \checkmark &  & \checkmark & \checkmark \\ 
2 & (16.97, 0.53) &   & \checkmark & \checkmark & \checkmark & \checkmark & \checkmark &  \\ 
3 & (110.43, 1.89) &   & \checkmark &  & \checkmark &  & \checkmark &  \\ 
4 & (21.97, -0.29) &   & \checkmark & \checkmark & \checkmark &  & \checkmark &  \\ 
5 & (336.73, -0.98) &   & \checkmark & \checkmark & \checkmark &  & \checkmark &  \\ 
6 & (34.99, -0.96) &   &   & \checkmark & \checkmark &  & \checkmark &  \\ 
7 & (337.84, -0.4) &   &   & \checkmark & \checkmark &  & \checkmark &  \\ 
8 & (345.57, 0.79) &   &   & \checkmark &  &  & \checkmark &  \\ 
9 & (341.34, 0.21) &   &   & \checkmark & \checkmark & \checkmark & \checkmark &  \\ 
10 & (36.1, -0.14) &   &   &   & \checkmark &  & \checkmark &  \\ 
11 & (286.62, -0.36) &   &   &   & \checkmark &  & \checkmark &  \\ 
12 & (39.33, -0.32) &   &   &   & \checkmark &  &  &  \\ 
13 & (328.91, -0.13) &   &   &   & \checkmark &  & \checkmark &  \\ 
14 & (14.14, -0.59) &   &   &   & \checkmark &  & \checkmark &  \\ 
15 & (16.24, -1.02) &   &   &   & \checkmark &  & \checkmark &  \\ 
16 & (309.2, -0.96) &   &   &   & \checkmark &  & \checkmark &  \\ 
17 & (18.81, -0.46) &   &   &   & \checkmark &  &  &  \\ 
18 & (331.21, -0.25) &   &   &   & \checkmark &  & \checkmark &  \\ 
19 & (106.66, 1.16) &   &   &   & \checkmark &  & \checkmark &  \\ 
20 & (322.51, 0.17) &   &   &   & \checkmark &  & \checkmark &  \\ 
21 & (22.1, -1.06) &   &   &   & \checkmark &  & \checkmark &  \\ 
22 & (32.73, -0.42) &   &   &   & \checkmark &  & \checkmark &  \\ 
23 & (34.87, 0.38) &   &   &   & \checkmark &  & \checkmark &  \\ 
24 & (21.78, -0.4) &   &   &   & \checkmark &  & \checkmark &  \\ 
25 & (35.64, 0.01) &   &   &   & \checkmark &  & \checkmark &  \\ 
26 & (18.4, -0.27) &   &   &   & \checkmark &  &  &  \\ 
27 & (19.96, -0.69) &   &   &   & \checkmark &  & \checkmark &  \\ 
28 & (342.52, 0.26) &   &   &   & \checkmark &  & \checkmark &  \\ 
29 & (21.08, -0.55) &   &   &   & \checkmark &  &  &  \\ 
30 & (17.3, -1.4) &   &   &   & \checkmark &  & \checkmark &  \\ 
31 & (283.08, 1.26) &   &   &   & \checkmark &  &  &  \\ 
32 & (25.31, -0.37) &   &   &   & \checkmark &  & \checkmark &  \\ 
33 & (333.01, 0.78) &   &   &   & \checkmark &  &  &  \\ 
34 & (342.3, 0.19) &   &   &   & \checkmark &  & \checkmark &  \\ 
35 & (291.58, -1.25) &   &   &   & \checkmark &  &  &  \\ 
36 & (16.61, -0.38) &   &   &   & \checkmark &  &  &  \\ 
37 & (14.22, -0.2) &   &   &   & \checkmark &  &  &  \\ 
38 & (343.64, -0.54) &   &   &   & \checkmark &  &  &  \\ 
39 & (25.89, 1.0) &   &   &   & \checkmark &  &  &  \\ 
40 & (34.46, -0.81) &   &   &   & \checkmark &  & \checkmark &  \\ 
41 & (332.53, 0.93) &   &   &   & \checkmark &  &  &  \\ 
42 & (332.34, 0.04) &   &   &   & \checkmark &  &  &  \\ 
43 & (54.26, -0.07) &   &   &   & \checkmark &  &  &  \\ 
44 & (338.99, 0.63) &   &   &   & \checkmark &  & \checkmark &  \\ 
45 & (332.28, -0.07) &   &   &   & \checkmark &  &  &  \\ 
46 & (39.46, 0.97) &   &   &   & \checkmark &  &  &  \\ 
47 & (142.4, 1.38) &   &   &   & \checkmark &  &  &  \\ 
48 & (38.96, -0.42) &   &   &   & \checkmark &  &  &  \\ 
49 & (15.6, -0.39) &   &   &   & \checkmark &  &  &  \\ 
50 & (20.89, -1.39) &   &   &   & \checkmark &  &  &  \\ 
51 & (322.6, -0.63) &   &   &   & \checkmark &  &  &  \\ 
52 & (340.53, 1.02) &   &   &   & \checkmark &  &  &  \\ 
53 & (332.38, 0.6) &   &   &   & \checkmark &  & \checkmark &  \\ 
54 & (301.56, -0.37) &   &   &   & \checkmark &  &  &  \\ 
55 & (310.48, -0.24) &   &   &   & \checkmark &  &  &  \\ 
56 & (344.93, -0.26) &   &   &   & \checkmark &  &  &  \\ 
57 & (344.09, 0.12) &   &   &   & \checkmark &  &  &  \\ 
58 & (43.28, -0.28) &   &   &   & \checkmark &  &  &  \\ 
59 & (23.8, 1.58) &   &   &   & \checkmark &  &  &  \\ 
60 & (340.42, -0.97) &   &   &   &   &   & \checkmark &  \\ 
61 & (326.13, -0.65) &   &   &   &   &   & \checkmark &  \\ 
\hline
\end{tabular}
    \label{tab:cloud_detectability}
\end{table}


\bsp	
\label{lastpage}
\end{document}